\documentclass[preprint,12pt,authoryear,review]{elsarticle}

\usepackage{amssymb}
\usepackage{lineno}

\usepackage{tabularx}
\usepackage{rotating}

\usepackage{hyperref} 

\usepackage{changes}
\setdeletedmarkup{\textcolor{red}{\sout{#1}}}
\usepackage{caption}

\usepackage{bm}

\newcolumntype{Y}{>{\raggedright\arraybackslash}X}

\journal{Icarus}

\begin{document}

\begin{frontmatter}



%\title{Implications of Realistic Crustal Rheology and Intrusive Magmatism for the Tectonics of Venus}
\title{The Tectonics and Volcanism of Venus: New Modes Facilitated by Realistic Crustal Rheology and Intrusive Magmatism}


\author[inst1]{Jiacheng Tian\corref{cor1}}
\author[inst1]{Paul J. Tackley}
\author[inst1]{Diogo L. Louren\c{c}o}

\affiliation[inst1]{organization={Institute of Geophysics, Department of Earth Sciences, ETH Zurich},%Department and Organization
            city={Zurich},
            postcode={8092}, 
            country={Switzerland}}
\cortext[cor1]{Correspondence to: \href{jiacheng.tian@erdw.ethz.ch}{jiacheng.tian@erdw.ethz.ch}}

% \affiliation[inst2]{organization={Department Two},%Department and Organization
%             addressline={Address Two}, 
%             city={City Two},
%             postcode={22222}, 
%             state={State Two},
%             country={Country Two}}

\begin{abstract}

To explain Venus’ young surface age and lack of plate tectonics, Venus’ tectonic regime has often been proposed to be either an episodic-lid regime with global lithospheric overturns, or an equilibrium resurfacing regime with numerous volcanic and tectonic activities. Stratigraphic analysis suggests that Venus’ surface tectonics could be a combination of these two end-member models with a global resurfacing event that created most of the crust followed by tectonic and volcanic activities until now. Recent analyses of Venus’ satellite images also suggest widespread lithospheric mobility in the lowland basins that does not fit the episodic-lid tectonic regime. Here, we use global 2-D thermochemical convection models with realistic parameters, including rheology (dislocation creep, diffusion creep, and plastic yielding), an experiment-based plagioclase (An$_{75}$) crustal rheology, and intrusive magmatism, to investigate the tectonics and mantle evolution of Venus. We find that surface tectonics is strongly affected by crustal rheology. With a ``weak'' plagioclase-rheology crust, models exhibit episodic overturns but with continuously high surface mobility and high distributed surface strain rates between overturns, leading to a new tectonic regime that we name ``deformable episodic lid". On the other hand, olivine-crustal-rheology models exhibit either standard episodic-lid tectonics, i.e. with mobility that is high  during overturns and near zero between overturns, or stagnant-lid tectonics, i.e. with near-zero mobility over the entire model time. In models with plagioclase-rheology crust, magmatic resurfacings are short-lived and randomly located, leading to a rather uniform resurfacing rate. Also, a combination of plagioclase crustal rheology and dislocation creep can weaken the lithosphere sufficiently to facilitate lithospheric overturns without applying plastic yielding. Internally, the composition-dependent density profile results in a ``basalt barrier'' at the mantle transition zone, which strongly affects Venus’ mantle evolution. Only strong plumes can penetrate this basalt barrier and cause global lithospheric overturns. This basalt barrier also causes global internal episodic overturns that generate global volcanic resurfacing in stagnant-lid models, which suggests a new resurfacing mechanism that does not involve lithospheric overturns. We name this regime ``stagnant episodic-volcanic-resurfacing". Results in models with plagioclase crustal rheology and low eruption efficiencies best fit current estimates of crustal thickness, surface age, and tectonic structures based on observations of Venus. The crustal thickness in models with an olivine-rheology crust is limited by the basalt-eclogite phase transition and is higher than observation-based estimates. Higher intrusion rates better fit these estimates, as models with high eruption efficiency predict very low surface age ($<$ 100 Myr).

\end{abstract}


\begin{highlights}
\item Realistic rheology and intrusive magmatism influence the modelled tectonics of Venus.
\item A weak crustal rheology leads to thinner crust and more uniform resurfacing rates.
\item Dislocation creep and intrusions can cause global overturn without plasticity.
\item A ``basalt barrier'' below the MTZ leads to global internal episodic overturns.
\item Internal overturns can cause episodic global resurfacing in a stagnant-lid regime. 

\end{highlights}

\begin{keyword}
%Venus \sep mantle evolution \sep tectonics \sep compositional mantle stratification \sep geodynamic modelling
Venus \sep Venus interior \sep Tectonics \sep Volcanism \sep Geophysics
\end{keyword}

\end{frontmatter}

%\linenumbers

\section{Introduction}
\label{sec:intro}
Venus is of particular interest in planetary science because of its Earth-like size, density, distance from the Sun, and active, rocky surface, all of which suggest a comparable bulk composition between Venus and Earth. From images from NASA's Magellan mission, the observations of $\sim$1000 impact craters on the surface of Venus lead to an estimated surface age between 300 Myr and 1 Gyr \citep[e.g.][]{hauckVenusCraterDistribution1998, mckinnonCrateringVenusModels1997,nimmoVolcanismTectonicsVenus1998}. Some studies suggest an even younger surface age of 150 - 240 Myr \citep{herrickPostimpactModificationVolcanic2011, lefeuvreNonuniformCrateringMoon2011} based on new analyses of the crater population and a revised crater chronology of the Solar System. These estimates suggest that Venus is the only terrestrial planet in the Solar System that has a resurfacing age comparable to that of Earth. However, the nature of the current tectonics of Venus that leads to a young and relatively uniform surface remains unclear.

The near random distribution of the impact craters on Venus' surface suggests a fairly uniform surface age for Venus, with the age of about 80\% of the surface being not significantly different from the average surface age \citep{ivanovHistoryTectonismVenus2015}. Also most of these craters are not significantly altered by volcanism and tectonics \citep{stromGlobalResurfacingVenus1994}. Such a uniform surface can hardly be explained by Earth-like plate tectonics. Thus, two end-member models have been proposed to explain Venus' tectonics: the catastrophic resurfacing model and the equilibrium resurfacing model. The catastrophic resurfacing model suggests that a global catastrophic resurfacing event occurred in a relatively short period of time and reworked most of the surface of Venus \citep[e.g.][]{stromGlobalResurfacingVenus1994,romeoResurfacingVenus2010}. The equilibrium resurfacing model argues that the crater distribution results from numerous randomly distributed volcanic or tectonic events, and can also explain these cratering observations \citep[e.g.][]{bjonnesEquilibriumResurfacingVenus2012}. These two end-member models correspond to different evolution paths for volcanic and tectonic activities, which can be reflected in variations in the history of the resurfacing rate and surface mobility.

In the absence of plate tectonics, most of the plains on Venus' surface are believed to be resurfaced by volcanism \citep{stofanResurfacingStylesRates2005}, which favors the equilibrium resurfacing model. Some observations indicate that volcanism could remain active until today, such as the high surface emissivity at topographic rises \citep{smrekarRecentHotspotVolcanism2010} and lava flows \citep{brossierDistinctMineralogyAge2021}, a locally thinned lithosphere at Aramaiti Corona \citep{russellEvidenceLocallyThinned2021}, and spikes of sulfur dioxide in the atmosphere of Venus \citep[e.g.][]{marcqVariationsSulphurDioxide2013}. Numerical models of coronae formation also suggest that some coronae could be currently active, possibly due to ongoing plume activities \citep{gulcherCoronaStructuresDriven2020}. However, although these volcanic features on Venus' surface show evidence of regional-scale resurfacing, the formation of large volcanic plains on Venus can hardly be explained by the equilibrium resurfacing models \citep{romeoResurfacingVenus2010}.  

The evolution of Venus' surface mobility is reflected by tectonic features. From stratigraphic analyses of Venus' surface, the tectonic and volcanic units are interpreted to form at two different periods \citep{ivanovHistoryTectonismVenus2015,kreslavskyResurfacingHistoryVenus2015}, which suggests that Venus could have experienced both catastrophic and equilibrium resurfacing events. For most of these units, the crater densities are similar to the mean crater density of Venus' surface and are interpreted to form at a similar time with a global resurfacing event. Some units, such as those related to the network rifting-volcanism regime in \citet{ivanovHistoryTectonismVenus2015} and the post-regional-plain units in \citet{kreslavskyResurfacingHistoryVenus2015}, are significantly younger. In addition, there is evidence of high lithospheric mobility preserved on Venus' surface, including tesserae \citep[e.g.][]{hansenTesseraTerrainCrustal1999} and compressional structures around Lakshmi Planum \citep{harrisInteractionsContinentlikeDrift2015}. Recent analysis of satellite images of Venus' lowland plains also suggests that there are fragmented lithospheric blocks that collide and rotate relative to each other, indicating possibly widespread surface mobility over time \citep{byrneGloballyFragmentedMobile2021} which is not consistent with a purely stagnant-lid regime. These observations led to the inference that Venus could have been operating in an intermediate regime between catastrophic and equilibrium resurfacing, with one global overturn followed by continuous regional-scale resurfacing and tectonic events.

Numerical studies on the global tectonics of Venus have mainly focused on the two end-member models described above \citep[e.g.][]{orourkeThermalEvolutionVenus2015, armannSimulatingThermochemicalMagmatic2012,rolfInferencesMantleViscosity2018, uppalapatiDynamicsLithosphericOverturns2020}. These models often assume an olivine composition and diffusion-creep rheology for Venus' crust, and neglect intrusive magmatism. Furthermore, most of the two-dimensional (2-D) and three-dimensional (3-D) numerical models focused on the episodic-lid regime, with near-global lithospheric overturns and near-stagnant-lid phases between overturns. However, a near-stagnant-lid phase after the last global overturn cannot explain the presence of young tectonic units on Venus' surface. Pure stagnant-lid models also cannot explain Venus' surface age and crustal thickness, because when magmatism is purely eruptive, heat transfer is dominated by the magmatic heat-pipe mechanism, resulting in high volcanic resurfacing rates and very young surface ages that are inconsistent with estimates based on cratering density, as demonstrated by the stagnant-lid simulations of \citet{armannSimulatingThermochemicalMagmatic2012}. 
Recently, a new tectonic regime named ``plutonic-squishy lid'' was proposed by \citet{lourencoPlutonicSquishyLidNew2020}. In this regime, intrusive magmatism warms, weakens and thins the lithosphere and facilitates surface deformation and mobility. Interpretations that Venus' lowland plains are fragmented and moving through time based on satellite images \citep{byrneGloballyFragmentedMobile2021}, and that the lithospheric thickness and heat flow of Venus' coronae are similar to those at rifting zones on Earth \citep{smrekar2022}, are more consistent with this tectonic regime than with episodic or stagnant-lid regimes. For a review of Venus' mantle dynamics and tectonic regime see \citet{rolfDynamicsEvolutionVenus2022}. 

Although it is generally accepted that Venus' crust is basaltic \citep[e.g.][]{surkovNewDataComposition1984}, the rheology of this basaltic crust is uncertain. Basalt is mainly made up of two groups of minerals, plagioclase feldspar and pyroxene, which have quite different creep strengths. Compared to olivine, plagioclase is typically considered to be weaker, while pyroxene is typically considered to be stronger. \citet{mackwellHightemperatureDeformationDry1998} measured the rheology of dry diabase (the same composition as basalt but larger grain size), and found that its strength is dependent on the plagioclase fraction, but in any case is likely to result in a high viscosity crust that is strongly coupled to the mantle. On the other hand, \citet{azumaRheologicalDecouplingMoho2014} argued that the crust of Venus is likely weaker than the mantle. The observations of crustal deformation discussed above, as well as the low to moderate ($<$40 km) effective elastic thickness for extensive regions of Venus' lithosphere  \citep[e.g.][]{jimenez-diazLithosphericStructureVenus2015}, are also evidence for a relatively weak crustal rheology for the planet. Using regional numerical models, \citet{geryaPlumeinducedCrustalConvection2014} and \citet{gulcherCoronaStructuresDriven2020} were able to obtain a good match to the observed coronae and novae structures using a plagioclase An$_{75}$ crustal rheology, rather than the olivine crustal rheology used by previous numerical models. Dislocation creep was also included in their models, which has a further weakening effect.

In this work, we aim to (1) constrain Venus' crustal rheological parameters by comparing the results of numerical experiments using different crustal rheologies to observations, and (2) understand how crustal rheology and intrusive magmatism affect Venus' tectonics and evolution. 

\section{Methodology}
\label{sec:method}

\subsection{Numerical Model and Boundary Conditions}
The thermo-chemical evolution of the interior of Venus is modelled using the finite-volume code StagYY \citep{tackleyModellingCompressibleMantle2008} in a 2-D spherical annulus geometry \citep{hernlundModelingMantleConvection2008}. The physical parameters and the numerical model are based on those of \citet{armannSimulatingThermochemicalMagmatic2012} and \citet{lourencoPlutonicSquishyLidNew2020}, with parameters listed in Table \ref{tab:modelpar}. The conservation equations for a compressible fluid with an infinite Prandtl number are solved on a fully staggered grid for 4.5 Gyr model time, i.e. from 0 Gyr to the present day at 4.5 Gyr. The model domain is resolved by $512\times 96$ cells with radial grid refinement up to a factor of two at the surface and the core-mantle boundary (CMB), giving $\sim$ 13 km grid spacing at the surface and $\sim$ 25 km at the CMB in the radial direction. The grid spacing in the azimuthal direction ranges from $\sim$ 38 km at the CMB to $\sim$ 74 km at the surface. About half a million tracers (about 10 tracers per cell) are used to track the non-diffusive advection of composition, temperature, and heat-producing elements (HPE) within the domain. Solid and molten tracers can be created or merged during partial melting and freezing such that arbitrarily small melt fractions can be generated or frozen even with very few initial tracers per cell. Tests indicate that increasing the number of tracers makes no qualitative difference to the results (see Supplementary Material).

Free-slip mechanical boundary conditions are applied both on the upper and lower boundaries of the domain. Thermally, a constant temperature of 740 K is applied at the top boundary, while the bottom boundary is isothermal and cools with time due to removal of heat from the core. The parameterisation of core cooling is similar to that in \citet{nakagawaEffectsThermochemicalMantle2004, nakagawaInfluenceInitialCMB2010}, which takes into account gravitational energy released by inner core growth and latent heat release. Internal heating is produced by HPE that initially have a uniform concentration over the whole domain but subsequently partition between solid and melt. The partition coefficient for HPE is set to be 0.001, which means that the concentration of HPE in the melt is 1000 times that in the coexisting solid. The initial temperature profile is adiabatic with a potential temperature of 1800 K, thermal boundary layers 50 km thick at top and bottom, and random temperature perturbations of peak amplitude 20 K throughout the domain. The initial CMB temperature of 4025 K is identical or similar to that used in previous models \citep{armannSimulatingThermochemicalMagmatic2012,gillmannAtmosphereMantleCoupling2014,rolfInferencesMantleViscosity2018,uppalapatiDynamicsLithosphericOverturns2020}; tests with higher values of up to 6025 K indicate that the initial CMB temperature only influences the early part of the evolution and not the long-term dynamics  (see Supplementary Material), consistent with previous models for Earth \citep{nakagawaInfluenceInitialCMB2010}.

\begin{table}[!thbp]
    \centering
    \scriptsize
    \begin{tabular}{ p{5.5cm} p{2.25cm}p{2.25cm}p{2cm}}
        \hline
        Parameter                        & Symbol                    & Value                   & Units                    \\
        \hline
        Mantle domain depth              & $D$                       & 2942                    & km                       \\
        CMB radius                       & $R_{\mathrm{CMB}}$        & 3110                    & km                       \\
        Gravitational acceleration       & g                         & 8.87                    & $\mathrm{m/s^2}$         \\
        Surface temperature              & $T_{\mathrm{surf}}$       & 740                     & K                        \\
        Initial CMB temperature          & $T_{\mathrm{CMB\_ init}}$ & 4025                    & K                        \\
        Specific heat capacity           & $C_p$                     & 1000                    & J/($\mathrm{kg\cdot K})$ \\
        Initial internal heating rate    & $ H_{\mathrm{init}}$      & $18.77 \times 10^{-12}$ & W/kg                     \\
        Half life of heat-producing elements (HPE) & $t_{\mathrm{half}}$     & 2.43                    &Gyr\\
        Partition coefficient of HPE     & $D_{\mathrm{HPE}}$        & 0.001                    &                          \\
        Initial mantle potential temperature     & $T_{\mathrm{potential}}$  & 1800                    & K                        \\
        
        \hline
    \end{tabular}
    \caption{Physical parameters used in this study, mostly following \cite{armannSimulatingThermochemicalMagmatic2012}. $R_{\mathrm{CMB}}$ is from \cite{stevensonMagnetism1983}, $C_p$ from \cite{gulcherCoronaStructuresDriven2020}; heat-producing isotopes of K, U and Th are lumped together into a single ``HPE'' component with $D_{\mathrm{HPE}}$ being the order-of-magnitude partition coefficient (e.g. \cite{xieEvolutionUPbSmNd2004}), $t_{\mathrm{half}}$ an average half-life \cite{nakagawaCoreEvo2005} and the heating rate $H$ is based on Bulk Silicate Earth models \citep{palmeTreatise2003}.}
    \label{tab:modelpar}
\end{table}

\subsection{Composition and Phase Transitions}
\begin{figure}[!htbp]
    \centering
    \includegraphics[width=\textwidth]{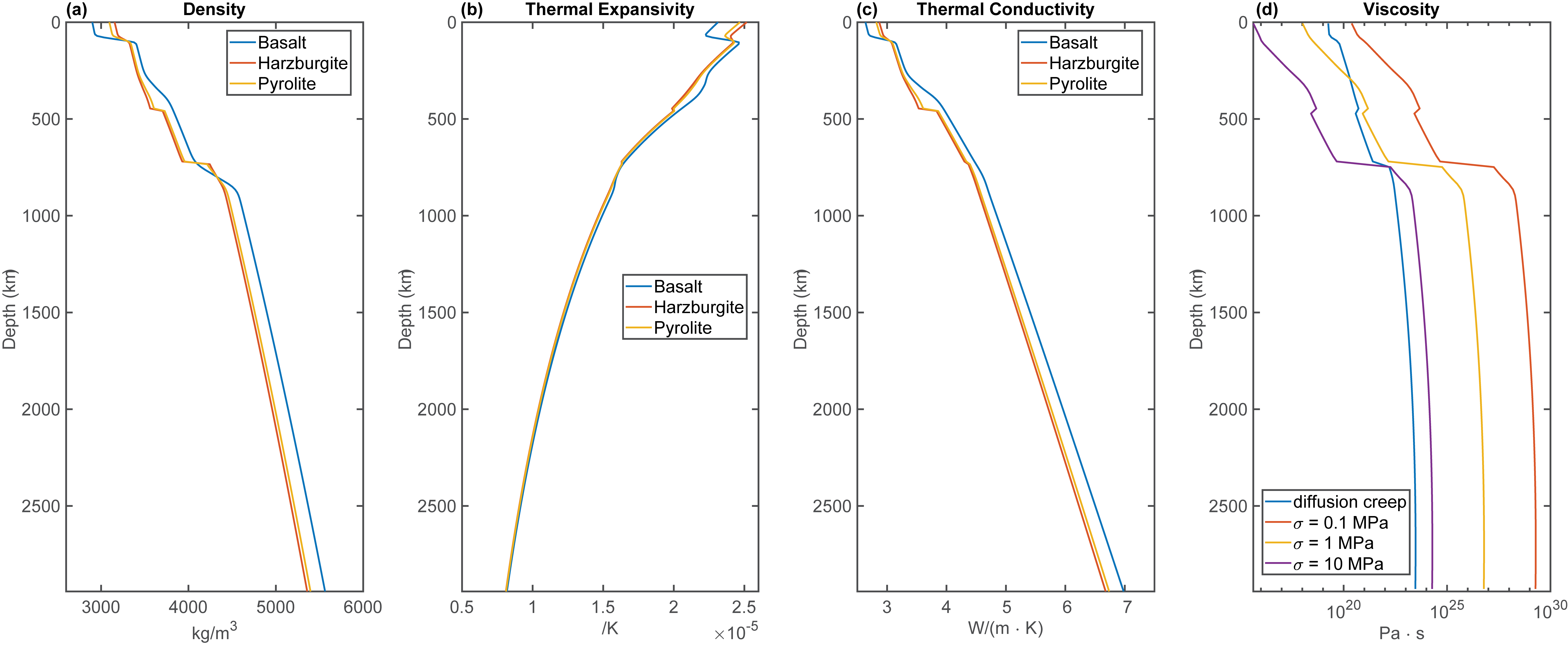}
    \caption{(a - c) Profiles of density, thermal expansivity, and thermal conductivity for basalt (100\% pyroxene-garnet), harzburgite (75\% olivine + 25\% pyroxene-garnet) and pyrolite. (d) Viscosity profile for olivine along the 1600 K reference adiabat, including diffusion creep viscosity and dislocation creep viscosity under different stresses ($\sigma$). }
    \label{fig:refdensity}
\end{figure}

The parameterisation of composition is based on mineralogy: material is divided into olivine (ol) and pyroxene-garnet (px-gt) phase systems. The composition varies between two end members, basalt and harzburgite, which are linear mixtures of the ol and px-gt mineral systems. Basalt is assumed to be $100\%$ pyroxene-garnet, and harzburgite is assumed to be $75\%$ olivine and $25\%$ pyroxene-garnet. An initial composition of pyrolite ($20\%$ basalt and $80\%$ harzburgite) is assumed for the entire mantle. 
Both the ol and px-gt systems undergo pressure- and temperature-dependent solid-solid phase transitions. The depth, reference temperatures, density jumps, and Clapeyron slopes for the phase transitions (see Appendix, Table \ref{tab:phasechange}) are identical to those used in \citet{lourencoPlutonicSquishyLidNew2020} except for those of the basalt-eclogite transition (in the px-gt system), which has parameters based on those in \citet{gulcherCoronaStructuresDriven2020}. For each phase, the density is calculated using a third-order Birch-Murnaghan equation of state. As detailed in \citet{tackleyMantleDynamicsSuperEarths2013}, thermal expansivity is calculated from the bulk modulus and the Gruneisen parameter using a standard relationship, while the thermal conductivity has a power-law dependence on density. The overall properties of each grid cell are the arithmetic average of the properties of phases within that cell. The resulting depth-dependences of density, thermal expansivity, and thermal conductivity with the reference adiabat are shown in Figure \ref{fig:refdensity}.

\subsection{Rheology}
A visco-plastic rheology is assumed (see Appendix, Table \ref{tab:rheo}). Viscous deformation includes both diffusion creep and dislocation creep, with the composite viscosity ($\eta_{\mathrm{viscous}}$) given by 
\begin{equation}
    \frac{1}{\eta_{\mathrm{viscous}}}=\frac{1}{\eta_{\mathrm{diffusion}}}+\frac{1}{\eta_{\mathrm{dislocation}}}.
\end{equation}

The diffusion creep viscosity ($\eta_{\mathrm{diffusion}}$) is governed by an Arrhenius-type law.
\begin{equation}
    \eta_{\mathrm{diffusion }}(T, P)=\eta_{0} \Delta \eta_i \exp \left(\frac{E_{a,i}+P V_i}{R T}-\frac{E_{a,i}}{R T_{0}}\right), 
\end{equation}
where $T$ and $P$ are temperature and pressure, the subscript $i$ represents different phases, $\eta_0$ is the reference viscosity at $T_0 = 1600$ K and zero pressure, $\Delta \eta_i$ is a phase-dependent viscosity offset (calculated at phase change T and P to give the relative viscosities listed in Table \ref{tab:rheo}), $E_a$ and $V$ are the activation energy and activation volume for diffusion creep, and R = 8.314 $\mathrm{J\ mol^{-1}K^{-1}}$ is the gas constant. The dislocation-creep viscosity, on the other hand, is calculated as: 
\begin{equation}
    \eta_{\mathrm{dislocation}}(T, P, \sigma)=\eta_0 \Delta \eta_i \left(\frac{\sigma_0}{\sigma}\right)^{n_i-1} \exp \left(\frac{E^{\prime}_{a,i}+P V^{\prime}_i}{R T}-\frac{E^{\prime}_{a,i}}{R T_{0}}\right),
\end{equation}
where $E^{\prime}_{a,i}$, $V^{\prime}_i$, and $n_i$ correspond to the activation energies and activation volumes for dislocation creep, and stress exponents for different phases. $\sigma$ is the second invariant of the deviatoric stress tensor and $\sigma_0$ is the crossover stress between diffusion creep and dislocation creep: $\sigma_0 = 0.3 \ \textrm{MPa}$ for upper mantle phases (based on the below-cited laboratory measurements and for a grain size of 0.825 mm), and $\sigma_0 = 3\ \textrm{MPa}$ for lower mantle phases (slightly higher than typical convective stresses based on the common view that Earth's lower mantle is dominated by diffusion creep, e.g. \cite{ammannFirstprinciplesConstraintsDiffusion2010}) and consistent with experimental results \citep{tsujinoRheology2022}. The activation volumes for perovskite phases are pressure-dependent \citep{tackleyMantleDynamicsSuperEarths2013} and follow:
\begin{equation}
    V(P) = V_0 \exp \left(- \frac{P}{P_{\mathrm{decay}}}\right).
\end{equation}

For the shallowest phase in the px-gt system, which represents basalt before the basalt-eclogite transition, a plagioclase (An$_{75}$) rheology from \citet{ranalliRheologyEarth1995} is applied, as in \citet{gulcherCoronaStructuresDriven2020}. For other phases in the ol and px-gt systems, the rheological parameters are similar to the parameters used in \citet{lourencoMeltinginducedCrustalProduction2016a}, which are based on data for olivine from \citet{karatoRheologyUpperMantle1993} for the upper mantle and bridgmanite (\citet{yamazakiMineralPhysicsConstraints2001,ammannDFTStudyMigration2009}) for the lower mantle. Two viscosity jumps are applied in our models (Table \ref{tab:rheo}): a viscosity jump of 3.0 for the transition from upper-mantle phases to lower-mantle phases (similar to that in current Earth viscosity models \citep{cizkova2012} and compatible with geoid and admittance ratios on Venus \citep{rolfInferencesMantleViscosity2018}, and a factor of 0.0133 for the transition from eclogite to plagioclase when the plagioclase rheology is applied to crust (calculated from the above-cited laboratory measurements). These viscosity jumps are applied at the relevant phase transition (T,P) and result in different $\Delta \eta_i$ in Equation (2) applicable at the reference (T,P). A reference viscosity of $1.0 \times 10^{20} \  \mathrm {Pa \cdot s}$ is used for all our models. In a grid cell that contains different phases, the overall viscosity of the cell is calculated as the geometric mean of the viscosities of the individual phases; the geometric mean is used because it is in-between the limiting cases \citep{yamazakiMineralPhysicsConstraints2001} of arithmetic mean (strong phase dominates by forming a load-bearing framework) and harmonic mean (weak phase dominates by forming interconnected weak layers).

Plastic yielding is also included in some of our models and can act to weaken and mobilise the lithosphere \citep{moresiMantleConvectionBrittle1998,tackleySelfconsistentGenerationTectonic2000}. The yield stress in our models follows a Drucker-Prager yield criterion:
\begin{equation}
    \sigma_y = C +\phi P
\end{equation}
where $\phi$ is the friction coefficient and $C$ is the cohesion. The effective viscosity $\eta_\mathrm{effective}$ is calculated as:
\begin{equation}
    \eta_{\mathrm{effective}} = \min \left(\eta_{\mathrm{viscous}}, \: \frac{\sigma_y}{2 \dot{\epsilon}}\right),
\end{equation}
where $\dot{\epsilon}$ is the second invariant of the strain rate tensor.
Finally, the viscosity is truncated between the lower and upper limits of $10^{18}$ and $10^{26}$ $\mathrm{Pa\cdot s}$, respectively.

\subsection{Melting}
The parameterisation of melting and crust production in the model is extended from that in \citet{lourencoPlutonicSquishyLidNew2020}. Depth-dependent solidus and liquidus curves are used to calculate melt production, with the solidus applying to pyrolite (20\% basalt) and increasing linearly as basalt is depleted. The solidus is fitted to \citet{herzbergNewExperimentalObservations2000} for the upper mantle and \citet{zerrSolidusEarthDeep1998} for the lower mantle. Melt that forms at depths shallower than the depth of neutral density ($\sim$330 km on Venus) is either erupted or intruded. A fraction of the produced melt given by the ``eruption efficiency'' input parameter is placed on the surface and the rest is intruded below the crust, the base of which is detected for each vertical column of the grid cells. Any composition (from basalt to harzburgite) can melt, with the basaltic component melting first before any harzburgitic component melts. Only melt with a pure basaltic composition can be erupted or intruded. To simulate fast cooling due to magma convection in regions where the melt fraction exceeds the rheological transition (i.e. solid disaggregates), we use an effective thermal conductivity for regions that have large melt fractions with a formulation as in \citet{lourencoPlutonicSquishyLidNew2020} with a maximum value of $1000 \ \mathrm{W/(m\cdot K)}$.

In reality, the viscosity of molten rocks is much lower than corresponding solid rocks, for example $10^2-10^{13} \ \mathrm{Pa\cdot s}$ for hydrous leucogranitic melts \citep{hessViscositiesHydrousLeucogranitic1996}. However, due to computational limitations, the viscosity of melts in our models is artificially set to be the minimum viscosity in the domain ($10^{18} \ \mathrm{Pa \cdot s}$). Melts that are not erupted or intruded, like those deeper than 330 km, are advected with the flow and may later freeze.

\subsection{Diagnostics}
In this study, we computed the surface age and crustal thickness to compare our results with inferences from satellite observations and previous numerical models. The surface age is calculated as the average time since the eruption of the tracers within the outermost grid cells of the computational domain. This average age depends on the depth extent (radial thickness) of these outermost grid cells, as younger, shallower tracers are averaged together with older, deeper tracers. That is, there can be sub-grid stratification of tracer age within each outermost grid cell due to different eruptions, with newer tracers being placed on the top of the older tracers, and these sub-grid stratifications are ignored by averaging in grid cells. This also neglects the possibility that non-erupted tracers are present in the top cells. Therefore, radial grid refinement is applied near the top boundary to minimise the influence of grid resolution. The crustal thickness, on the other hand, is the thickness of solid basaltic material at the top of the domain, and is detected during each time step in each radial column of the grid. Eruption rates are calculated from model outputs to measure the rate of crustal production. As all models have a 2-D geometry, eruption rates are calculated as the thickness of crust erupted at the surface within a given time, with units $\mathrm{km/Gyr}$.

We also calculate surface mobility to characterise the tectonic activity at the surface. The surface mobility $M$ is defined as the ratio of the root mean square (rms) of surface velocity to the rms velocity averaged over the entire domain, in the no-net-lithospheric-rotation frame of reference \citep[][]{tackleySelfconsistentGenerationTectonic2000}:
\begin{equation}
    M=\frac{\left(v_{\mathrm{rms}}\right)_{\mathrm{surface }}}{\left(v_{\mathrm{rms}}\right)_{\mathrm{domain}}}.
\end{equation}

For stagnant-lid tectonics, $M$ is near zero, due to negligible surface velocities. For mobile-lid and plutonic-squishy-lid tectonics, M is larger than zero, reflecting either a moving lithosphere mostly pulled by subduction \citep{nakagawaInfluencePlateTectonic2015} or lithospheric movements facilitated by intrusions \citep{lourencoPlutonicSquishyLidNew2020}. Episodic-lid tectonics, on the other hand, is characterised by occasional lithospheric overturns with high surface velocities (mobile-lid phase, $M > 1$), and low surface velocities between overturns (stagnant-lid phase, $M \approx 0$). Many numerical studies focussing on the catastrophic resurfacing model suggest that Venus is in the episodic-lid regime \citep[e.g.][]{armannSimulatingThermochemicalMagmatic2012,rolfInferencesMantleViscosity2018, uppalapatiDynamicsLithosphericOverturns2020} or transitioning from mobile-lid to stagnant-lid \citep[][]{wellerPhysicsChangingTectonic2020}, whereas \citet{byrneGloballyFragmentedMobile2021} suggests that Venus could be in a plutonic-squishy-lid regime \citep{lourencoPlutonicSquishyLidNew2020} from satellite images of lowland plains. For a review see \citet{rolfDynamicsEvolutionVenus2022}.

\subsection{Parameter Study}
In this study, the two main properties that we tested are crustal rheology and eruption efficiency. As discussed in the introduction, previous studies \citep[e.g.][]{gulcherCoronaStructuresDriven2020, jimenez-diazLithosphericStructureVenus2015} suggest that a relatively weak plagioclase (An$_{75}$) rheology \citep{ranalliRheologyEarth1995} may be appropriate for modelling Venus' crust. Therefore, in terms of rheology, we tested (1) the influence of this ``weak'' crustal rheology compared to the previously-used olivine-based ``strong'' rheology \citep[e.g.][]{armannSimulatingThermochemicalMagmatic2012}, (2) the influence of including dislocation creep, as most previous global models have typically assumed only the diffusion creep mechanism, and (3) the influence of increasing the brittle failure strength (yield strength). 

In addition, different eruption efficiencies ($E$) are tested in our models. Eruption efficiency is defined as the percentage of melt that is erupted to the surface, with the remaining melt being intruded below the crust. Previous geodynamic models that included melting often assumed that all melt erupts to the surface (i.e., $E = 100\%$). However, for Earth's magmatism, the estimate of eruption efficiency is only 10 - 20\% on average, depending on tectonic setting \citep{crispRatesMagmaEmplacement1984}. The highest estimate for eruption efficiency in \citet{crispRatesMagmaEmplacement1984} is about $60 \%$. Also, values of 20 \% extrusion efficiency have been found to better match observations for the early Earth \citep[e.g.][]{rozelContinentalCrustFormation2017}, which is thought to be an analogue for modern-day Venus. Therefore, pure eruptive magmatism would not be a realistic approximation. Recent studies \citep{lourencoEfficientCoolingRocky2018, lourencoPlutonicSquishyLidNew2020} found that a low eruption efficiency (high intrusion efficiency) could lead to a new tectonic regime, the plutonic-squishy-lid tectonic regime, which could explain several Venus' observations \citep{byrneGloballyFragmentedMobile2021,smrekar2022}. Therefore, we vary the eruption efficiency from 20\% to 100\% to test how it influences volcanic and tectonic activity in our models. 

\section{Results}
\label{sec:result}
Here we present 14 numerical models with different rheological parameters and eruption efficiencies (summarised in Table \ref{tab:result}). We first present the reference model, E20P03Pl, which has a ``weak'' plagioclase crustal rheology, dislocation creep, standard yielding parameters similar to \citet{gulcherCoronaStructuresDriven2020}, and 20\% eruption efficiency (80\% intrusion). We then examine the influence of differences in rheology and the effect of eruption vs. intrusion of melt on the surface and mantle dynamics of Venus.

\begin{sidewaystable}
%\begin{table}[!thbp]
        \centering
        \tiny
        \begin{tabularx}{\textwidth}{XXXXXXXXXXX}       
            \hline
            Model&Crustal Rheology&Dislocation Creep & Cohesion (MPa) & Eruption Efficiency & Number of Global Overturns&Surface Age (Myr)&Crustal Thickness (km) & Topography Std. Dev. (km) & Average Eruption Rate in last 10 Myr) (km/Gyr) & Surface Conductive Heat Flux ($\mathrm{mW/m^2}$)\\
            \hline
            \textbf{E20P03Pl}    & An$_{75}$  & yes   & 0.3 & $20\%$&10 & $\bm{285 \pm 141}$ & $\bm{36.7 \pm 12.4}$  & $ \bm{\pm 0.92}$& $3.46$ & $44.37$\\
            E20P03Ol    & Ol    & yes  & 0.3 & $20\%$& 6  & $\bm{510 \pm 119}$ & $69.7 \pm 8.7$  & $ \pm 3.32$ & $5.22$ & $26.60$\\
            E20P03Pl\_D & An$_{75}$  & no  & 0.3 & $20\%$ & $12^*$ & $25 \pm 23$ & $85.3 \pm 62.3$  &$ \pm 3.61$& $331.46$ & $90.59$\\
            E20P03Ol\_D & Ol    & no  & 0.3 & $20\%$ & 5 & $45 \pm 32$ & $110.4 \pm 41.1$  & $ \pm 3.16$& $316.90$ & $17.48$\\
            E20P00Pl    & An$_{75}$  & yes   & no plasticity  &$20\%$& 8  & $\bm{194\pm 71}$ & $\bm{42.8 \pm 15.7}$&$ \pm 1.50$& $11.62$ & $37.73$\\
            E20P50Pl$^a$& An$_{75}$  & yes   & 50 & $20\%$ & 10  & $\bm{220 \pm 120}$     & $\bm{35.3 \pm 17.6}$ & $ \pm 1.20$ & $134.42$ & $48.00$\\
            E20P00Ol    & Ol    & yes     & no plasticity   & $20\%$& 0  & $\bm{576 \pm 370}$   & $74.9 \pm 8.1$   & $\bm{\pm0.85}$ & $3.38$ & $35.72$ \\
            E20P50Ol    & Ol    & yes      & 50 & $20\%$& 6      & $\bm{263 \pm 79}$& $76.7 \pm 7.2$   & $ \pm 1.65$& $2.05$ & $30.32$\\
            E40P03Pl    & An$_{75}$  & yes    & 0.3 & $40\%$ &$10^*$ & $87 \pm 45$ & $\bm{34.2 \pm 20.0}$ & $\bm{ \pm 0.98}$ & $36.20$  &$47.98$\\
            E80P03Pl    & An$_{75}$  & yes    & 0.3 & $80\%$ &$9^*$ & $70 \pm 47$  & $\bm{18.9 \pm 14.3}$ & $\bm{\pm1.09}$ & $43.26$ & $51.63$\\
            E100P03Pl   & An$_{75}$  & yes    & 0.3 & $100\%$ &$11^*$ & $36 \pm 21$ & $\bm{16.2 \pm 15.2}$ & $ \bm{\pm0.94}$ & $109.34$  &$61.19$\\
            E40P03Ol    & Ol    & yes  & 0.3 & $40\%$& $7^*$  & $\bm{248 \pm 61}$   & $70.1 \pm 7.2$   &$ \bm{\pm 0.82}$& $9.14$ & $31.18$\\
            E80P03Ol    & Ol    & yes  & 0.3 & $80\%$& $12^*$  & $65 \pm 28$   & $\bm{50.3 \pm 22.0}$  & $ \pm 1.45$&  $191.48$  &$36.62$\\
            E100P03Ol   & Ol    & yes  &0.3 & $100\%$& $14^*$  & $27 \pm 23$   & $\bm{41.9 \pm 30.4}$  & $ \pm2.05$& $472.26$  & $48.16$\\

            \hline
            \multicolumn{10}{l}{$^a$: an overturn occurs at 4.41 Gyr, so the results are taken from 4.39 Gyr} \\
        \end{tabularx}
        \caption{Results of number of global overturns, average surface age, crustal thickness, topography and surface conductive heat flux at 4.5 Gyr, and average eruption rate in last 10 Myr for models in this study. Values that fit current estimates of average surface age (150 - 1000 Myr), average crustal thickness (10 - 60 km), and observed topography (standard deviation = 0.995 km, model outputs within $20 \%$ difference) are marked with bold fonts. The preferred model that matches current estimates is E20P03Pl. For some models (marked with asterisks ($^*$)), the number of global overturns is hard to count as there could be a series of regional overturns that reset the average surface age to zero in a relatively short period of time. The results for surface age and crustal thickness include the average values and standard deviations at the end of simulation (4.5 Gyr), apart from model E20P50Pl (which is before the last global overturn).}
        \label{tab:result}
%\end{table}
\end{sidewaystable}

\subsection{Behaviour of the Reference Model}
\begin{figure}[!p]
    \centering
    \includegraphics[width=\textwidth,height=0.9\textheight,keepaspectratio]{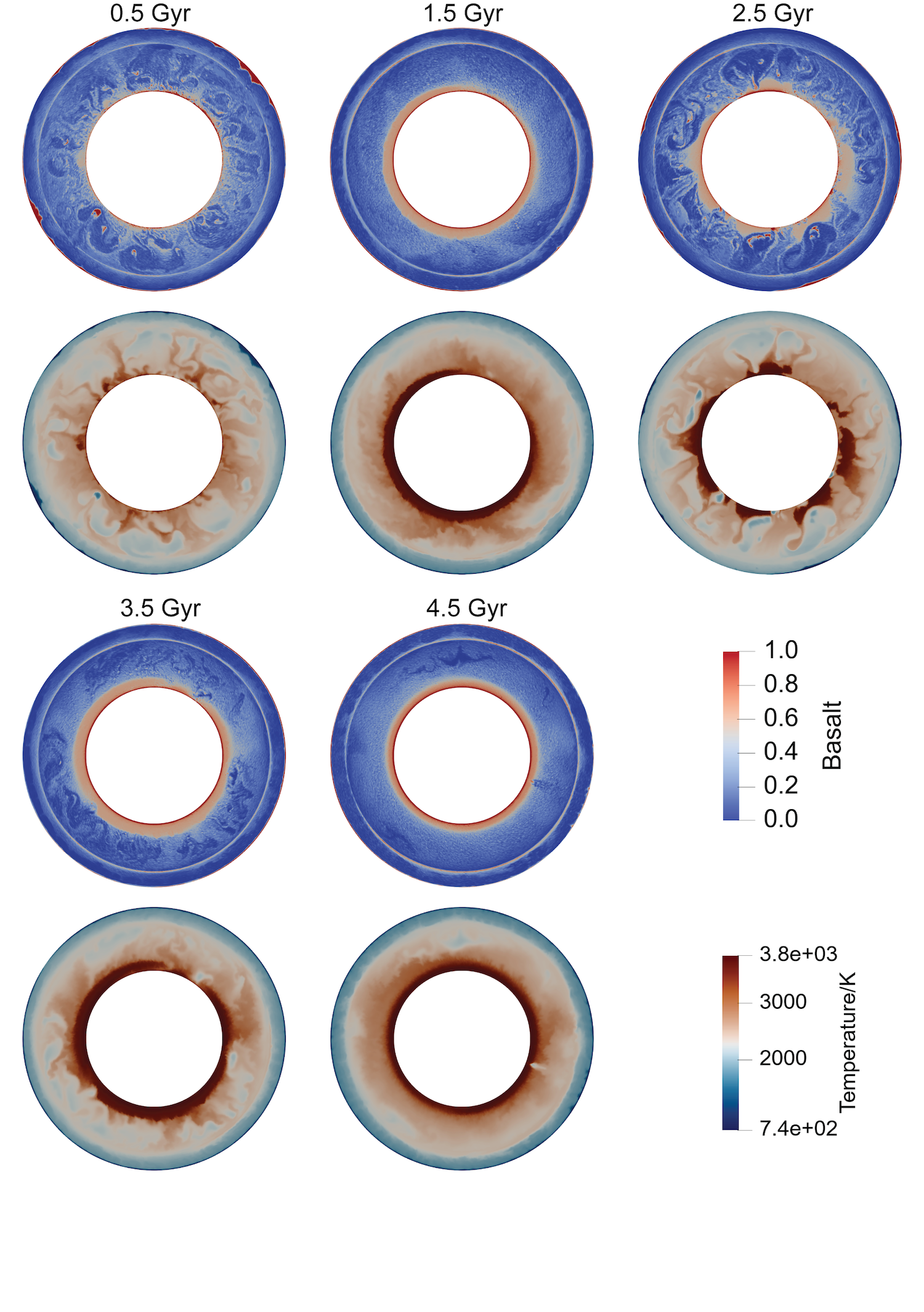}
    \caption{Evolution of basalt fraction and mantle temperature fields for the reference model E20P03Pl.}
    \label{fig:M0}
\end{figure}

\begin{figure}[!htbp]
    \centering
    \includegraphics[width=\textwidth]{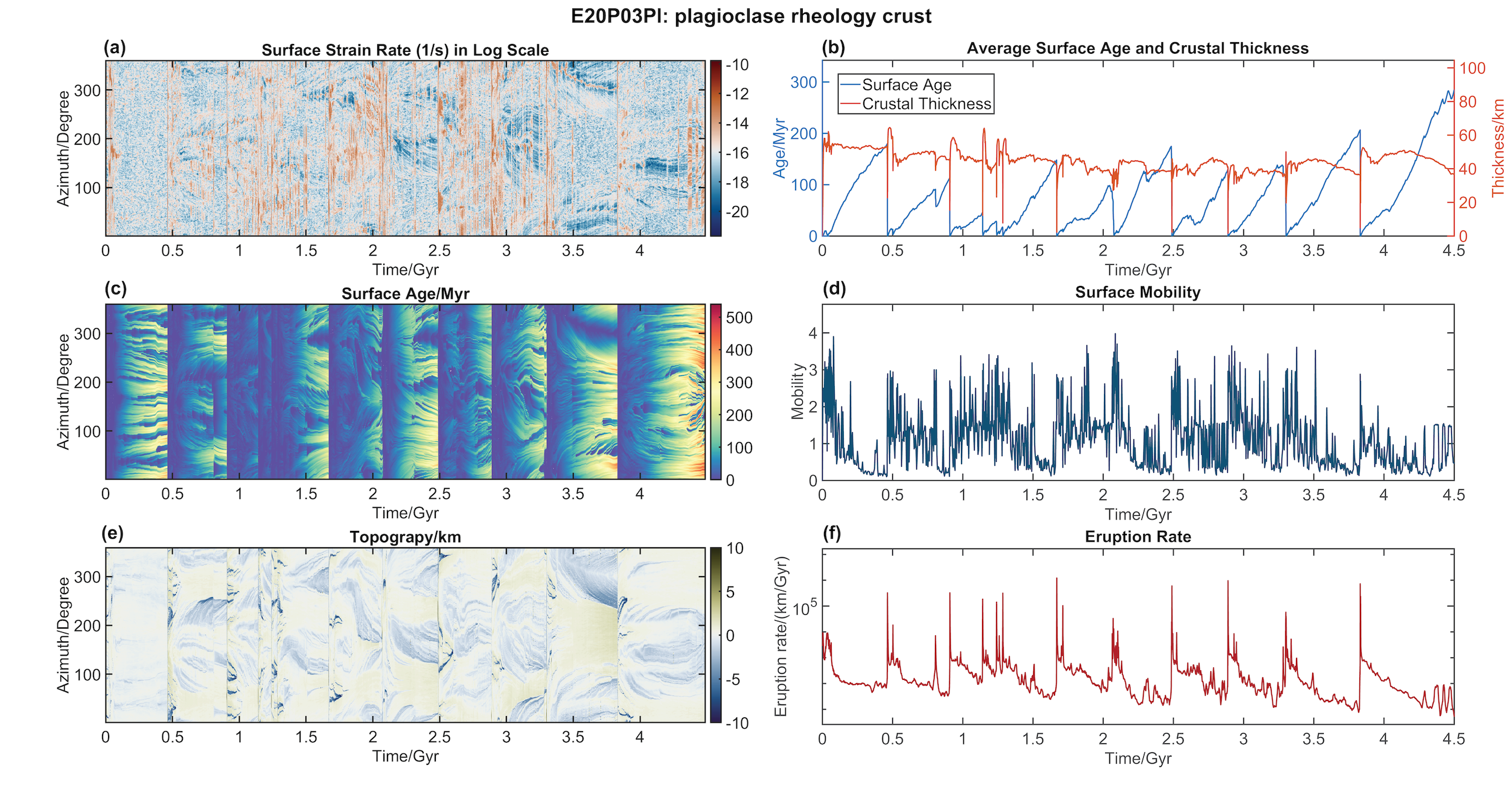}
    \caption{For reference model E20P03Pl, the evolution of (a) surface strain rate, (b) average surface age and crustal thickness, (c) surface age, (d) surface mobility, (e) topography, and (f) eruption rates over 4.5 Gyr. Eruption rate is shown as the thickness of crust erupted per Gyr.}
    \label{fig:M0_Surf}
\end{figure}

\begin{figure}[!htbp]
    \centering
    \includegraphics[width=\textwidth]{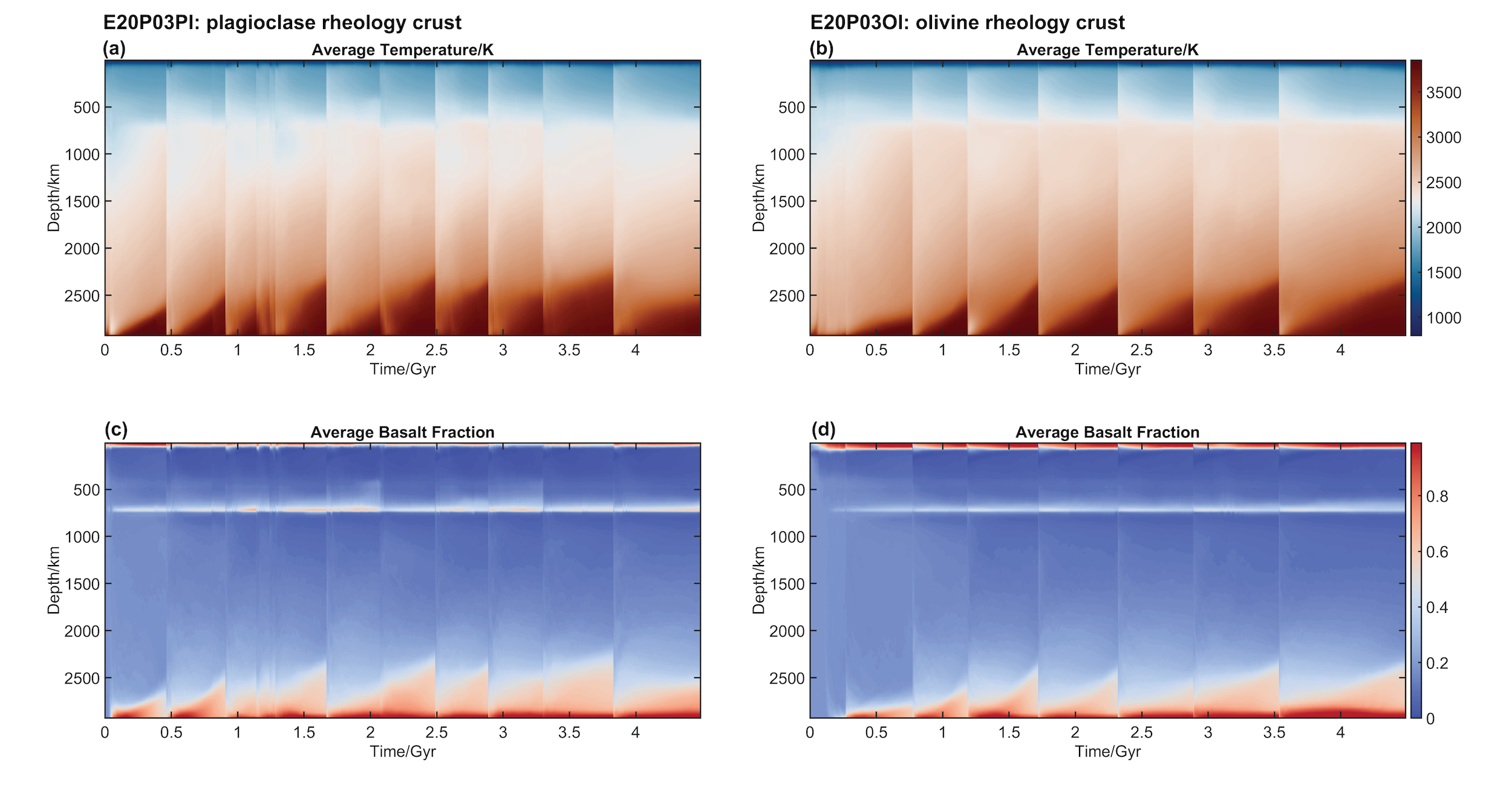}
    \caption{The evolution of depth profiles of azimuthally-averaged (a)(b) temperature and (c)(d) basalt fraction over 4.5 Gyr. (a) and (c) are for reference model E20P03Pl, while (b) and (d) are for E20P03Ol, which is the same but with a strong olivine crustal rheology. When a global lithospheric overturn occurs, the temperature difference between upper and lower mantles is effectively homogenized. The basalt fraction above the CMB also significantly decreases during global overturns. The basaltic layer at the base of the MTZ forms within the first 0.5 Gyr and exists for the whole 4.5 Gyr model time.}
    \label{fig:M0_M1_BST}
\end{figure}

The evolution of the reference model E20P03Pl can be seen in Figure \ref{fig:M0} and is characterised by global overturns, but with continuous high surface mobility in between and high surface strain rates through 4.5 Gyr (Figure \ref{fig:M0_Surf}). The evolution of the composition field (Figures \ref{fig:M0} and \ref{fig:M0_M1_BST}) indicates that the basaltic crust remained relatively thin over 4.5 Gyr, reflecting an equilibrium between crust production from decompression melting and crust removal due to the basalt-eclogite phase transition and convective stresses from mantle flow. A part of the delaminated crust remains at the base of the mantle transition zone (MTZ) and forms a basaltic layer at about 730 km depth, due to the positive buoyancy of basalt in the depth range between the olivine-system phase transition to perovskite and the pyroxene-garnet system phase transition to perovskite, as previously found in other simulations \citep[][]{ogawaChemicalStratificationTwodimensional2003,tackleyNumericalLaboratoryStudies2005,nakagawaMassTransportMechanism2005,nakagawaSelfConsistentMineralPhysics2010} and named the ``basalt barrier'' \citep[][]{daviesBasaltBarrier2008,papucBasaltBarrierVenus2012}. Convective flows in the mantle are disrupted by this basalt barrier, resulting in a thermally and compositionally stratified mantle (Figure \ref{fig:M0_M1_BST}) in our models.

\begin{figure}[!htbp]
    \centering
    \includegraphics[width=\textwidth]{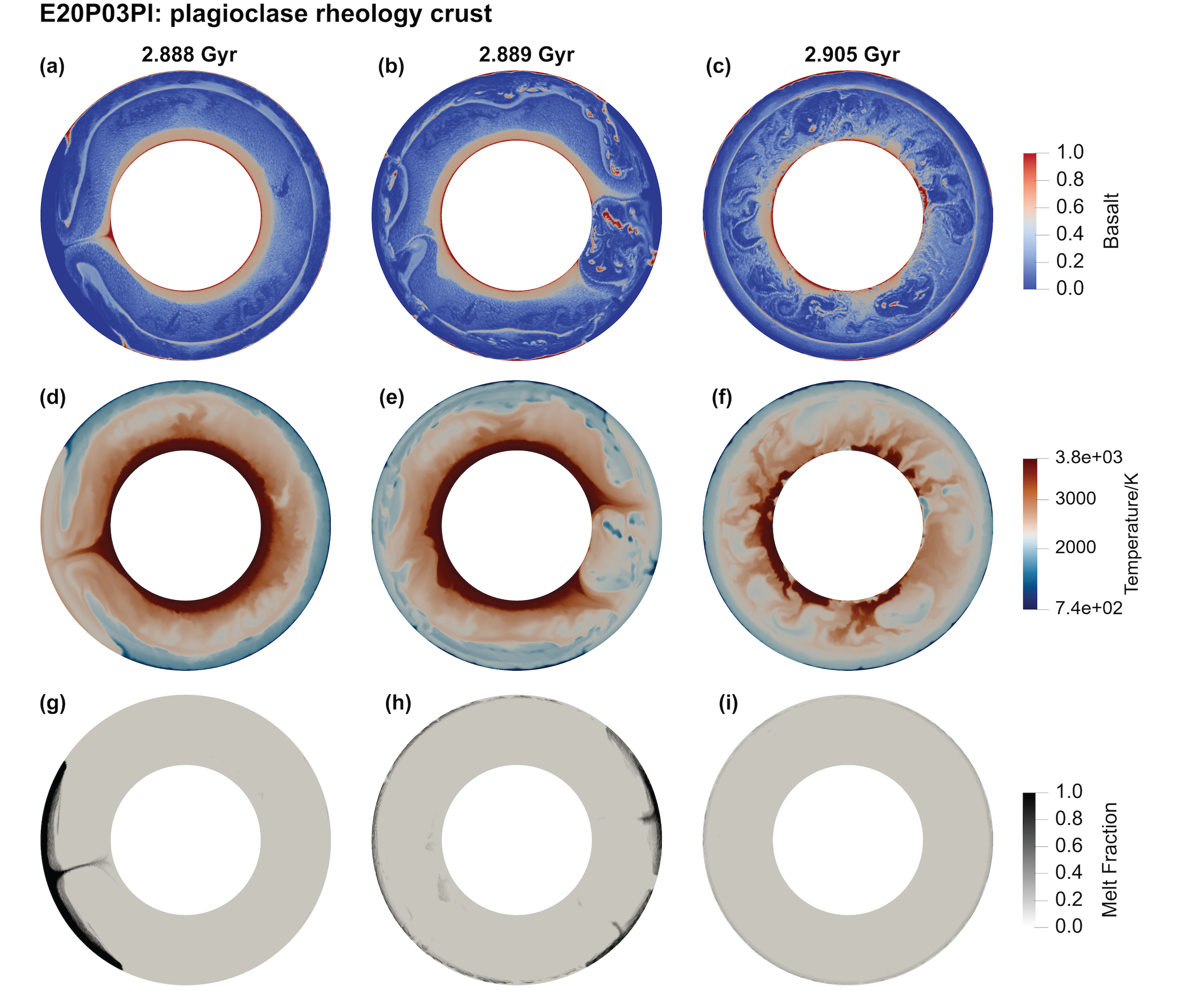}
    \caption{Snapshots of composition (basalt) (a - c), temperature (d - f) and melt fraction (g - i) for a global overturn at 2.89 Gyr for reference model E20P03Pl. The global overturn is triggered by a mantle plume from the CMB that can break through the basalt barrier at the MTZ. The overturn is followed by extensive melt production at the surface, which creates new basaltic crust within a few million years after the overturn.}
    \label{fig:M0_overturn}
\end{figure}

Global lithospheric overturns are triggered only by mantle flows that can break through the basalt barrier at the base of the MTZ, and most of these are mantle upwellings from the CMB (Figure \ref{fig:M0_overturn}). When a global lithospheric overturn occurs, the whole mantle becomes mixed, as shown by the relatively homogeneous temperature field immediately after overturns, as well as the reduced basalt fraction above the CMB (Figure \ref{fig:M0_overturn}). The average interval between global overturns for model E20P03Pl is approximately 0.4 Gyr.

Between overturns, both surface mobility and surface strain rate (Figure \ref{fig:M0_Surf}) indicate continuous surface deformation, which is contrary to the near-zero surface mobility between overturns in previous episodic-lid models. The average surface strain rate between overturns in model E20P03Pl ($\mathrm{\sim 10^{-16} \ s^{-1}}$) is higher than the strain rate estimated from impact craters ($\mathrm{10^{-18} - 10^{-17} \ s^{-1}}$) on Venus' surface \citep{grimmRecentDeformationRates1994}. At 4.5 Gyr, we observe small old blocks separated by regions with low surface age and high strain rates (Figure \ref{fig:M0_Surf} (a) and (c)), which could agree with the fragmented lowland basins observed on Venus' surface \citep{byrneGloballyFragmentedMobile2021}.

Magmatism also occurs between overturns, as indicated by regions with high strain rates and near-zero surface age in Figure \ref{fig:M0_Surf} (a) and (c). However, as these features do not persist over time, there are no persistent mantle upwellings in model E20P03Pl, but rather randomly-located, short-lived magmatism. 

\begin{figure}[!htbp]
    \centering
    \includegraphics[width=\textwidth]{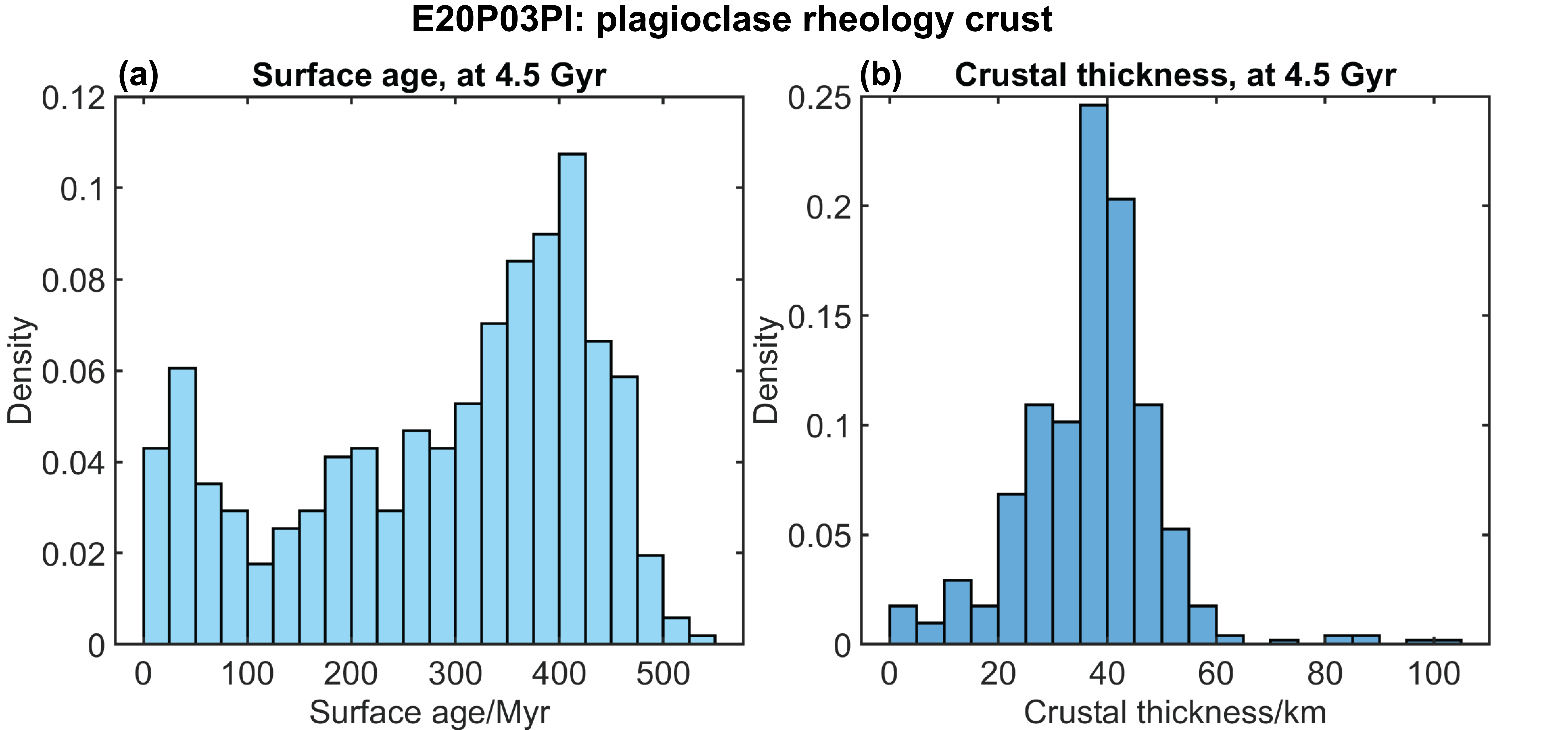}
    \caption{Histograms of (a) surface age and (b) crustal thickness at 4.5 Gyr, for model E20P03Pl.}
    \label{fig:M0_histo}
\end{figure}

The evolution of surface age and crustal thickness of model E20P03Pl is shown in Figure \ref{fig:M0_Surf} (b). The average surface age decreases to zero when global lithospheric overturn occurs, but the average crustal thickness remains nonzero for some global overturn events, such as the one at 2.1 Gyr. This is because these global overturns occur as a series of regional overturn events that cover parts of the surface over several million years. The increase in crustal thickness after overturns indicates an increase in melting and crustal production during and immediately after these overturns. Between global overturns, there are regional magmatic resurfacings shown by low-age regions in Figure \ref{fig:M0_Surf} (c). As radiogenic heat production in the mantle and CMB heat flow decrease over time, the interval between global overturns and the occurrence of regional magmatism decreases. As a result, the surface age distribution of model E20P03Pl at 4.5 Gyr is bimodal, with $\sim20\%$ of the surface younger than 100 Myr, and the rest being around 400 Myr old, corresponding to the last global overturn at $\sim 3.9$ Gyr (Figure \ref{fig:M0_histo}). 

\subsection{Influence of rheology}
\subsubsection{``Weak'' vs ``Strong'' Crust}
\begin{figure}[!htbp]
    \centering
    \includegraphics[width=\textwidth]{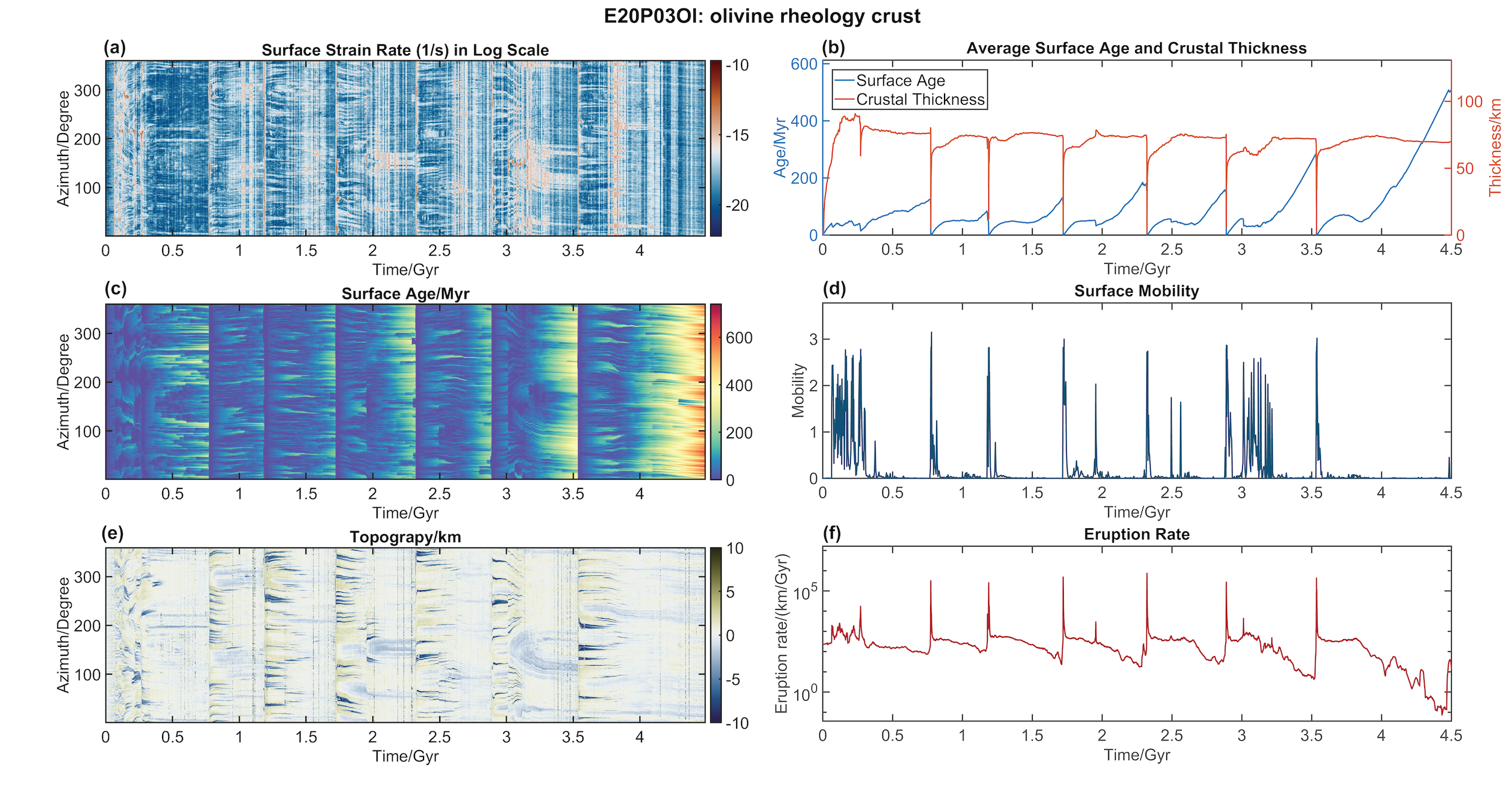}
    \caption{For model E20P03Ol, the evolution of (a) surface strain rate, (b) average surface age and crustal thickness, (c) surface age, (d) surface mobility, (e) topography and (f) eruption rate over 4.5 Gyr.}
    \label{fig:M1_Surf}
\end{figure}

\begin{figure}[!htbp]
    \centering
    \includegraphics[width=\textwidth]{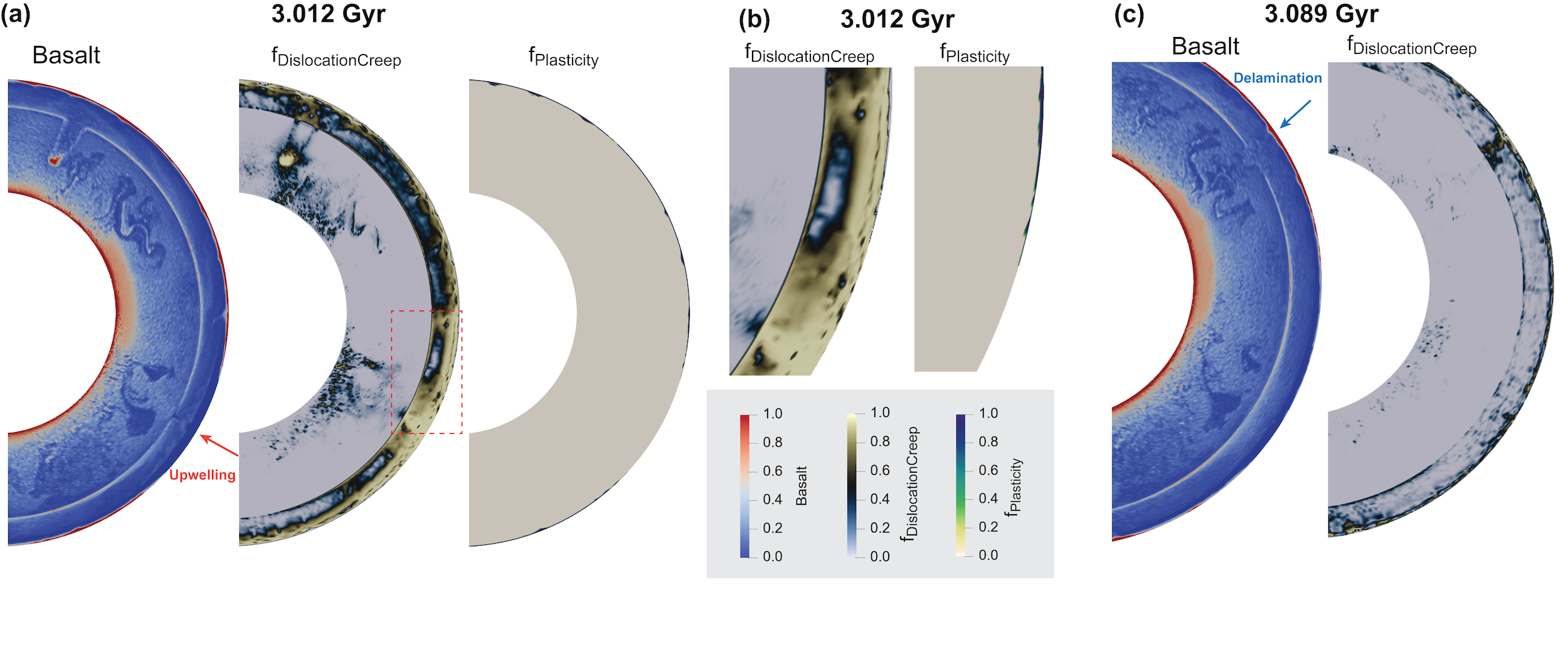}
    \caption{A regional overturn in the ``strong crust'' model E20P03Ol shown by fields of basalt fraction, plasticity fraction, and dislocation creep fraction at 3.012 Gyr and 3.089 Gyr. (a) The mantle upwelling at 3.012 Gyr is indicated by the orange arrow. Plastic yielding dominates the top-most lithosphere ($f_{\mathrm{Plasticity}} = 1$).  Dislocation creep dominates in regions around the mantle plume, shown by $f_{\mathrm{DislocationCreep}} \approx 1$. $f_{\mathrm{DislocationCreep}}$ is also high beneath the top-most lithosphere. (b) Zoom-in of the region with dashed red line. (c) At 3.089 Gyr, the mantle upwelling causes crustal delamination (blue arrow), rather than a global lithospheric overturn The strain rates of dislocation creep are also high around delaminated crust. The rheological fractions plotted are defined as the fraction of the total strain rate that is accommodated by a particular deformation mechanism:  $f_i = \dot{\epsilon}_{i}/\dot{\epsilon}_{\mathrm{total}}$, where $i$ represents plasticity or dislocation creep.}
    \label{fig:regionaloverturn}
\end{figure}

To investigate the impact of crustal rheology, we compare the model E20P03Ol (with ``strong'' olivine crustal rheology) to the reference model E20P03Pl (with ``weak'' plagioclase crustal rheology). The differences can be seen in Figure \ref{fig:M0_M1_BST} and by comparing Figures \ref{fig:M0_Surf} and \ref{fig:M1_Surf}. As in the reference case, the ``strong'' crust model E20P03Ol is also characterised by global overturns and a basalt barrier at the base of the MTZ; however, the basalt fraction at the base of the MTZ is lower when compared to the reference case E20P03Pl, while crustal thicknesses are larger. In addition to the global lithospheric overturns observed in the reference model E20P03Pl, regional resurfacing events are observed in E20P03Ol (Figure \ref{fig:regionaloverturn}). Magmatism associated with these regional resurfacing events is generally weaker and restricted to hemispheric or sub-hemispheric extent. Such regional resurfacing events are indicated by small peaks in the average crustal thickness, or periods with high surface mobility but small changes in mantle temperature (e.g. $\sim$ 1.9 Gyr and $\sim$ 3.0 Gyr in Figure \ref{fig:M1_Surf} (b)). The existence of such regional resurfacing events is in contrast to some previous episodic-lid models in which lithospheric overturn occurs over the whole surface \citep[e.g.][]{uppalapatiDynamicsLithosphericOverturns2020}. In our simulations, plastic yielding still propagates through the entire lithosphere (Figure \ref{fig:regionaloverturn}), however, at the boundaries of mantle plumes, the viscosity of the lithosphere and the underlying mantle asthenosphere is further reduced by dislocation creep due to high strain rates in this region. As a result, crustal production by mantle plumes is balanced by crustal delamination, rather than destabilising the lithosphere and triggering a global overturn. 

Regarding surface mobility, unlike the continuous surface mobility observed in reference case E20P03Pl, the surface mobility in the ``strong'' crust case E20P03Ol is similar to that in previously-published episodic-lid models \citep[e.g.][]{armannSimulatingThermochemicalMagmatic2012}: high during global overturns and near-zero between overturns (Figure \ref{fig:M1_Surf} (d)). The interval between the overturns is also longer than that of E20P03Pl, possibly due to the higher viscosity of the crust. Due to the long intervals between overturns, the average surface age in case E20P03Ol is higher than that in reference case E20P03Pl. The average crustal thickness (60 - 75 km between overturns) is also higher than that in E20P03Pl, and higher than the current estimated crustal thickness for Venus. 

In terms of magmatism and surface deformation, the mean surface strain rate between overturns decreases to $\sim 10^{-18}\ \mathrm{s^{-1}}$, or two orders of magnitude lower than that of E20P03Pl. Unlike E20P03Pl, there are regions with long-lived relatively high strain rates and low surface ages in E20P03Ol (Figure \ref{fig:M1_Surf}). As these regions indicate magmatic crust production, the mantle upwellings that produced these melts could be stationary for hundreds of millions of years in E20P03Ol. Such persistent mantle upwellings lead to non-uniform resurfacing rates at the surface, which are also observed in episodic-lid models from other studies \citep{uppalapatiDynamicsLithosphericOverturns2020}. 

\subsubsection{Dislocation Creep}
As shown by the fractions of deformation accommodated by different rheological mechanisms shown in Figure \ref{fig:regionaloverturn}, the viscosity of the upper mantle is mainly controlled by dislocation creep, while diffusion creep dominates in the lower mantle. This balance between diffusion and dislocation creep is expected to be dependent on the specified transition stress values in both regions, which are themselves affected by the grain sizes in the upper and lower mantles, the evolution of which is not modelled here. To investigate the effect of dislocation creep on mantle and lithosphere dynamics, we switched off dislocation creep in models E20P03Pl\_D and E20P03Ol\_D (which are otherwise identical to E20P03Pl and E20P03Ol, respectively). Thus, these models contain only diffusion creep and plastic yielding, as have most previously published Venus models \citep[e.g.][]{armannSimulatingThermochemicalMagmatic2012,uppalapatiDynamicsLithosphericOverturns2020}.

\begin{figure}[!htbp]
    \centering
    \includegraphics[width=\textwidth]{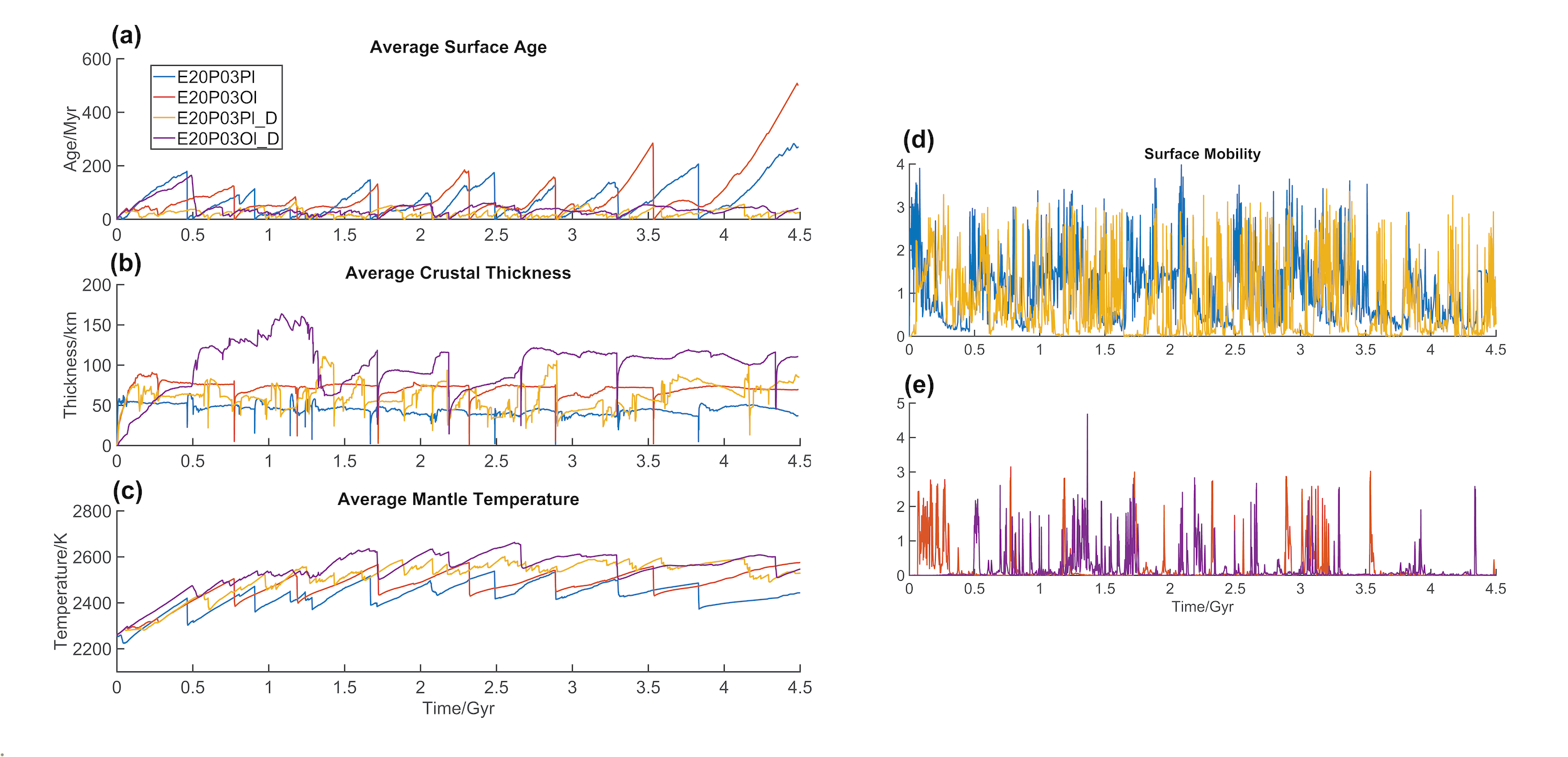}
    \caption{(a - e) The evolution of surface age, crustal thickness, mantle temperature, and surface mobility of models with (E20P03Pl, E20P03Ol) and without (E20P03Pl\_D, E20P03Ol\_D) dislocation creep.}
    \label{fig:diffdisl}
\end{figure}

\begin{figure}[!htbp]
    \centering
    \includegraphics[width=\textwidth]{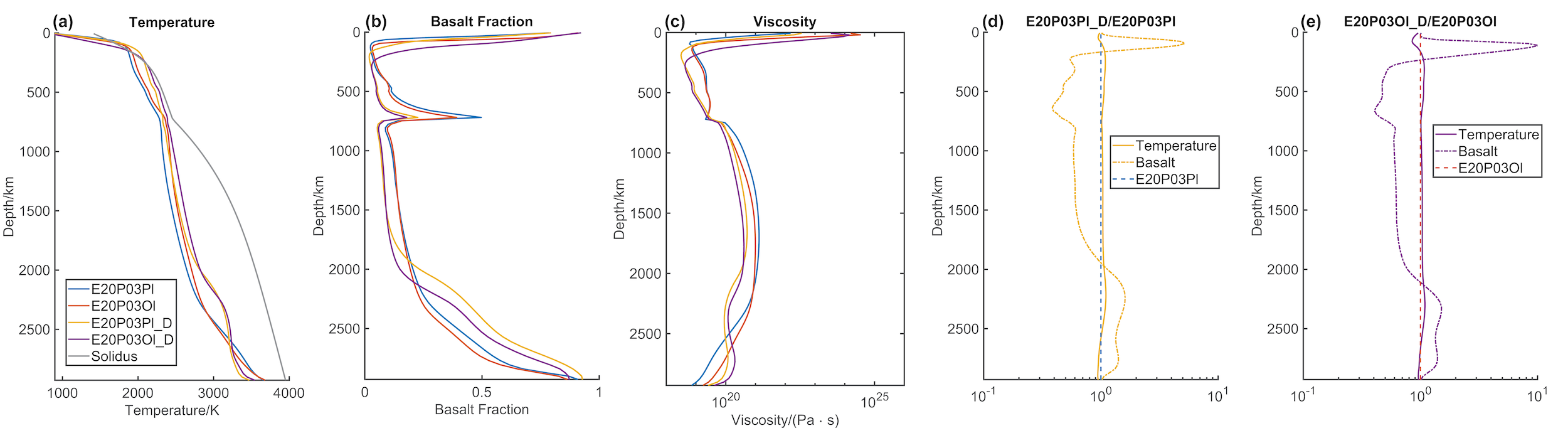}
    \caption{(a - c) Temperature, basalt fraction, and viscosity averaged over both azimuthal direction and 4.5 Gyr for models with (E20P03Pl, E20P03Ol) and without (E20P03Pl\_D, E20P03Ol\_D) dislocation creep. (d, e) Relative average temperature and basalt fraction between models with and without dislocation creep.}
    \label{fig:diffdisl2}
\end{figure}

The difference caused by dislocation creep is clearly seen in average mantle temperatures, shown in Figure \ref{fig:diffdisl} and Figure \ref{fig:diffdisl2}. Mantle temperatures are higher in models without dislocation creep, presumably because including dislocation creep lowers the viscosity at a particular pressure and temperature, thus increasing the vigor of convection and the heat transport efficiency. Higher mantle temperatures enhance magmatic crustal production, leading to much lower average surface ages in models E20P03Pl\_D and E20P03Ol\_D compared to E20P03Pl and E20P03Ol. Quantitatively, the non-dislocation-creep models have average surface ages that are always less than 100 Myr whereas in with-dislocation-creep cases surface age can reach several hundreds of Myr, thus better matching estimates for Venus' surface. The high magmatic crustal production rates in non-dislocation-creep models also lead to a thicker crust and a very depleted mantle at 4.5 Gyr for E20P03Pl\_D and E20P03Ol\_D, including lower basalt concentrations at the base of MTZ but large amounts of delaminated basalt accumulated above the CMB (Figure \ref{fig:diffdisl2}). 

The frequency of global overturns does not show a clear trend. More global overturns are observed in E20P03Pl\_D, but not in E20P03Ol\_D, when compared with E20P30Pl and E20P30Ol. Surface mobility depends mostly on crustal rheology (``strong'' vs. ``weak'') rather than the inclusion of dislocation creep. Both models with ``weak'' plagioclase crustal rheology have continuous surface mobility, whereas both models with ``strong'' olivine crustal rheology have episodic-lid surface mobility (Figure \ref{fig:diffdisl} (d) and (e)). Regional resurfacing is also observed in E20P03Ol\_D at 0.8 Gyr (indicated by high surface mobility and crustal thickness but low surface age in Figure \ref{fig:diffdisl}). The mobility peaks at 3.9 Gyr are caused by several small-scale mantle upwellings starting at 3.7 Gyr that cause crust delamination at 3.9 Gyr.

\subsubsection{Plasticity}
As in previous geodynamic models of Venus, plastic yielding is applied in our models as a weakening mechanism for the lithosphere. Plasticity represents brittle failure/fault slip, and in this study we use a friction coefficient that is fixed at 0.2 and three different values of cohesion: $C = 0.3$ MPa, $C = 50$ MPa, and infinity (no plastic yielding). This ``brittle'' yield stress formulation, as commonly used in the lithospheric dynamics community, is different from the formulation of ``ductile'' yield stress commonly used in the mantle convection community in that the latter is constant or increases only slowly with pressure (depth). Previous global numerical models often used both formulations \citep[e.g.][]{armannSimulatingThermochemicalMagmatic2012}, but in this study only brittle yield stress is applied in our models, which is similar to the regional models of coronae formation \citep{gulcherCoronaStructuresDriven2020}.  Nevertheless, both formulations lead to similar dynamical regimes in global models \citep[][]{moresiMantleConvectionBrittle1998,tackleySelfconsistentGenerationTectonic2000}. How different ``ductile'' yield stresses or ``brittle'' friction coefficients affect the overturn frequency and crustal thickness has been discussed in previous numerical studies \citep[e.g.][]{nakagawaInfluencePlateTectonic2015,rolfInferencesMantleViscosity2018,uppalapatiDynamicsLithosphericOverturns2020}. Here, the two models with $C = 50$ MPa exhibit similar tectonics compared to models with $C = 0.3$ MPa. Therefore, we focus on the two models without plastic yielding E20P00Pl and E20P00Ol, which differ in having a ``weak'' vs. a ``strong'' crust.  

\begin{figure}[!htbp]
    \centering
    \includegraphics[width=\textwidth]{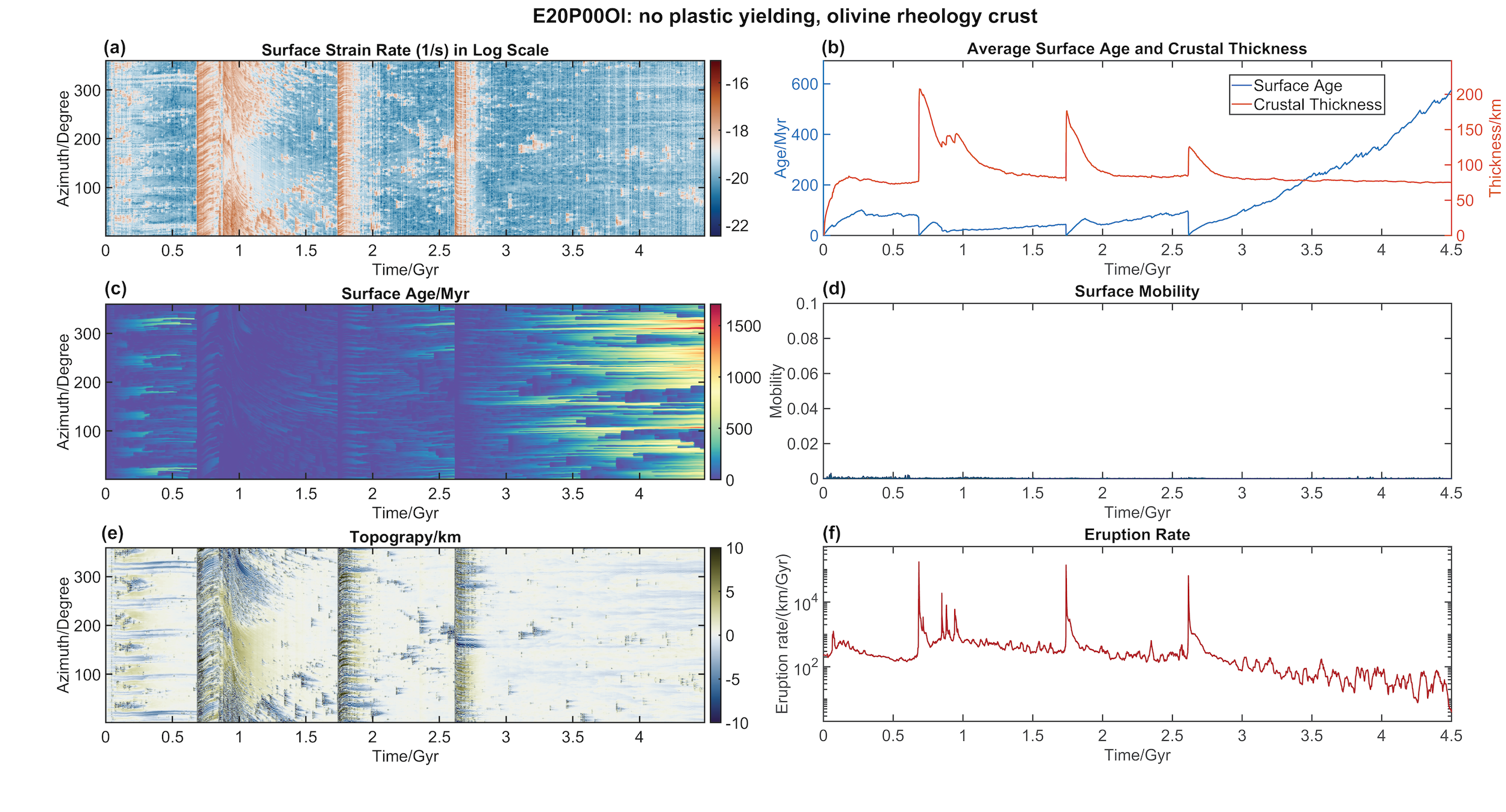}
    \caption{The evolution of (a) surface strain rate, (b) surface age, (c) average crustal thickness and surface age, (d) surface mobility ,(e) topography, and (f) eruption rate for model E20P00Ol with no plasticity and a ``strong'' olivine-rheology crust. The model displays stagnant-lid tectonics, as surface mobility is near zero throughout 4.5 Gyr. However, as shown by the evolution of surface strain rate and surface age, internal mantle overturns beneath the lithosphere can still cause global resurfacings and reset the surface age via melting-induced crustal production. Topography is truncated to [-10, 10] km.}
    \label{fig:plas_ol}
\end{figure}

The model without plasticity and a ``strong'' crust (E20P00Ol) is the only model presented in this study that exhibits stagnant-lid tectonics, indicated by near-zero surface mobility for the whole 4.5 Gyr model time (Figure \ref{fig:plas_ol} (d)). This shows that the weakening from dislocation creep and intrusive magmatism is less effective than plastic yielding when the crustal creep rheology is ``strong'' (olivine-like).

\begin{figure}[!htbp]
    \centering
    \includegraphics[width=\textwidth]{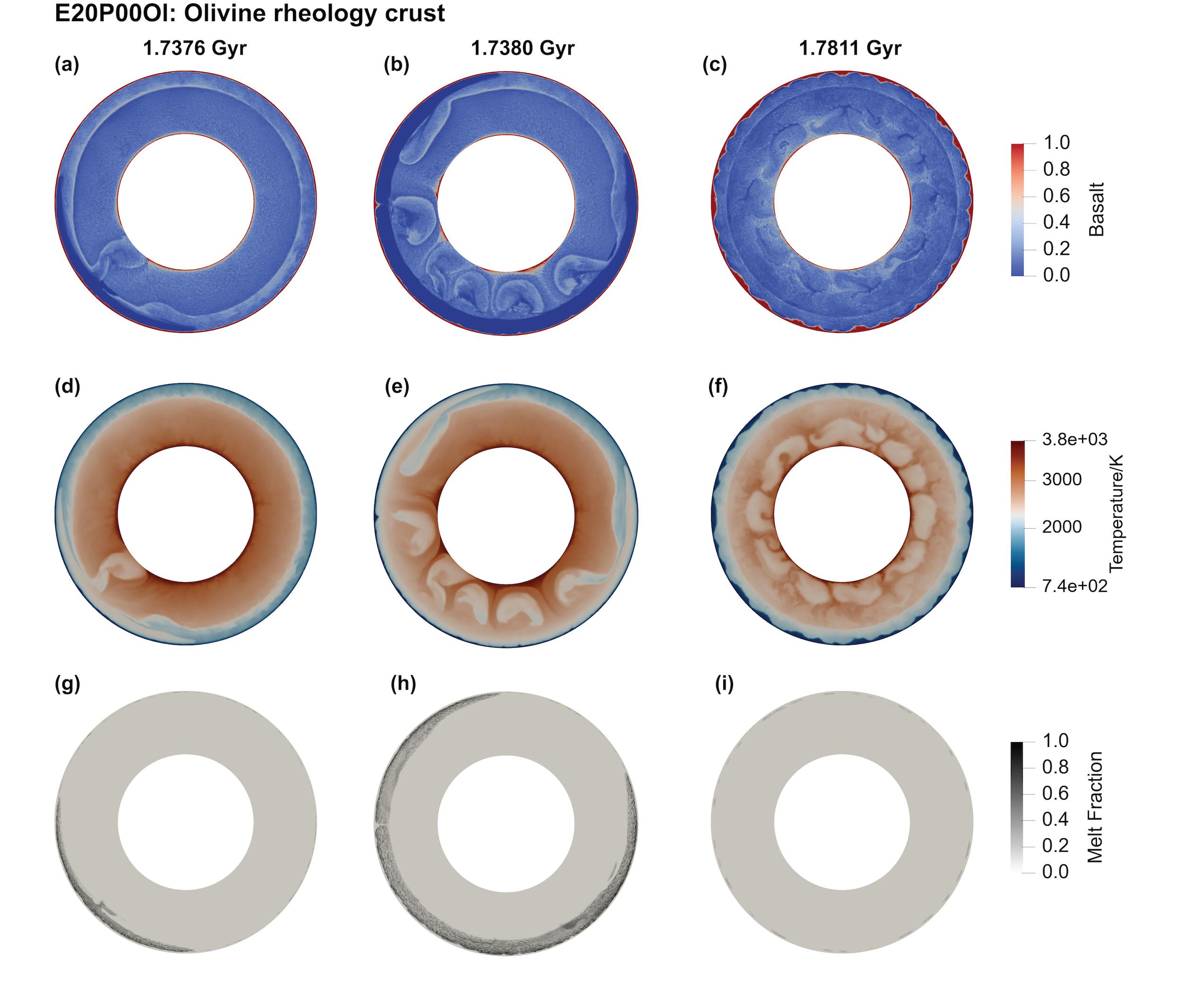}
    \caption{Snapshots of composition (basalt) (a - c), temperature (d - f), and melt fraction (g - i) for a internal mantle overturn at 1.74 Gyr for stagnant-lid model E20P00Ol.}
    \label{fig:inter_overturn}
\end{figure}

Interestingly, this model exhibits three global resurfacing events, as indicated by zero surface age and a sudden increase in crustal thickness (Figure \ref{fig:plas_ol} (b), e.g. at $\sim$ 1.8 Gyr). However, these are not due to lithospheric overturn - they are instead due to internal overturn of the mantle (``avalanches'') modulated by the ``basalt barrier'' mechanism. Hot lower-mantle material enters the upper mantle and causes a global pulse of magmatism. Heat transfer through the lithosphere is dominated by the magmatic heat-pipe mechanism \citep[][]{oreillyIo1981,solomonVenusTectonics1982,armannSimulatingThermochemicalMagmatic2012} and new eruptions to the surface lead to zero surface age and an increase in crustal thicknesses. Between these global internal mantle overturns, there are also regional magmatic resurfacing events (Figure \ref{fig:plas_ol}). As a result, although there is no lithospheric overturn, the evolution of surface age in this stagnant-lid model is similar to in episodic-lid models, and reaches around 550 Myr at present-day. On the other hand, the average crustal thickness in E20P00Ol remains nonzero throughout the model time. After mantle overturns, the crustal thickness first increases to larger than 100 km, as global mantle overturn triggers global magma production beneath the existing crust (Figure \ref{fig:inter_overturn}). Then, the crustal thickness starts to decrease and reaches $\sim 75 \mathrm{km}$ within several hundred Myr due to crustal delamination, which is similar to in other models with olivine crustal rheology and possibly limited by the basalt to eclogite phase transition (reference depth is 66.4 km, see Table \ref{tab:phasechange}). 

\begin{figure}[!htbp]
    \centering
    \includegraphics[width=\textwidth]{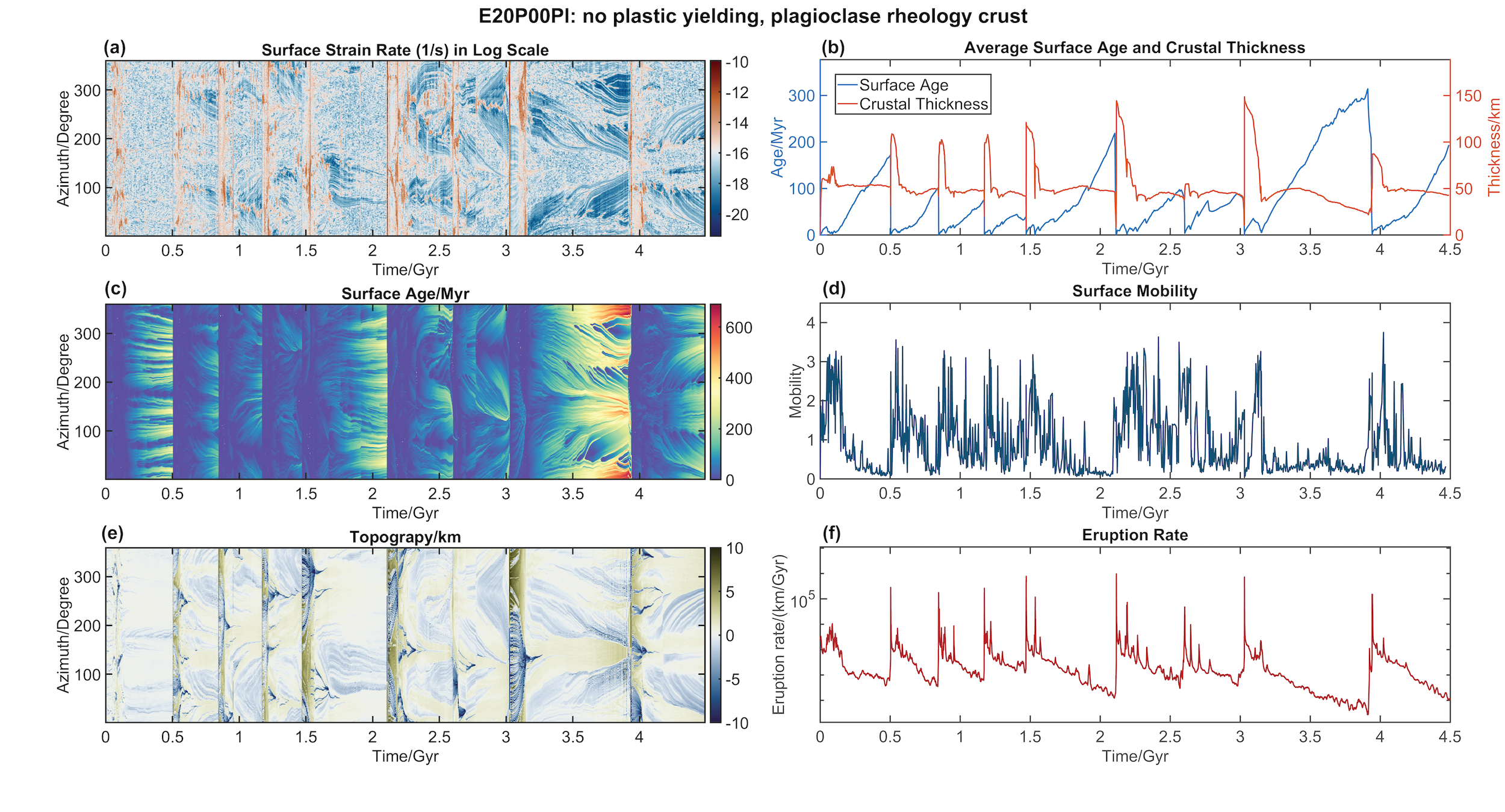}
    \caption{The evolution of (a) surface strain rate, (b) surface age, (c) average crustal thickness and surface age, (d) surface mobility, (e) topography and (f) eruption rate for model E20P00Pl with no plasticity and a ``weak'' plagioclase crust. The model displays lithospheric overturn events without plastic yielding. Topography is truncated to [-10, 10] km.}
    \label{fig:plas_plag}
\end{figure}

With a ``weak'' plagioclase crustal rheology (model E20P00Pl), global lithospheric overturns are observed despite the lack of plastic yielding (Figure \ref{fig:plas_plag}), indicating that the combination of a plagioclase crustal rheology, dislocation creep and intrusive magmatism is sufficient to weaken the lithosphere and facilitate overturns. The evolution of this model is generally similar to other models with ``weak'' crustal rheology, with global overturns that reduce the average surface age to near zero, and regional resurfacing between global overturns. The number of global overturns is less than in models with plastic yielding (see Table \ref{tab:result}), and the surface mobility is higher than zero through the 4.5 Gyr model time. The absence of plastic yielding is also reflected in large increases in crustal thickness (100-200 km) after global overturns. However, delamination by mantle convective flow effectively reduces the average crustal thickness to $\sim$ 50 km in less than 100 Myr.

\subsection{Influence of Eruption Efficiency}
In order to systematically investigate the effect of varying eruption efficiencies on the tectonic regime, models with $20\%$, $40\%$, $80\%$, and $100\%$ eruption efficiencies are tested in this study. For models with a plagioclase crustal rheology, the surface mobility is continuously high. On the other hand, for models with olivine crustal rheology and $>80\%$ eruption efficiency, the intervals between overturns are too short for stagnant-lid phases. Therefore, the surface mobility is also continuous over the 4.5 Gyr model time.

\begin{figure}[!htbp]
    \centering
    \includegraphics[width=\textwidth]{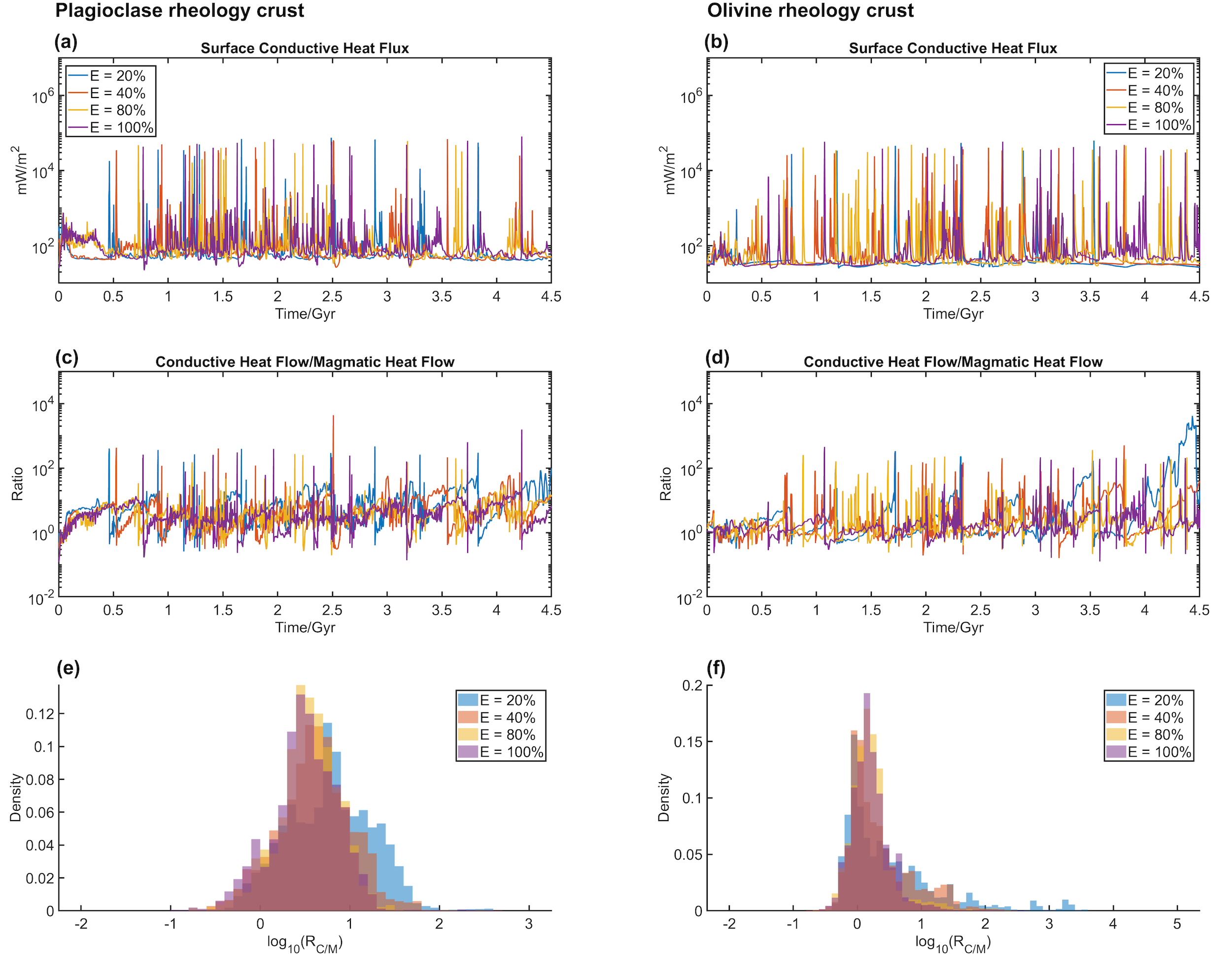}
    \caption{(a - d) The evolution of the surface conductive heat flux and the ratio between conductive and magmatic heat flow ($R_{C/M}$) for models with different eruption efficiencies. (c, d) Histograms for $\log_{10}{R_{C/M}}$, after interpolating the evolution of $R_{C/M}$ with 0.1 Myr time period.}
    \label{fig:erupt_2}
\end{figure}

\begin{figure}[!htbp]
    \centering
    \includegraphics[width=\textwidth]{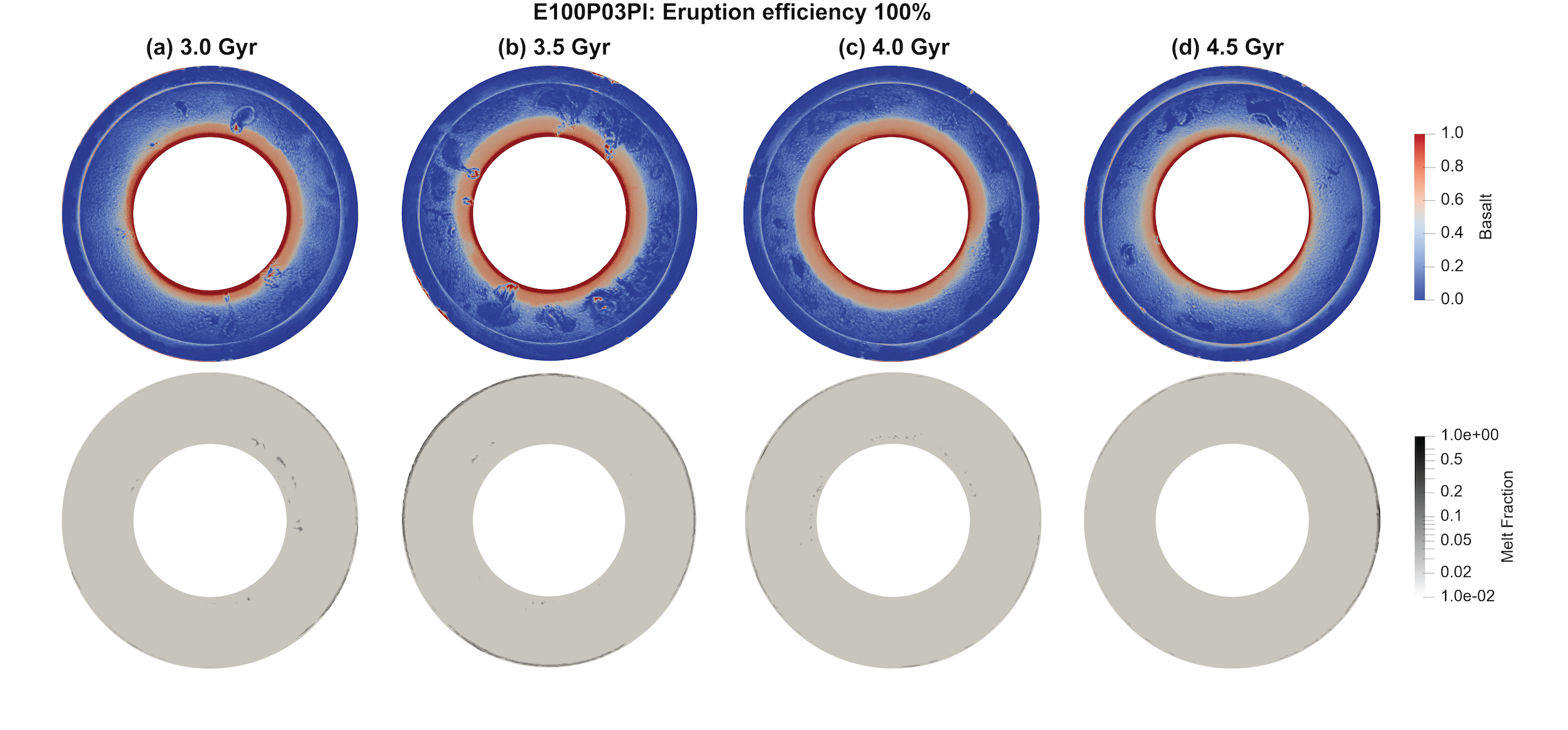}
    \caption{Snapshots of basalt fraction and melt fraction (in log scale) for the model with 100\% eruption and ``weak'' crust (E100P03Pl) from 3.5 Gyr to 4.5 Gyr. The melt fraction plots show a partially-molten zone below the crust that can remain for a long time.}
    \label{fig:erupt100}
\end{figure}

In models with a strong (olivine rheology) crust, higher eruption efficiency leads to more frequent overturns. These models also have lower ratios between the conductive and magmatic heat flow, or $R_{C/M}$ (Figure \ref{fig:erupt_2}). As the heat-pipe mechanism is less efficient at cooling down the mantle over time than intrusive magmatism \citep{lourencoEfficientCoolingRocky2018}, the heat transfer efficiencies in mantle are lower with higher $E$, resulting in more frequent overturns to release heat. On the other hand, for models with a ``weak'' plagioclase-rheology crust, the numbers of global overturns are similar between models with different $E$, indicating that crustal rheology might be more important than heat transfer efficiency in terms of overturn dynamics for these models. The average surface ages are lower with higher eruption efficiency, due to extensive partially molten zones (Figure \ref{fig:erupt100}). The average $R_{C/M}$ is similar between different models, and is larger than 1 for most of the model time (Figure \ref{fig:erupt_2}). When comparing models with the same eruption efficiency and different crustal rheology, $R_{C/M}$ is generally larger for models with plagioclase rheology, suggesting that heat transfer by intrusive magmatism, which facilitates conductive heat transfer by forming a thin lithosphere \citep{lourencoEfficientCoolingRocky2018}, is enhanced by a weaker crustal rheology.

\begin{figure}[!htbp]
    \centering
    \includegraphics[width=\textwidth]{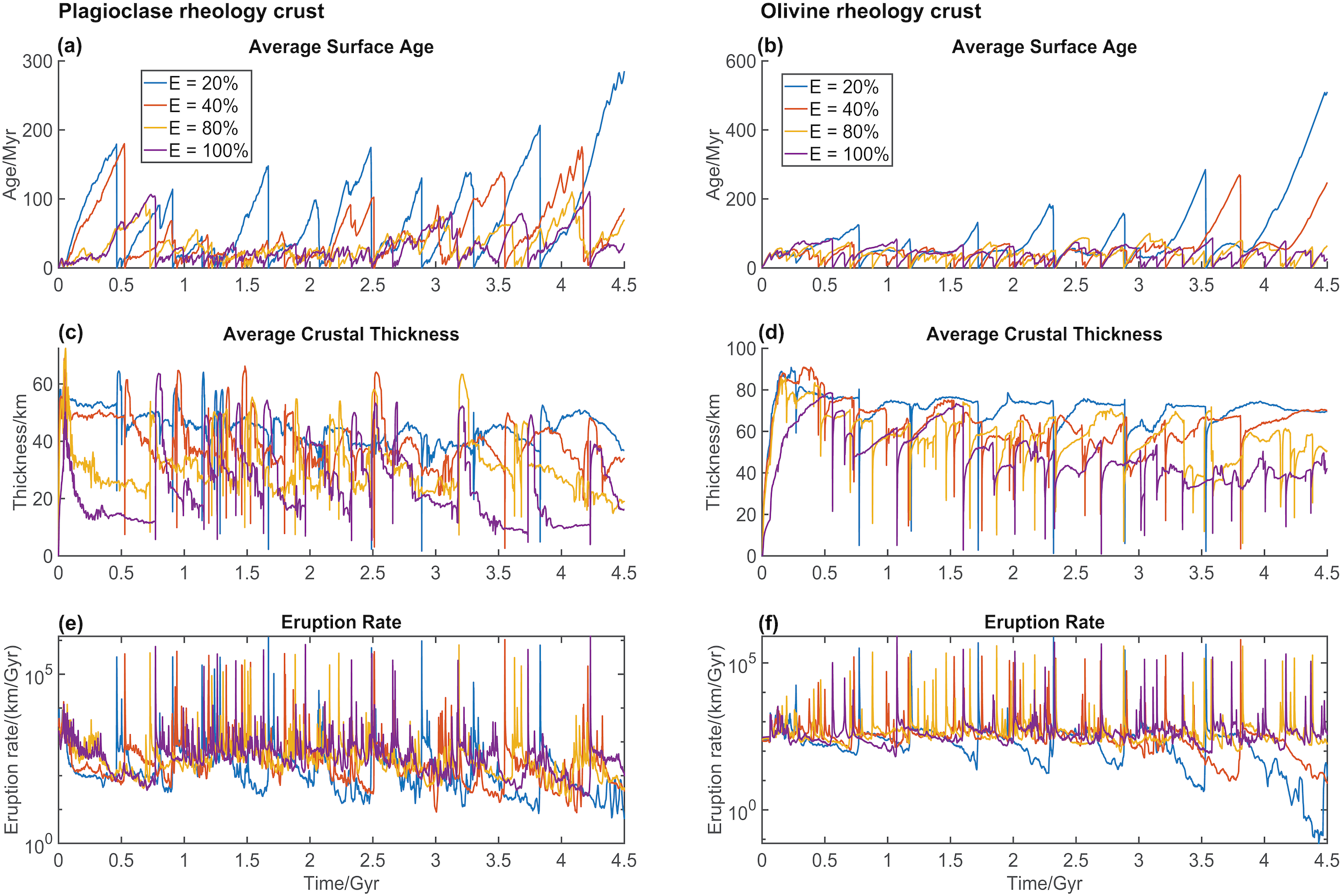}
    \caption{The evolution of average crustal thickness, average surface age, and eruption rates for models with different eruption efficiencies. (a, c, e) for models with ``weak'' (plagioclase An$_{75}$) crustal rheology, and (b, d, f) for models with ``strong'' (olivine) crustal rheology. }
    \label{fig:erupt_1}
\end{figure}

\begin{figure}[!htbp]
    \centering
    \includegraphics[width=0.95\textwidth]{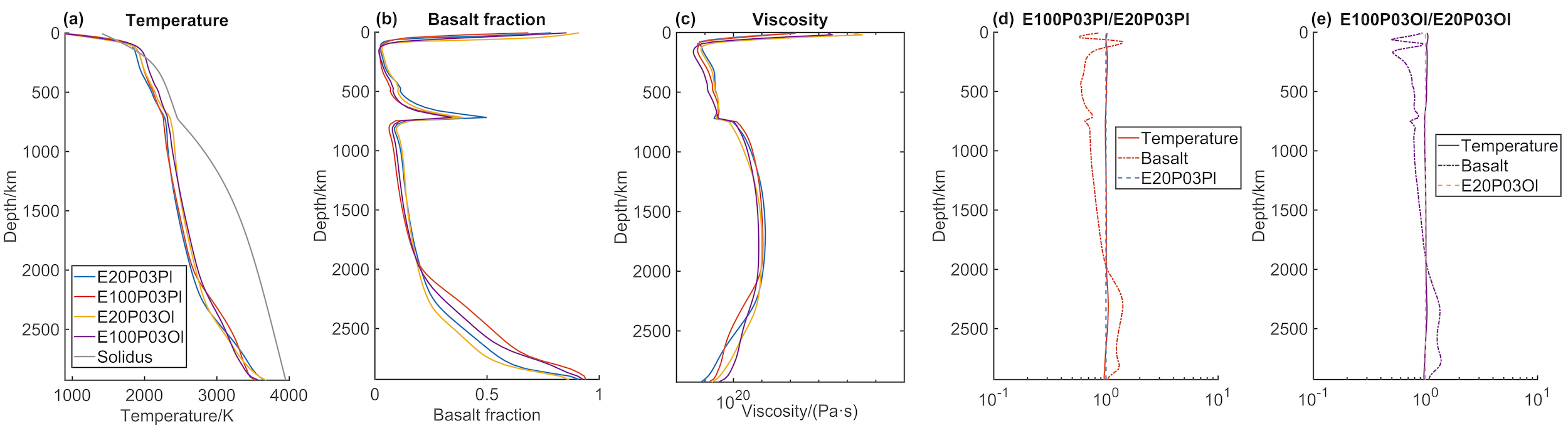}
    \caption{(a - c) Temperature, basalt fraction, and viscosity averaged over both azimuthal direction and time for 4.5 Gyr. Comparing models with different eruption efficiencies but the same crustal rheology (d)(e), the upper mantle in models with $E = 100\%$ is hotter and more depleted than models with $E = 20\%$}
    \label{fig:avgTBS}
\end{figure}

Due to more frequent overturns and extensive volcanism, models with higher eruption efficiencies have a lower average surface age (Figure \ref{fig:erupt_1}). Unlike previous numerical studies, in which increasing $E$ leads to a thicker crust \citep{lourencoEfficientCoolingRocky2018}, the crust is thinner with higher $E$ in our models. Also, there are partially molten zones beneath the crust. With higher $E$, the upper mantle is hotter and more depleted (Figure \ref{fig:avgTBS}), resulting in thinner crust at the end of the model.

\section{Discussion}
\label{sec:discussion}
\subsection{Comparing Model Results with Observations for Venus}
\subsubsection{Tectonics and Volcanism}
Observations of Venus' surface suggest that its resurfacing history could be a combination of two end-member models: global or near-global lithospheric overturns reworking most of the planet's surface and leading to a unimodal crater distribution, with small-scale tectonic and magmatic resurfacing events that created some of the highly deformed regions and young lava flows active in-between. Previous numerical models of Venus used olivine diffusion-creep rheology for all materials and focused mostly on an episodic-lid tectonics regime in which the surface mobility between global overturns is near zero. However, in our current models with composite crustal rheology, dislocation creep, and intrusive magmatism, we observe both global overturns and regional resurfacing, continuous surface mobility, as well as short-term magmatic events. This study thus illustrates the importance of applying a composite crustal rheology and intrusive magmatism to numerical models that focus on the tectonics of Venus. 

The introduction of more realistic rheologies and intrusive magmatism changes two main aspects of the tectonic regime: (1) continuous surface mobility, and (2) the occurrence of regional resurfacing. The surface mobility is affected mainly by crustal rheology: models with ``weak'' plagioclase crustal rheology exhibit continuous surface mobility in-between overturns, while models with ``strong'' olivine rheology and low eruption efficiency have episodic-lid surface mobility. Regarding regional resurfacing, plastic yielding still controls the viscosity in the uppermost lithosphere, which can lead to near-global overturns that are similar to those in previous numerical models \citep[e.g.][]{crameriSubductionInitiationStagnant2016}. However, as dislocation creep dominates in the upper mantle, the high strain rates around the mantle plumes and lower crustal eclogite (due to its high density) could facilitate the formation of small-scale convective cells, which can break the lithosphere above them and cause regional resurfacing that covers less than a hemisphere when small upwellings occur. These regional resurfacing events could explain how the local tectonic features on Venus survived lithospheric overturns, one example being the ``drifting'' of Lakshmi Planum \citep{harrisInteractionsContinentlikeDrift2015}. 

Models with different brittle cohesion values ($C = 0.3$ MPa and $C = 50$ MPa) display similar tectonic evolutions: the number of overturns and crustal thicknesses are similar between these models (Table \ref{tab:result}). This seems consistent with the results of \citet{uppalapatiDynamicsLithosphericOverturns2020} in which ductile yield stresses between 45 MPa to 70 MPa result in the same number of overturns and similar average crustal thicknesses. Thus, overturn frequency and crustal thickness are not very sensitive to changes in yielding parameters within a certain range. 

Plastic yielding, even with a low brittle cohesion value, still results in a rigid, non-deforming lithosphere between global overturns (i.e. episodic lid mode) when the crustal rheology is strong (olivine). On the other hand, when the crustal rheology is weak (plagioclase), lithospheric overturns can occur without plastic yielding (model E20P00Pl). This suggests that crustal rheology may be equally or more important than plastic yielding in controlling the occurrence of lithospheric overturns.

Regarding volcanism, another new observation in these models is that crustal rheology can affect the duration of local magmatic resurfacing. With olivine crustal rheology, upwellings in the upper mantle are stationary between global overturns and can be active for tens of millions of years. Resurfacing from magmatic crustal production is focused in the regions above these plumes, and the remaining surface lacks tectonic and volcanic activity, resulting in a non-uniform resurfacing rate. However, with a ``weak'' plagioclase crustal rheology, the time span of magmatic resurfacing is much shorter, possibly due to surface movements. Regions with higher-than-average strain rates are active for only a few million years. There are numerous short-lived, randomly located volcanic events between overturns, resulting in a more uniform resurfacing rate for Venus' surface, which is consistent with the cratering observations. The tectonics just described are also similar to the plutonic-squishy lid regime \citep{lourencoPlutonicSquishyLidNew2020}, but with some differences as discussed later.

Another observation is that in models with olivine rheology, there are more regional events than in models with plagioclase rheology, which could be counterintuitive. This can be explained by the fact that a plagioclase-rheology crust is (relatively) weak everywhere, leading to lithospheric blocks that are not well defined because, due to their relatively fast movement, weak zones caused by intrusions cannot survive for a long time. Therefore, a plagioclase-rheology crust leads to a lithosphere that can be considered as a mobile lid for most of the model time. Regarding overturn events, multiple regional resurfacing events tend to occur within a short period and cover the entire surface (similar to but less vigorous than a global overturn) and reset the age of the entire surface. In contrast, in models with an olivine-rheology crust, as the surface is mostly immobile during the near-stagnant-lid phase between overturns, the weakening from intrusions forms a compositionally continuous but rheologically discontinuous lid, which can lead to regional overturns.

\subsubsection{Crustal Thickness, Surface Age, and Topography}
Estimates of the crustal thickness of Venus vary between 20 - 60 km, depending on the geological province \citep{breuerTerrestrialPlanetsTreatise2015, wieczorekGravityTopographyTerrestrial2015}. As the crustal thickness is estimated by inverting gravity and topography observations, the result is highly dependent on the assumptions made, as the analysis by \citet{orthConstraintsVenusianCrustal2012} shows. Their preferred model was consistent with an average crustal thickness of 50 - 60 km. Other inversions, making other assumptions, have obtained smaller values (e.g. $<$ 25 km in \citet{jamesCrustalThicknessSupport2013} and \citet{jimenez-diazLithosphericStructureVenus2015}). Additionally, the maximum crustal thickness (under highland plateaus) is limited by the basalt-eclogite phase transition, which occurs at depths of 30 - 70 km, depending on temperature \citep{breuerTerrestrialPlanetsTreatise2015}. 
In our models, the crustal thickness is controlled by a balance between magmatic crustal production and crustal delamination. The depth of crustal delamination probably depends on the depth of the basalt-eclogite phase transition and the convective stress in the underlying mantle. For models with olivine crustal rheology and low eruption efficiency, the crustal thickness is limited by basalt-eclogite phase transition (66 km, crustal thicknesses are 69.7 km for E20P03Ol, 74.9 km for E20P00Ol, and 76.7 km for E20P50Ol). But for models with a ``weak'' plagioclase-rheology crust, the crust is thinner ($<$45 km in average) and can better fit current estimates. Recent analysis of Venusian crustal plateaus suggests that the basalt-eclogite phase transition can only be reached at the base of the two highest plateaus (Ovda and Thesis) and not in other plateaus if the heat flow there is higher than about 20 $\mathrm{mW/m^2}$ \citep{maiaLithosphericStructureVenusian2022}, which is consistent with models with a ``weak'' plagioclase crustal rheology.

Unlike in our previous models in which only the basaltic component can melt \citep{armannSimulatingThermochemicalMagmatic2012}, the melt composition can be anywhere in the basalt-harzburgite spectrum, although only pure basaltic melt can be erupted or intruded. Previous numerical studies show that basalt-only melting with 100\% eruption is too simplified for magmatic crustal production for Venus because it often results in very thick ($>$100 km) crust, especially in stagnant-lid cases as crustal production is enhanced by higher mantle temperature \citep[e.g.][]{armannSimulatingThermochemicalMagmatic2012}. More recent modelling studies, in which only basalt can melt, applied different eruption fractions in Earth-like models, and obtained thinner crusts with lower eruption fractions \citep{lourencoEfficientCoolingRocky2018,lourencoPlutonicSquishyLidNew2020}. However, if Venus' interior is hotter than Earth's, harzburgite could also melt, as its solidus is 150-200K higher than the solidus of lherzolite \citep{maaloeSolidusHarzburgiteGPa2004}, resulting in komatiitic-composition magma. In this study, a partially molten zone below the crust is often observed (Figure \ref{fig:erupt100}, also movies in the Supplementary Material). Whether such partially molten zones exist for Venus remains uncertain, as there are large variations with the estimates on Venus surface heat flow and temperature gradient in the lithosphere. Modelling of the formation of the Mead impact basin suggests the temperature gradient to be less than 14 K/km, which corresponds to surface heat fluxes of less than $28 \ \mathrm{mW/m^2}$ \citep{bjonnesEstimatingVenusianThermal2021}. However, in other deformation models, the formation of felsic tessera requires a much steeper geotherm ($>$ 40 K / km) so that deformed craters are not preserved when folds are generated \citep{resorFelsicTesseraeVenus2021}. A recent study also supports a thin lithosphere and high surface heat flow for Venus based on the topographic flexure at 65 coronae \citep{smrekar2022}. Such a steep geotherm could cause major melting below the base of the lithosphere \citep{ghailRheologicalPetrologicalImplications2015}. Recent analysis of Venusian crustal plateaus also indicates that the melting of deep crustal material could occur when crustal plateaus formed \citep{maiaLithosphericStructureVenusian2022}.

Previous numerical studies suggest that increasing eruption efficiency results in a thicker crust \citep{lourencoEfficientCoolingRocky2018}. In the current simulations, however, models with high eruption efficiencies display thin crusts and partially molten zones beneath the crust. The reasons are: (1) the temperature in the upper mantle is higher for models with higher eruption efficiency, causing more magmatic crustal production (Figure \ref{fig:avgTBS}); then (2) mantle overturns bring the thick basaltic crust to the CMB, where much of it remains. Therefore, in models with higher eruption efficiency, the basaltic crust is quickly produced and removed from the upper mantle, resulting in a more depleted upper mantle. Thus, at the end of these models, the melting process can only create a thin crust, with large, partially molten zones beneath the crust containing harzburgitic melt that cannot erupt (Figure \ref{fig:erupt100}). For models with lower eruption efficiencies and plagioclase crust rheology, as crust dripping is easier with a ``weak'' crust, the maximum crustal thickness is thinner and better fits estimates (e.g. 36.7 km for E20P03Pl). As for the average surface age, in models with a weak plagioclase-rheology crust, the average surface age is mainly controlled by the time of last overturn. Although a wide range of surface ages between $150-1000$ Myr has been suggested for Venus\citep[e.g.][]{hauckVenusCraterDistribution1998, herrickPostimpactModificationVolcanic2011}, only models with a low eruption efficiency ($E<40\%$) can fit the estimates. For models with high eruption efficiency, the average surface age is less than 100 Myr due to rapid magmatic crustal production.

As for topography, it is not straightforward to compare the model results to the current topography of Venus, as the topographic variation in the models varies with time. The standard deviations of topography from models with plagioclase crustal rheology are smaller than those of models with olivine crustal rheology, apart from the stagnant-lid cases E20P00Ol. However, most of the model results have higher topography standard deviations than that of the current Venus topography (std. = 0.995 km, calculated from ``Venus Magellan Global Topography 4641m (v2)" available on the USGS website), which could be explained by factors not included in this study, like flowing lava and surface erosion.

\subsection{Interior Evolution}
It is now well established that including compositional variations and realistic composition-dependent density changes in numerical models has a first-order effect on the model behaviour. This can be seen in the simulations presented here, previous simulations for Venus \citep{armannSimulatingThermochemicalMagmatic2012} and Earth (e.g. \citet{nakagawaInfluenceInitialCMB2010, lourencoMeltinginducedCrustalProduction2016a, yanEvolutionDistributionRecycled2020}) using the same numerical model StagYY, and simulations that include such density variations using other numerical models \citep[e.g.][]{papucBasaltBarrierVenus2012, ogawaMantleEvolutionVenus2014, vesterholtMantleOverturnThermochemical2021}. Compared to purely thermal models \citep[e.g.][]{solomatovVenusStagnantLid1996,huangConstraintsTopographyGravity2013}, applying this composition-dependent density leads to significantly different mantle structure and dynamics, including:

\textbf{1. A basaltic layer at the base of the mantle transition zone coupled with strong flow inhibition between the upper and lower mantle.} Seismic observations and laboratory experiments have provided evidence for such a basalt layer at the MTZ on Earth, including sound-velocity measurements of $\mathrm{CaSiO_3}$ perovskite \citep{greauxSoundVelocityCaSiO32019} and seismic inversions at subduction zones around the Pacific ocean \citep{bissigEvidenceBasaltEnrichment2022}. This ``basalt barrier'' effect \citep{daviesBasaltBarrier2008} leads to compositional and thermal stratification between the upper and lower mantle, which can only be penetrated by strong mantle plumes. Similar dynamics is observed in models of Venus by \citet{ogawaMantleEvolutionVenus2014}, with mantle bursts that destroy the basalt barrier, strongly stir the mantle, and result in widespread magmatism at the surface. The episodic variation of upper mantle temperature and spikes in basaltic crustal thickness that are observed in our simulations, were also observed in \citet{papucBasaltBarrierVenus2012}, although none of their models had a proper stagnant lid because they applied a viscosity cutoff at 100 times the reference viscosity. The strength of the basalt barrier is controlled by the ratio between compositional density differences and thermal density differences in the MTZ. Thus, this effect is expected to be stronger in a stagnant-lid mode than in a mobile-lid mode because compositional density differences (between basalt and harzburgite or pyrolite) are independent of tectonic mode, whereas temperature variations (and therefore thermal density differences) inside the mantle, are lower in stagnant-lid mode because the lithosphere does not participate in convection. Indeed, the stagnant lid case (E20P00Ol) does exhibit stronger layering, with only three internal overturns, than the episodic-lid cases.

\textbf{2. Basalt settling above the core-mantle boundary.} Such a basaltic layer is also observed in Earth models and its structure is strongly dependent on the tectonic mode \citep{nakagawaInfluencePlateTectonic2015, lourencoPlutonicSquishyLidNew2020}. The present simulations obtain a continuous layer of basalt around the CMB, which is consistent with previous episodic-lid models; more Earth-like continuous subduction can result in isolated piles instead. The basalt layer accumulated above the CMB strongly reduces the CMB heat flux because this layer is enriched in HPE \citep[][]{nakagawaCoreEvo2005,nakagawaCoreEvo2014}. Global overturns cause rapid thinning of this layer thickness and pulses of heat from the core. Therefore, it is possible that Venus' magnetic field could be episodic over time and be recorded by ferromagnetic minerals in its crust \citep{orourkeDetectabilityRemanentMagnetism2019}. The amount of basalt settling above the CMB is strongly sensitive to its density contrast with the ambient mantle, which is moderately uncertain; for example, the relevant density contrasts depend strongly on the assumed basalt composition (e.g. \cite{nakagawaSelfConsistentMineralPhysics2010} and are smaller when calculated using a recent mineralogical database \citep[][]{stixrude2022} than when calculated using an older version of the database \citep[][]{Xu2008}.

\citet{rolfInferencesMantleViscosity2018} and \citet{uppalapatiDynamicsLithosphericOverturns2020} included magmatic crustal production in their models; however, the above compositional effects are not observed in their simulations. This is because their models did not include realistic composition-dependent density contrasts, as the density differences are only about 10 $\mathrm{kg/m^3}$ between basalt and harzburgite in most of the mantle. Also, the density changes around the upper mantle - lower mantle transition zone are greatly reduced in their studies. Thus, the compositional density difference is too low to support a basalt barrier and form a separate basalt layer above the CMB. When applying a more realistic density profile to numerical models, for example by calculating phase assemblages and densities by using free energy minimisation \citep[e.g.][]{nakagawaMineralPhysics2009,nakagawaSelfConsistentMineralPhysics2010, vesterholtMantleOverturnThermochemical2021}, the aforementioned strong compositional effects are observed.

\subsection{New ``deformable episodic lid" tectonic regime}
Models with plagioclase crustal rheology and low eruption efficiencies show different behaviors compared to previously proposed stagnant-, episodic-, or plutonic-squishy-lid regimes, leading to a new regime that we name ``deformable episodic lid". This regime is characterised by a generally high surface mobility plus episodic lithospheric overturns (in one model (E20P00Pl) even without plastic yielding). Between overturns this tectonic regime shares some similarities with a plutonic-squishy lid \citep{lourencoPlutonicSquishyLidNew2020} regime: downwellings are dominated by crustal delamination and/or drips rather than subduction. However, these two regimes are different in that in a plutonic-squishy lid the lithosphere is split into small rigid plates or blocks separated by weak boundaries due to intrusions, while in the ``deformable episodic lid" lithospheric deformation is not localized into ``plate boundaries'' but instead high almost everywhere, making the lithosphere rheologically continuous for most of the time. Models with olivine crustal rheology are more similar to the plutonic-squishy lid regime \citep[][]{lourencoPlutonicSquishyLidNew2020} and show more long-lived volcanism when compared to models with a plagioclase rheology crust. These differences arise because the primary weakening mechanism in a plutonic squishy lid is magma intrusion, while the primary weakening mechanism in a deformable episodic lid is a relatively weak and non-linear crustal rheology.

\subsection{Global volcanic resurfacing events}
Here we find that global volcanic resurfacing events may occur in a stagnant-lid regime due to the ``basalt barrier" effect at the MTZ (Figure \ref{fig:plas_ol}) discussed above. This volcano-tectonic regime can be described as stagnant episodic-global-volcanic-resurfacing. Similar resurfacing behavior is observed in \citet{ogawaMantleEvolutionVenus2014} and \citet{papucBasaltBarrierVenus2012}, but both studies were performed in relatively narrow Cartesian domains; this is the first observation of such global events in a full spherical annulus. Furthermore, \citet{papucBasaltBarrierVenus2012} applied a mobile surface rather than stagnant lid in their model. In between global resurfacing events, local or regional volcanic resurfacing events can occur. Our model results show that this stagnant episodic-volcanic-resurfacing regime can fit the equilibrium resurfacing models for Venus, with both global and regional magmatic resurfacing in a stagnant-lid regime, which leads to a modelled surface age that can fit present-day estimates of Venus. However, these models also lead to a crustal thickness that is significantly higher than current estimates. Models with a weaker crustal rheology lead to a better fit to crustal thickness estimates, but the tectonic regime in those models changes from stagnant- to a deformable episodic lid. 

Curiously, global volcanic resurfacing events were not observed in the stagnant-lid models of \citet{armannSimulatingThermochemicalMagmatic2012} despite including the necessary composition- and depth-dependent density contrasts. This may be due to the relatively high lower-mantle viscosity in those models compared to the current ones: lower viscosity (higher effective Rayleigh number) results in phase-change-related density contrasts having more effect \citep[e.g.][]{christensen1985}. The viscosity contrast between upper and lower mantles chosen for our models here is consistent with modern viscosity models of Earth's mantle (e.g. \cite{cizkova2012} and is also more consistent with the geoid and the correlation between geoid and topography of Venus \citep[e.g.][]{rolfInferencesMantleViscosity2018}.

\subsection{Limitations of this study}
Several parameters and assumptions in this study are subject to significant uncertainties and discussion. Regarding rheology, the diffusion-creep rheology in this study assumes a uniform grain size giving $\eta_0 = 1\times 10^{20} \ \mathrm{Pa \cdot s}$ for all models. However, a wide range of grain sizes is expected in both Earth's and Venus' mantles and grain size evolution can have important effects on the dynamics \citep[e.g.][]{rozelGrainSize2012,dannbergGrainSize2017,schierjottGrainSize2020}. The evolution of the grain sizes for mantle minerals could also affect the development of lithospheric shear zones through strain weakening and shear localization \citep[e.g.][]{bercoviciPlateTectonicsDamage2014}. Additionally, the influence of the viscosity contrast between upper and lower mantles (here assumed to be small for diffusion creep and large for dislocation creep, based on a combination of Earth-based constraints and Venus' admittance ratios) is worth exploring in the future.

The parameterisation of melting in our models is based on melting processes on Earth. Venus' mantle solidus and HPE concentration could be different due to compositional variations \citep[e.g.][]{kieferEffectsMantleComposition2015,shahVenus2022}. The melting and eruption behaviour also needs further analysis, as it strongly affects magmatic crustal production. 

Finally, no three-dimensional (3-D) model is included in this study due to their computational expense. In 2-D, mantle plumes are infinite sheet-like features, which may exaggerate their resurfacing effects. On the other hand, models in \citet{uppalapatiDynamicsLithosphericOverturns2020} comparing 2-D and 3-D geometry show similar crustal thickness and number of major overturns, indicating similar tectonics between models with different model geometries, as did the Earth models of \citet{nakagawaInfluenceInitialCMB2010}. Internal overturns modulated by the ``basalt barrier'' mechanism might also manifest differently in 3-D than in 2-D, perhaps being more regional as are ``avalanches'' driven by the ringwoodite to bridgmanite+ferropericlase transition (e.g. \cite{tackleyavalanches1993}.

\section{Conclusions}
\label{sec:conclusion}
Using 2-D mantle convection models, we carry out a systematic modelling study of Venus' tectonics and interior evolution, considering the effects of a weak, experiment-based crustal rheology and intrusive magmatism. Unlike previous 2-D or 3-D models, where an episodic-lid or stagnant-lid tectonic mode were obtained, the results from this study show that Venus' tectonics can be a combination of global overturns, regional overturns, and short-term, randomly-located magmatic resurfacing events over the entire surface. Global resurfacing events can be caused either by lithospheric overturns or by internal overturns driven by breakdown of a ``basalt barrier'' between the upper and lower mantles, what leads to hotter lower-mantle material flooding the upper mantle and causing widespread volcanism.

The effect of applying a weak crustal rheology is both reflected in surface mobility and in upwelling activity. Models with a ``weak'' plagioclase crustal rheology exhibit both relatively high, continuous surface mobility and episodic lithospheric overturns, a new mode that we term deformable episodic lid, while models with a ``strong'' olivine crustal rheology display standard episodic or stagnant-lid mobility. For models with with a ``strong'' olivine crustal rheology, surface motion between global overturn events can sometimes occur due to regional overturns, rather than having a completely rigid lid. The duration of regional magmatic resurfacing is also affected by crustal rheology. For models with an olivine rheology, upper mantle upwellings are near-stationary between overturns and can last for tens of millions of years, resulting in non-uniform magmatic resurfacing rates at the surface. However, for models with a ``weak'' plagioclase rheology, magmatic resurfacing occurs more randomly and exists only for a short time, leading to a more homogeneous resurfacing rates in these models.

Regarding lithospheric weakening mechanisms, a combination of weak crustal rheology, including dislocation creep and intrusive magmatism, can weaken the lithosphere sufficiently for global lithospheric overturns to occur without plastic yielding. When included, plastic yielding still dominates in the top-most lithosphere, but dislocation creep can further weaken the lithosphere around mantle plumes, resulting in regional overturns.

The interior dynamics is strongly affected by the composition- and phase-dependent density and dislocation creep. Composition/phase-dependent density leads to a strong ``basalt barrier'' at the base of the MTZ - stronger than occurs in Earth-like models, creating thermal and compositional stratification between the upper and lower mantle, plus a layer of basalt above the CMB. The formation of this basalt barrier can also cause internal global mantle overturns in stagnant-lid models, resulting in global magmatic resurfacing events without lithospheric overturn. Tectonics is strongly influenced by dislocation creep, as this creep mechanism dominates in the upper mantle. Dislocation creep lowers viscosity, resulting in more efficient heat transfer, lower temperatures, and less magmatic resurfacing compared to identical models without dislocation creep.  

Because of the different interpretations of tectonic features on Venus' surface, it is unclear which tectonic mode Venus experiences, but in general our models in a deformable episodic lid regime better match current observations and estimates, including crustal thickness, surface age, and the presence of both global resurfacing events and regional tectonic and volcanic activity between these events. Also, the episodic-volcanic resurfacing mechanism described for stagnant-lid models suggests a new global resurfacing mechanism for Venus, where global episodes of eruption create new crust that covers the whole planet. This mechanism can match the same observations and estimates as episodic-lid models with olivine crustal rheology, but without the need for global lithospheric overturns.

\section{Acknowledgements}
JT was partially supported by the ERC project NONUNE (Grant agreement ID: 885531). All simulations in this study were performed using the Euler cluster at ETH Zurich. The open source software Paraview (\url{https://www.paraview.org/}) is used for 2-D visualisation. Various colour maps from \citet{cramerifabioScientificColourMaps2021} and colorbrewer (\url{https://colorbrewer2.org/}) are used in
this study to visualise the data and prevent visual distortion \citep{crameriMisuseColourScience2020}. We thank Tobias Rolf and an anonymous reviewer for helpful comments that improved the manuscript, and Doris Breuer for the editorial work.

\section{Data Availability}
Datasets related to this article can be downloaded at \url{https://doi.org/10.5281/zenodo.7746396}. StagYY is the property of PJT and ETH Zurich, and is available for collaborative studies upon request (\url{paul.tackley@erdw.ethz.ch}).

\clearpage

\appendix
\section{Parameters for Phase Change and Rheology}
\setcounter{table}{0}
\setcounter{figure}{0}
\label{sec:sample:appendix}
\begin{table}[htbp]
\centering
\scriptsize
\begin{tabular}{ p{1.75cm} p{5cm}p{1.75cm}p{1.75cm}}
    \hline
    Depth(km) & Temperature (K)                                              & $\Delta \rho$ ($\mathrm{kg/m^{3}}$) & $\gamma$ (MPa/K) \\
    \hline
                & \textit{Olivine} ($\rho_s=3240$ $\mathrm{kg/m^{3}}$)         &                                     &
                \\
    453       & 1600                                                         & 180                                 & +2.5             \\
    730       & 1900                                                         & 400                                 & -2.5             \\

                & \textit{Pyroxene-Garnet} ($\rho_s=2900$ $\mathrm{kg/m^{3}}$) &                                     &
                \\
                & \textit{Plagioclase crustal rheology}                                 &                                     &
                \\
    $83.0^*$    & 1600                                                         & 450                                 & +1.76                 \\
                & \textit{Olivine crustal rheology}                           &                                     &
                \\
    $66.4^*$    & 1600                                                         & 350                                 & 0          \\
    442       & 1600                                                         & 180                                 & +1.0             \\
    796       & 1900                                                         & 400                                 & +1.0             \\
    \hline
    \multicolumn{4}{l}{$^*$ Eclogite phase transition}
                \\
\end{tabular}
\caption{Phase transition parameters for olivine and pyroxene-garnet systems, where $\rho_s$ stands for surface density at zero pressure, $\Delta \rho$ for density jump across a phase transition, and $\gamma$ for Clapeyron slope. The depths of phase transition are adjusted for Venus's gravity.}
\label{tab:phasechange}
\end{table}

\begin{table}[htbp]
\scriptsize
\centering
\begin{tabularx}{\textwidth}{p{2cm}>{\raggedright\arraybackslash}X>{\raggedright\arraybackslash}X>{\raggedright\arraybackslash}X>{\raggedright\arraybackslash}X>{\raggedright\arraybackslash}X}
    \hline
    Depth (km) & Activation energy (kJ/mol) & Activation volume ($\mathrm{cm^3/mol}$) & $P_{\mathrm{decay}}$ (GPa) & Stress exponent $n_i$ & $\Delta \eta_{phase}$\\
    \hline
                   & \multicolumn{5}{l}{\textit{Diffusion creep}}    \\                    
                   & \multicolumn{5}{l}{\textit{Olivine}}            \\
    $\leq$ 453     & 300  & 5         & $\infty$   &      & 1.0      \\
    453-730        & 300  & 5         & $\infty$   &      & 1.0      \\
    $>$730           & 370  & 3.65      & 200        &      & 3.0      \\

                     & \multicolumn{5}{l}{\textit{Pyroxene-Garnet}}      \\
    $\leq$ 66.4$^*$  & 300  & 5         & $\infty$   &   &1.0            \\
    66.4-442         & 300  & 5         & $\infty$   &   &1.0            \\
    442-796          & 300  & 5         & $\infty$   &   &1.0            \\
    $>$796             & 370  & 3.65      & 200        &   &3.0            \\

                     & \multicolumn{5}{l}{\textit{Plagioclase}}        \\
    $\leq$ 83.0$^*$  & 238  & 0         & $\infty$   &   & 0.0133      \\
    
                     & \multicolumn{5}{l}{\textit{Dislocation creep}}    \\
                     & \multicolumn{5}{l}{\textit{Olivine}}              \\
    $\leq$ 453       & 532  & 14       & $\infty$    & 3.5     &1.0       \\
    453-730          & 532  & 14       & $\infty$    & 3.5     &1.0       \\
    $>$730             & 370  & 3.65     & 200         & 3.5     &3.0       \\

                     & \multicolumn{5}{l}{\textit{Pyroxene-Garnet} }     \\
    $\leq$ 66.4$^*$  & 532  & 14       & $\infty$    & 3.5     &1.0      \\
    66.4-442         & 532  & 14       & $\infty$    & 3.5     &1.0      \\
    442-796          & 532  & 14       & $\infty$    & 3.5     &1.0      \\
    $>$796             & 370  & 3.65     & 200         & 3.5     &3.0      \\

                    & \multicolumn{5}{l}{\textit{Plagioclase}}   \\
    $\leq$ 83.0$^*$  & 238  & 0        & $\infty$    & 3.2    &0.0133     \\

    \hline
\end{tabularx}
\caption{Rheological parameters for olivine and pyroxene-garnet phase system that are used in our models. The basalt-eclogite phase transition is marked with asterisks ($^*$). If the plagioclase rheology is applied, its rheological parameters will replace parameters for basalt facies in the pyroxene-garnet system. Depths of phase changes have been scaled from Earth according to Venus' slightly lower gravity.}
\label{tab:rheo}

\end{table}

\newpage

\section{Supplementary Material: additional tests}
\subsection{Testing different numbers of tracers}

We performed tests for our reference case (E20P03Pl, initial 10 tracers per cell) with 20 or 30 initial tracers/cell (2 or 3 times that used in the standard cases) to check whether the initial number of tracers affects model evolution. The results show that the differences between the reference case run using 10, 20 or 30 tracers per cell are not qualitatively significant. The differences observed are due to small random perturbations in the initial conditions that can lead to overturns occurring at different times. The figures below may be compared to Figure \ref{fig:M0_Surf} in the main paper.

\begin{figure}[!p]
    \centering
    \includegraphics[width=\textwidth]{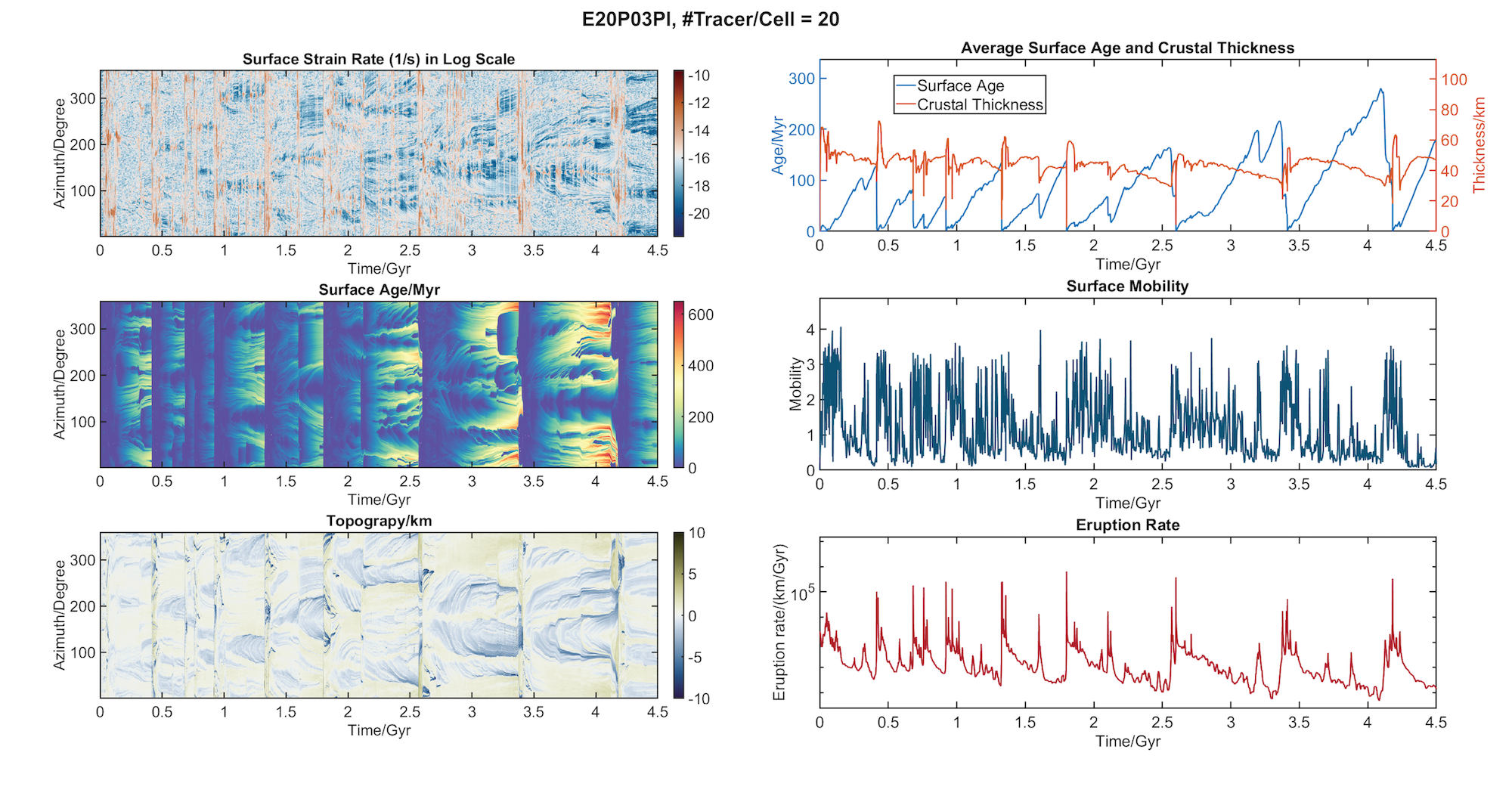}
    \caption{The evolution of E20P03Pl, with 20 tracers per cell}
    \label{fig:M0_t20}
\end{figure}

\begin{figure}[!p]
    \centering
    \includegraphics[width=\textwidth]{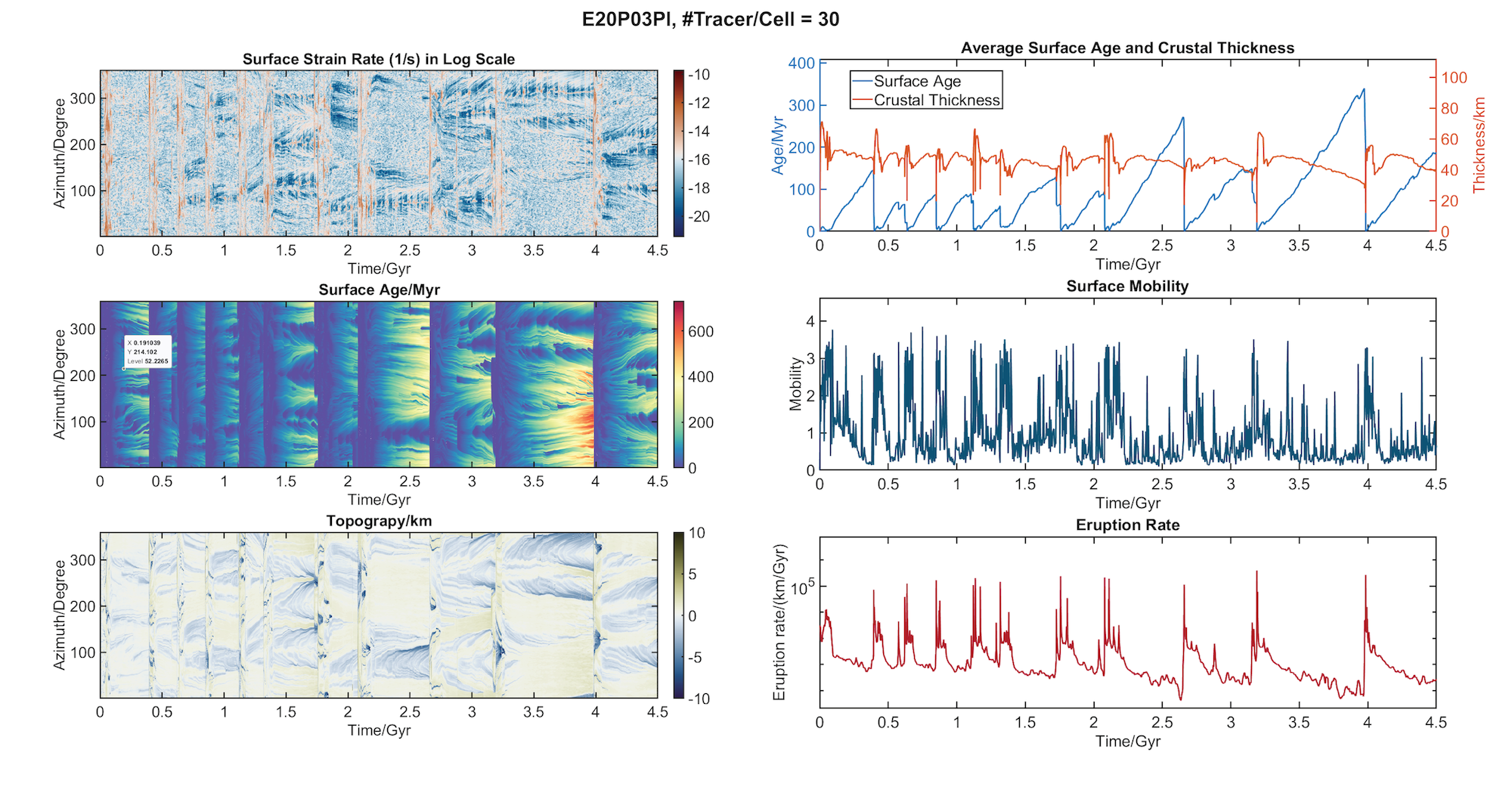}
    \caption{The evolution of E20P03Pl, with 30 tracers per cell}
    \label{fig:M0_t30}
\end{figure}

\newpage
\subsection{Testing different CMB temperature}
We performed tests for our reference case (E20P03Pl, initial CMB temperature = 4025 K) with different initial CMB temperature ($\mathrm{T_{CMB}}$ = 4525 / 5025 / 5525 / 6025 K). While near the beginning, models with higher initial $\mathrm{T_{CMB}}$ show more frequent overturns, after $\sim 2.5$ Gyr the dynamics are similar to the reference model in terms of surface mobility, frequency of global overturns, occurrence of regional resurfacings, and average surface age and crustal thickness between global overturns. Therefore, the initial $\mathrm{T_{CMB}}$ only affects the early evolution, not the long-term behaviour. The figures below may be compared to Figure \ref{fig:M0_Surf} in the main paper.

\begin{figure}[!p]
    \centering
    \includegraphics[width=\textwidth]{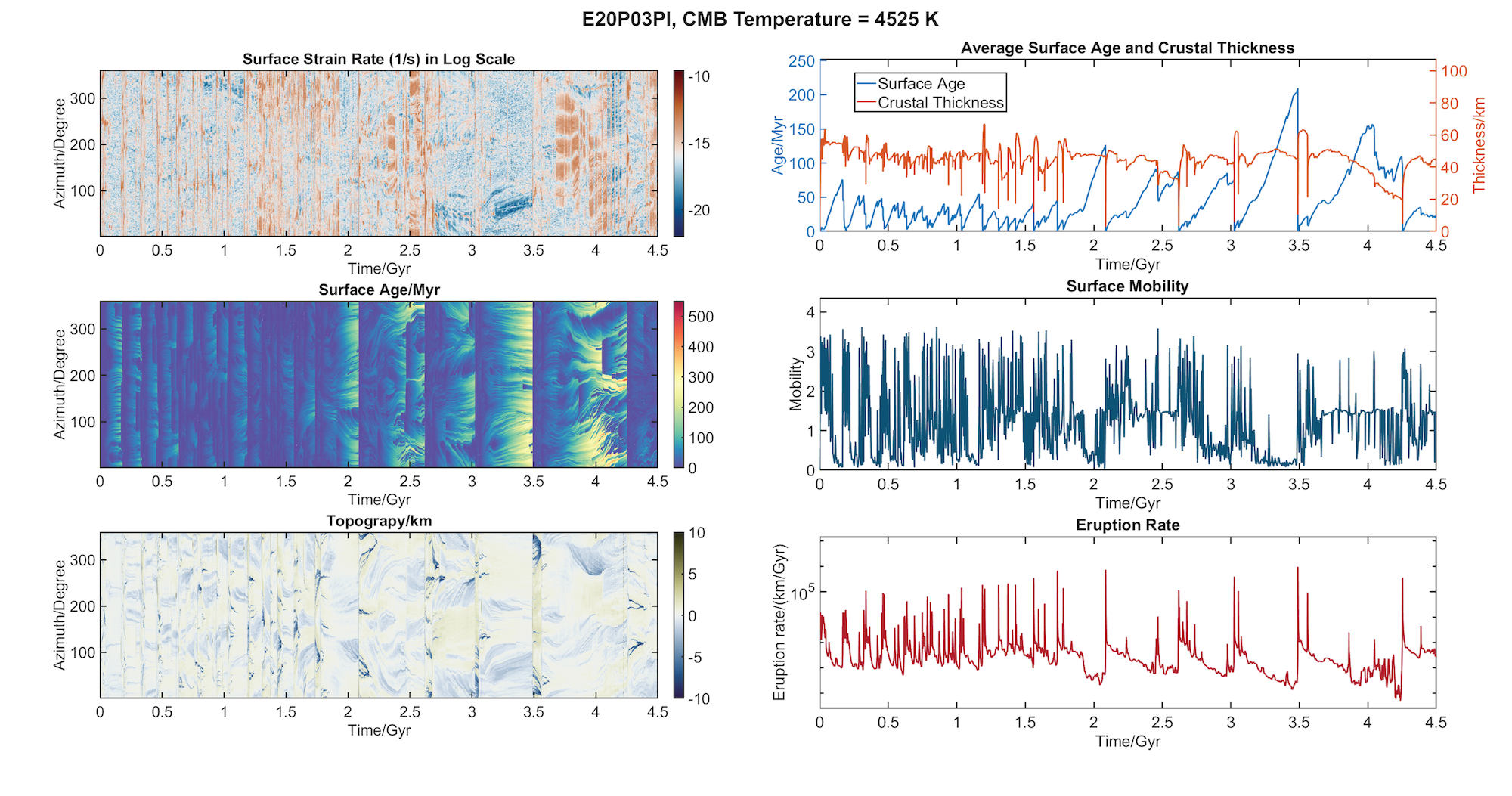}
    \caption{The evolution of E20P03Pl with $\mathrm{T_{CMB}}$ = 4525 K.}
    \label{fig:M0_4525}
\end{figure}

\begin{figure}[!p]
    \centering
    \includegraphics[width=\textwidth]{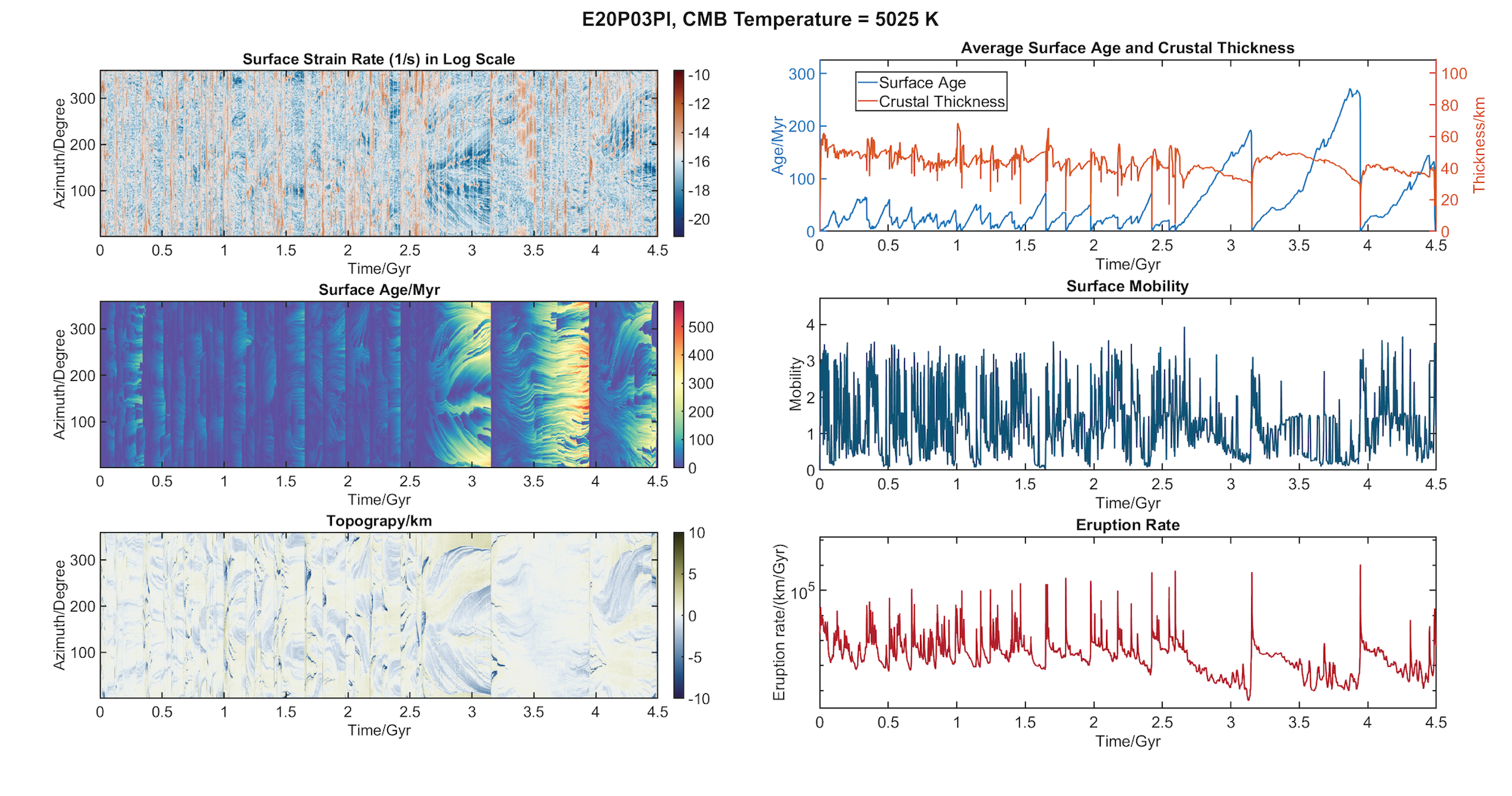}
    \caption{The evolution of E20P03Pl with $\mathrm{T_{CMB}}$ = 5025 K.}
    \label{fig:M0_5025}
\end{figure}

\begin{figure}[!p]
    \centering
    \includegraphics[width=\textwidth]{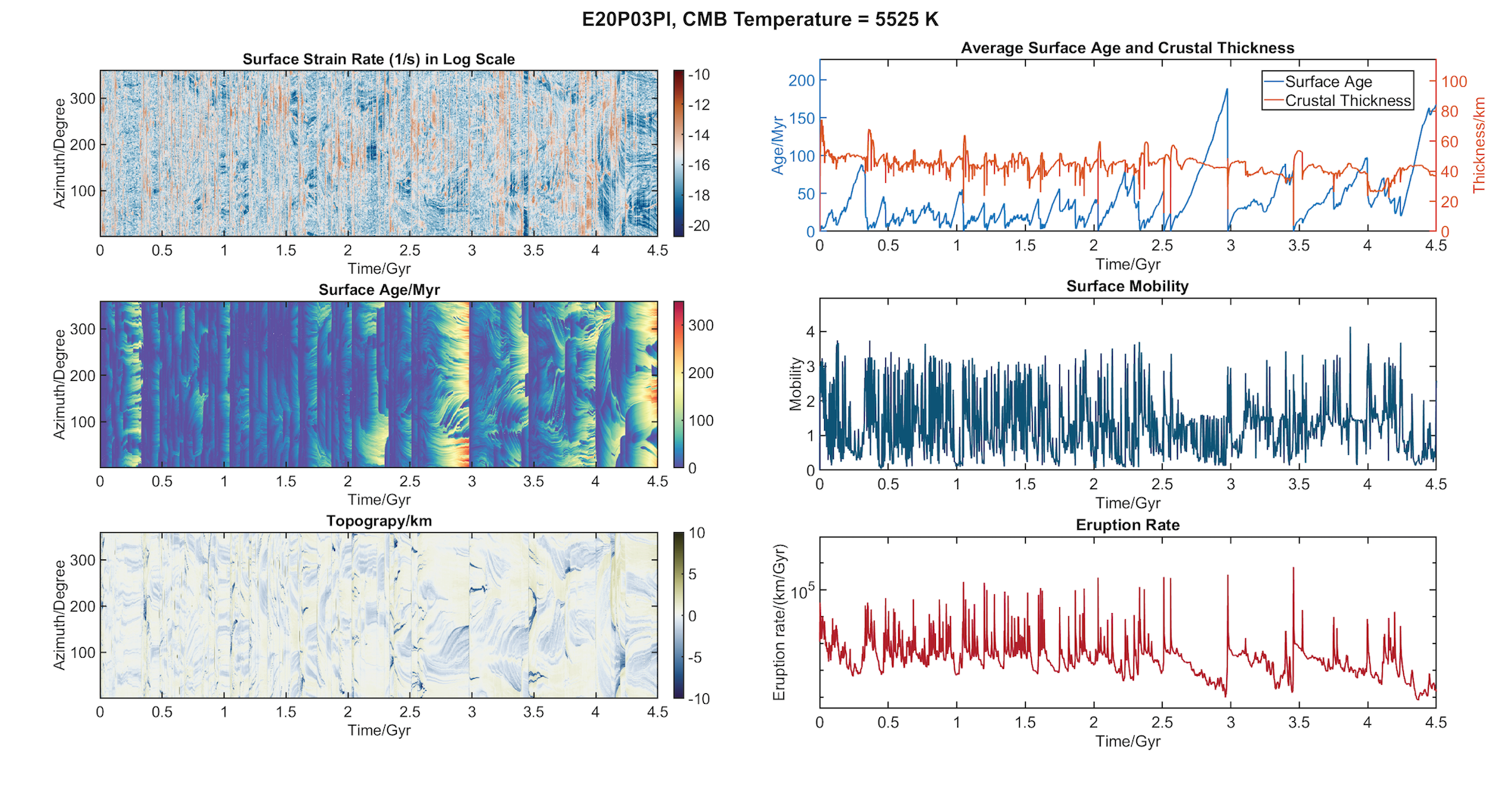}
    \caption{The evolution of E20P03Pl with $\mathrm{T_{CMB}}$ = 5525 K.}
    \label{fig:M0_5525}
\end{figure}

\begin{figure}[!p]
    \centering
    \includegraphics[width=\textwidth]{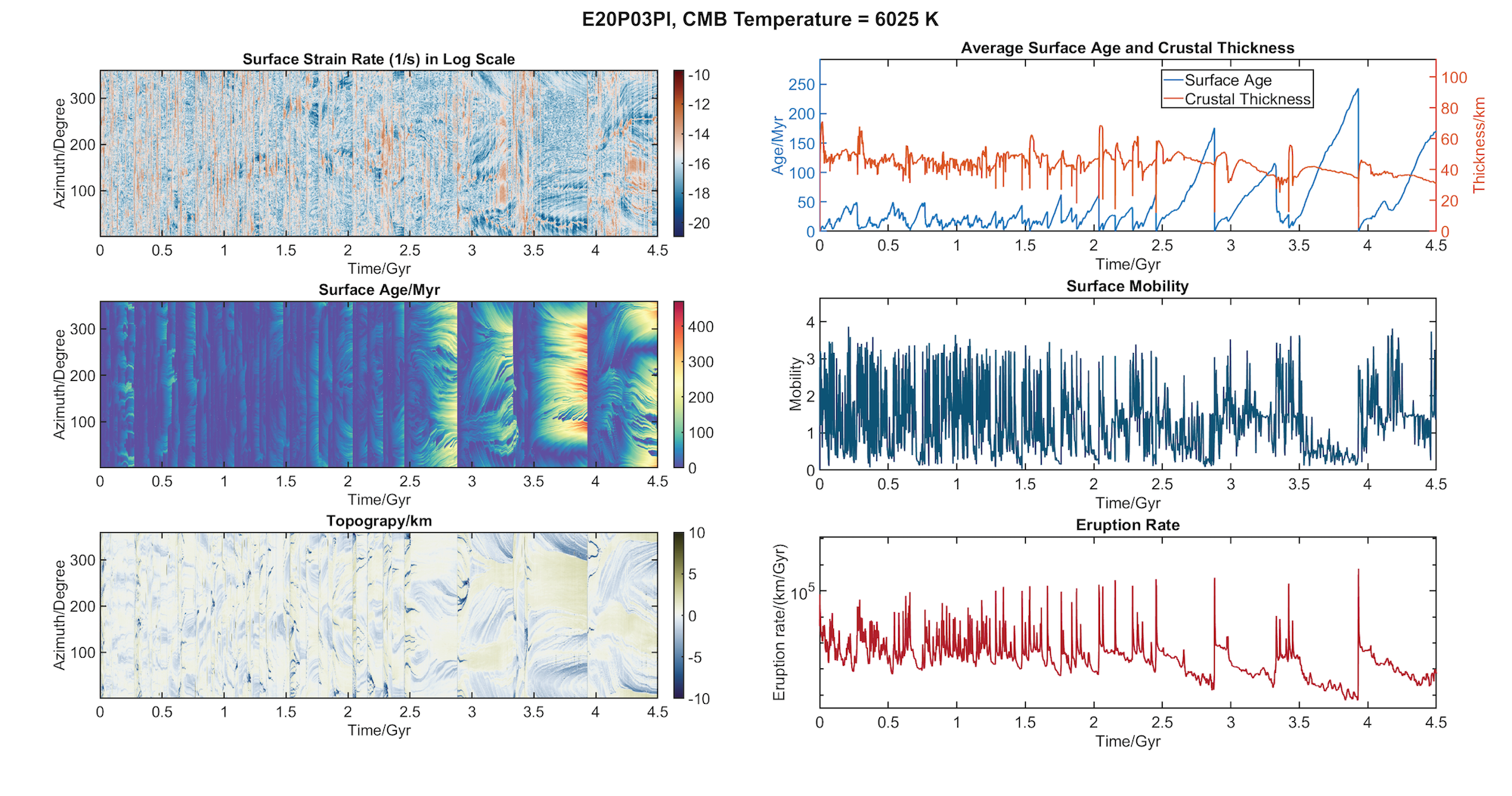}
    \caption{The evolution of E20P03Pl with $\mathrm{T_{CMB}}$ = 6025 K.}
    \label{fig:M0_6025}
\end{figure}

\newpage
\section{Supplementary Material: movies for models}
%Movies with 6 fields can be downloaded at \url{https://polybox.ethz.ch/index.php/s/GhtmzKYuYx4wRH7}

Movie S1: The evolution of model E20P03Pl

Movie S2: The evolution of model E20P03Ol

Movie S3: The evolution of model E20P00Pl

Movie S4: The evolution of model E20P00Ol

Movie S5: The evolution of model E20P03Pl\_D

Movie S6: The evolution of model E100P03Pl

\let\thefigure\thefigureSAVED
\let\thetable\thetableSAVED

\newpage
\bibliographystyle{elsarticle-harv} 
\bibliography{Venus}

\begin{thebibliography}{100}
\expandafter\ifx\csname natexlab\endcsname\relax\def\natexlab#1{#1}\fi
\providecommand{\url}[1]{\texttt{#1}}
\providecommand{\href}[2]{#2}
\providecommand{\path}[1]{#1}
\providecommand{\DOIprefix}{doi:}
\providecommand{\ArXivprefix}{arXiv:}
\providecommand{\URLprefix}{URL: }
\providecommand{\Pubmedprefix}{pmid:}
\providecommand{\doi}[1]{\href{http://dx.doi.org/#1}{\path{#1}}}
\providecommand{\Pubmed}[1]{\href{pmid:#1}{\path{#1}}}
\providecommand{\bibinfo}[2]{#2}
\ifx\xfnm\relax \def\xfnm[#1]{\unskip,\space#1}\fi
%Type = Article
\bibitem[{Ammann et~al.(2009)Ammann, Brodholt and
  Dobson}]{ammannDFTStudyMigration2009}
\bibinfo{author}{Ammann, M.W.}, \bibinfo{author}{Brodholt, J.P.},
  \bibinfo{author}{Dobson, D.P.}, \bibinfo{year}{2009}.
\newblock \bibinfo{title}{{{DFT}} study of migration enthalpies in {{MgSiO3}}
  perovskite}.
\newblock \bibinfo{journal}{Physics and Chemistry of Minerals}
  \bibinfo{volume}{36}, \bibinfo{pages}{151--158}.
\newblock \DOIprefix\doi{10.1007/s00269-008-0265-z}.
%Type = Article
\bibitem[{Ammann et~al.(2010)Ammann, Brodholt, Wookey and
  Dobson}]{ammannFirstprinciplesConstraintsDiffusion2010}
\bibinfo{author}{Ammann, M.W.}, \bibinfo{author}{Brodholt, J.P.},
  \bibinfo{author}{Wookey, J.}, \bibinfo{author}{Dobson, D.P.},
  \bibinfo{year}{2010}.
\newblock \bibinfo{title}{First-principles constraints on diffusion in
  lower-mantle minerals and a weak {{D}}{${'}{'}$} layer}.
\newblock \bibinfo{journal}{Nature} \bibinfo{volume}{465},
  \bibinfo{pages}{462--465}.
\newblock \DOIprefix\doi{10.1038/nature09052}.
%Type = Article
\bibitem[{Armann and
  Tackley(2012)}]{armannSimulatingThermochemicalMagmatic2012}
\bibinfo{author}{Armann, M.}, \bibinfo{author}{Tackley, P.J.},
  \bibinfo{year}{2012}.
\newblock \bibinfo{title}{Simulating the thermochemical magmatic and tectonic
  evolution of {{Venus}}'s mantle and lithosphere: {{Two-dimensional}} models}.
\newblock \bibinfo{journal}{Journal of Geophysical Research: Planets}
  \bibinfo{volume}{117}.
\newblock \DOIprefix\doi{10.1029/2012JE004231}.
%Type = Article
\bibitem[{Azuma et~al.(2014)Azuma, Katayama and
  Nakakuki}]{azumaRheologicalDecouplingMoho2014}
\bibinfo{author}{Azuma, S.}, \bibinfo{author}{Katayama, I.},
  \bibinfo{author}{Nakakuki, T.}, \bibinfo{year}{2014}.
\newblock \bibinfo{title}{Rheological decoupling at the {{Moho}} and
  implication to {{Venusian}} tectonics}.
\newblock \bibinfo{journal}{Scientific Reports} \bibinfo{volume}{4},
  \bibinfo{pages}{4403}.
\newblock \DOIprefix\doi{10.1038/srep04403}.
%Type = Article
\bibitem[{Bercovici and Ricard(2014)}]{bercoviciPlateTectonicsDamage2014}
\bibinfo{author}{Bercovici, D.}, \bibinfo{author}{Ricard, Y.},
  \bibinfo{year}{2014}.
\newblock \bibinfo{title}{Plate tectonics, damage and inheritance}.
\newblock \bibinfo{journal}{Nature} \bibinfo{volume}{508},
  \bibinfo{pages}{513--516}.
\newblock \DOIprefix\doi{10.1038/nature13072}.
%Type = Article
\bibitem[{Bissig et~al.(2022)Bissig, Khan and
  Giardini}]{bissigEvidenceBasaltEnrichment2022}
\bibinfo{author}{Bissig, F.}, \bibinfo{author}{Khan, A.},
  \bibinfo{author}{Giardini, D.}, \bibinfo{year}{2022}.
\newblock \bibinfo{title}{Evidence for basalt enrichment in the mantle
  transition zone from inversion of triplicated {{P-}} and {{S-waveforms}}}.
\newblock \bibinfo{journal}{Earth and Planetary Science Letters}
  \bibinfo{volume}{580}, \bibinfo{pages}{117387}.
\newblock \DOIprefix\doi{10.1016/j.epsl.2022.117387}.
%Type = Article
\bibitem[{Bjonnes et~al.(2021)Bjonnes, Johnson and
  Evans}]{bjonnesEstimatingVenusianThermal2021}
\bibinfo{author}{Bjonnes, E.}, \bibinfo{author}{Johnson, B.C.},
  \bibinfo{author}{Evans, A.J.}, \bibinfo{year}{2021}.
\newblock \bibinfo{title}{Estimating {{Venusian}} thermal conditions using
  multiring basin morphology}.
\newblock \bibinfo{journal}{Nature Astronomy} ,
  \bibinfo{pages}{1--5}\DOIprefix\doi{10.1038/s41550-020-01289-6}.
%Type = Article
\bibitem[{Bjonnes et~al.(2012)Bjonnes, Hansen, James and
  Swenson}]{bjonnesEquilibriumResurfacingVenus2012}
\bibinfo{author}{Bjonnes, E.E.}, \bibinfo{author}{Hansen, V.L.},
  \bibinfo{author}{James, B.}, \bibinfo{author}{Swenson, J.B.},
  \bibinfo{year}{2012}.
\newblock \bibinfo{title}{Equilibrium resurfacing of {{Venus}}: {{Results}}
  from new {{Monte Carlo}} modeling and implications for {{Venus}} surface
  histories}.
\newblock \bibinfo{journal}{Icarus} \bibinfo{volume}{217},
  \bibinfo{pages}{451--461}.
\newblock \DOIprefix\doi{10.1016/j.icarus.2011.03.033}.
%Type = Incollection
\bibitem[{Breuer and Moore(2015)}]{breuerTerrestrialPlanetsTreatise2015}
\bibinfo{author}{Breuer, D.}, \bibinfo{author}{Moore, W.B.},
  \bibinfo{year}{2015}.
\newblock \bibinfo{title}{10.08 - {{Dynamics}} and {{Thermal History}} of the
  {{Terrestrial Planets}}, the {{Moon}}, and {{Io}}}, in:
  \bibinfo{editor}{Schubert, G.} (Ed.), \bibinfo{booktitle}{Treatise on
  {{Geophysics}} ({{Second Edition}})}. \bibinfo{publisher}{{Elsevier}},
  \bibinfo{address}{{Oxford}}, pp. \bibinfo{pages}{255--305}.
\newblock \DOIprefix\doi{10.1016/B978-0-444-53802-4.00173-1}.
%Type = Article
\bibitem[{Brossier et~al.(2021)Brossier, Gilmore, Toner and
  Stein}]{brossierDistinctMineralogyAge2021}
\bibinfo{author}{Brossier, J.}, \bibinfo{author}{Gilmore, M.S.},
  \bibinfo{author}{Toner, K.}, \bibinfo{author}{Stein, A.J.},
  \bibinfo{year}{2021}.
\newblock \bibinfo{title}{Distinct {{Mineralogy}} and {{Age}} of {{Individual
  Lava Flows}} in {{Atla Regio}}, {{Venus Derived From Magellan Radar
  Emissivity}}}.
\newblock \bibinfo{journal}{Journal of Geophysical Research: Planets}
  \bibinfo{volume}{126}, \bibinfo{pages}{e2020JE006722}.
\newblock \DOIprefix\doi{10.1029/2020JE006722}.
%Type = Article
\bibitem[{Byrne et~al.(2021)Byrne, Ghail, {\c S}eng{\"o}r, James, Klimczak and
  Solomon}]{byrneGloballyFragmentedMobile2021}
\bibinfo{author}{Byrne, P.K.}, \bibinfo{author}{Ghail, R.C.},
  \bibinfo{author}{{\c S}eng{\"o}r, A.M.C.}, \bibinfo{author}{James, P.B.},
  \bibinfo{author}{Klimczak, C.}, \bibinfo{author}{Solomon, S.C.},
  \bibinfo{year}{2021}.
\newblock \bibinfo{title}{A globally fragmented and mobile lithosphere on
  {{Venus}}}.
\newblock \bibinfo{journal}{Proceedings of the National Academy of Sciences}
  \bibinfo{volume}{118}.
\newblock \DOIprefix\doi{10.1073/pnas.2025919118}.
%Type = Article
\bibitem[{Christensen and Yuen(1985)}]{christensen1985}
\bibinfo{author}{Christensen, U.R.}, \bibinfo{author}{Yuen, D.A.},
  \bibinfo{year}{1985}.
\newblock \bibinfo{title}{Layered convection induced by phase transitions}.
\newblock \bibinfo{journal}{J. Geophys. Res.} \bibinfo{volume}{90},
  \bibinfo{pages}{10291--10300}.
%Type = Misc
\bibitem[{Crameri(2021)}]{cramerifabioScientificColourMaps2021}
\bibinfo{author}{Crameri, F.}, \bibinfo{year}{2021}.
\newblock \bibinfo{title}{Scientific colour maps}.
\newblock \bibinfo{howpublished}{Zenodo}.
\newblock \DOIprefix\doi{10.5281/ZENODO.1243862}.
%Type = Article
\bibitem[{Crameri et~al.(2020)Crameri, Shephard and
  Heron}]{crameriMisuseColourScience2020}
\bibinfo{author}{Crameri, F.}, \bibinfo{author}{Shephard, G.E.},
  \bibinfo{author}{Heron, P.J.}, \bibinfo{year}{2020}.
\newblock \bibinfo{title}{The misuse of colour in science communication}.
\newblock \bibinfo{journal}{Nature Communications} \bibinfo{volume}{11},
  \bibinfo{pages}{5444}.
\newblock \DOIprefix\doi{10.1038/s41467-020-19160-7}.
%Type = Article
\bibitem[{Crameri and Tackley(2016)}]{crameriSubductionInitiationStagnant2016}
\bibinfo{author}{Crameri, F.}, \bibinfo{author}{Tackley, P.J.},
  \bibinfo{year}{2016}.
\newblock \bibinfo{title}{Subduction initiation from a stagnant lid and global
  overturn: New insights from numerical models with a free surface}.
\newblock \bibinfo{journal}{Progress in Earth and Planetary Science}
  \bibinfo{volume}{3}, \bibinfo{pages}{30}.
\newblock \DOIprefix\doi{10.1186/s40645-016-0103-8}.
%Type = Article
\bibitem[{Crisp(1984)}]{crispRatesMagmaEmplacement1984}
\bibinfo{author}{Crisp, J.A.}, \bibinfo{year}{1984}.
\newblock \bibinfo{title}{Rates of magma emplacement and volcanic output}.
\newblock \bibinfo{journal}{Journal of Volcanology and Geothermal Research}
  \bibinfo{volume}{20}, \bibinfo{pages}{177--211}.
\newblock \DOIprefix\doi{10.1016/0377-0273(84)90039-8}.
%Type = Article
\bibitem[{Dannberg et~al.(2017)Dannberg, Eilon, Faul, Gassmöller, Moulik and
  Myhill}]{dannbergGrainSize2017}
\bibinfo{author}{Dannberg, J.}, \bibinfo{author}{Eilon, Z.},
  \bibinfo{author}{Faul, U.}, \bibinfo{author}{Gassmöller, R.},
  \bibinfo{author}{Moulik, P.}, \bibinfo{author}{Myhill, R.},
  \bibinfo{year}{2017}.
\newblock \bibinfo{title}{The importance of grain size to mantle dynamics and
  seismological observations}.
\newblock \bibinfo{journal}{Geochemistry, Geophysics, Geosystems}
  \bibinfo{volume}{18}, \bibinfo{pages}{3034--3061}.
\newblock \DOIprefix\doi{10.1002/2017gc006944}.
%Type = Article
\bibitem[{Davies(2008)}]{daviesBasaltBarrier2008}
\bibinfo{author}{Davies, G.F.}, \bibinfo{year}{2008}.
\newblock \bibinfo{title}{Episodic layering of the early mantle by the `basalt
  barrier' mechanism}.
\newblock \bibinfo{journal}{Earth and Planetary Science Letters}
  \bibinfo{volume}{275}, \bibinfo{pages}{382--392}.
\newblock \DOIprefix\doi{10.1016/j.epsl.2008.08.036}.
%Type = Article
\bibitem[{Gerya(2014)}]{geryaPlumeinducedCrustalConvection2014}
\bibinfo{author}{Gerya, T.V.}, \bibinfo{year}{2014}.
\newblock \bibinfo{title}{Plume-induced crustal convection: {{3D}}
  thermomechanical model and implications for the origin of novae and coronae
  on {{Venus}}}.
\newblock \bibinfo{journal}{Earth and Planetary Science Letters}
  \bibinfo{volume}{391}, \bibinfo{pages}{183--192}.
\newblock \DOIprefix\doi{10.1016/j.epsl.2014.02.005}.
%Type = Article
\bibitem[{Ghail(2015)}]{ghailRheologicalPetrologicalImplications2015}
\bibinfo{author}{Ghail, R.}, \bibinfo{year}{2015}.
\newblock \bibinfo{title}{Rheological and petrological implications for a
  stagnant lid regime on {{Venus}}}.
\newblock \bibinfo{journal}{Planetary and Space Science}
  \bibinfo{volume}{113--114}, \bibinfo{pages}{2--9}.
\newblock \DOIprefix\doi{10.1016/j.pss.2015.02.005}.
%Type = Article
\bibitem[{Gillmann and Tackley(2014)}]{gillmannAtmosphereMantleCoupling2014}
\bibinfo{author}{Gillmann, C.}, \bibinfo{author}{Tackley, P.},
  \bibinfo{year}{2014}.
\newblock \bibinfo{title}{Atmosphere/mantle coupling and feedbacks on
  {{Venus}}}.
\newblock \bibinfo{journal}{Journal of Geophysical Research: Planets}
  \bibinfo{volume}{119}, \bibinfo{pages}{1189--1217}.
\newblock \DOIprefix\doi{10.1002/2013JE004505}.
%Type = Article
\bibitem[{Gr{\'e}aux et~al.(2019)Gr{\'e}aux, Irifune, Higo, Tange, Arimoto, Liu
  and Yamada}]{greauxSoundVelocityCaSiO32019}
\bibinfo{author}{Gr{\'e}aux, S.}, \bibinfo{author}{Irifune, T.},
  \bibinfo{author}{Higo, Y.}, \bibinfo{author}{Tange, Y.},
  \bibinfo{author}{Arimoto, T.}, \bibinfo{author}{Liu, Z.},
  \bibinfo{author}{Yamada, A.}, \bibinfo{year}{2019}.
\newblock \bibinfo{title}{Sound velocity of {{CaSiO3}} perovskite suggests the
  presence of basaltic crust in the {{Earth}}'s lower mantle}.
\newblock \bibinfo{journal}{Nature} \bibinfo{volume}{565},
  \bibinfo{pages}{218--221}.
\newblock \DOIprefix\doi{10.1038/s41586-018-0816-5}.
%Type = Article
\bibitem[{Grimm(1994)}]{grimmRecentDeformationRates1994}
\bibinfo{author}{Grimm, R.E.}, \bibinfo{year}{1994}.
\newblock \bibinfo{title}{Recent deformation rates on {{Venus}}}.
\newblock \bibinfo{journal}{Journal of Geophysical Research: Planets}
  \bibinfo{volume}{99}, \bibinfo{pages}{23163--23171}.
\newblock \DOIprefix\doi{10.1029/94JE02196}.
%Type = Article
\bibitem[{G{\"u}lcher et~al.(2020)G{\"u}lcher, Gerya, Mont{\'e}si and
  Munch}]{gulcherCoronaStructuresDriven2020}
\bibinfo{author}{G{\"u}lcher, A.J.P.}, \bibinfo{author}{Gerya, T.V.},
  \bibinfo{author}{Mont{\'e}si, L.G.J.}, \bibinfo{author}{Munch, J.},
  \bibinfo{year}{2020}.
\newblock \bibinfo{title}{Corona structures driven by plume\textendash
  lithosphere interactions and evidence for ongoing plume activity on
  {{Venus}}}.
\newblock \bibinfo{journal}{Nature Geoscience} \bibinfo{volume}{13},
  \bibinfo{pages}{547--554}.
\newblock \DOIprefix\doi{10.1038/s41561-020-0606-1}.
%Type = Article
\bibitem[{Hansen et~al.(1999)Hansen, Banks and
  Ghent}]{hansenTesseraTerrainCrustal1999}
\bibinfo{author}{Hansen, V.L.}, \bibinfo{author}{Banks, B.K.},
  \bibinfo{author}{Ghent, R.R.}, \bibinfo{year}{1999}.
\newblock \bibinfo{title}{Tessera terrain and crustal plateaus, {{Venus}}}.
\newblock \bibinfo{journal}{Geology} \bibinfo{volume}{27},
  \bibinfo{pages}{1071--1074}.
\newblock \DOIprefix\doi{10.1130/0091-7613(1999)027<1071:TTACPV>2.3.CO;2}.
%Type = Article
\bibitem[{Harris and
  B{\'e}dard(2015)}]{harrisInteractionsContinentlikeDrift2015}
\bibinfo{author}{Harris, L.B.}, \bibinfo{author}{B{\'e}dard, J.H.},
  \bibinfo{year}{2015}.
\newblock \bibinfo{title}{Interactions between continent-like `drift', rifting
  and mantle flow on {{Venus}}: Gravity interpretations and {{Earth}}
  analogues}.
\newblock \bibinfo{journal}{Geological Society, London, Special Publications}
  \bibinfo{volume}{401}, \bibinfo{pages}{327--356}.
\newblock \DOIprefix\doi{10.1144/SP401.9}.
%Type = Article
\bibitem[{Hauck et~al.(1998)Hauck, Phillips and
  Price}]{hauckVenusCraterDistribution1998}
\bibinfo{author}{Hauck, S.A.}, \bibinfo{author}{Phillips, R.J.},
  \bibinfo{author}{Price, M.H.}, \bibinfo{year}{1998}.
\newblock \bibinfo{title}{Venus: {{Crater}} distribution and plains resurfacing
  models}.
\newblock \bibinfo{journal}{Journal of Geophysical Research: Planets}
  \bibinfo{volume}{103}, \bibinfo{pages}{13635--13642}.
\newblock \DOIprefix\doi{10.1029/98JE00400}.
%Type = Article
\bibitem[{Hernlund and Tackley(2008)}]{hernlundModelingMantleConvection2008}
\bibinfo{author}{Hernlund, J.W.}, \bibinfo{author}{Tackley, P.J.},
  \bibinfo{year}{2008}.
\newblock \bibinfo{title}{Modeling mantle convection in the spherical annulus}.
\newblock \bibinfo{journal}{Physics of the Earth and Planetary Interiors}
  \bibinfo{volume}{171}, \bibinfo{pages}{48--54}.
\newblock \DOIprefix\doi{10.1016/j.pepi.2008.07.037}.
%Type = Article
\bibitem[{Herrick and Rumpf(2011)}]{herrickPostimpactModificationVolcanic2011}
\bibinfo{author}{Herrick, R.R.}, \bibinfo{author}{Rumpf, M.E.},
  \bibinfo{year}{2011}.
\newblock \bibinfo{title}{Postimpact modification by volcanic or tectonic
  processes as the rule, not the exception, for {{Venusian}} craters}.
\newblock \bibinfo{journal}{Journal of Geophysical Research-Planets}
  \bibinfo{volume}{116}, \bibinfo{pages}{E02004}.
\newblock \DOIprefix\doi{10.1029/2010JE003722}.
%Type = Article
\bibitem[{Herzberg et~al.(2000)Herzberg, Raterron and
  Zhang}]{herzbergNewExperimentalObservations2000}
\bibinfo{author}{Herzberg, C.}, \bibinfo{author}{Raterron, P.},
  \bibinfo{author}{Zhang, J.}, \bibinfo{year}{2000}.
\newblock \bibinfo{title}{New experimental observations on the anhydrous
  solidus for peridotite {{KLB-1}}}.
\newblock \bibinfo{journal}{Geochemistry, Geophysics, Geosystems}
  \bibinfo{volume}{1}.
\newblock \DOIprefix\doi{10.1029/2000GC000089}.
%Type = Article
\bibitem[{Hess and Dingwell(1996)}]{hessViscositiesHydrousLeucogranitic1996}
\bibinfo{author}{Hess, K.U.}, \bibinfo{author}{Dingwell, D.B.},
  \bibinfo{year}{1996}.
\newblock \bibinfo{title}{Viscosities of hydrous leucogranitic melts: {{A}}
  non-{{Arrhenian}} model}.
\newblock \bibinfo{journal}{American Mineralogist} \bibinfo{volume}{81},
  \bibinfo{pages}{1297--1300}.
%Type = Article
\bibitem[{Huang et~al.(2013)Huang, Yang and
  Zhong}]{huangConstraintsTopographyGravity2013}
\bibinfo{author}{Huang, J.}, \bibinfo{author}{Yang, A.},
  \bibinfo{author}{Zhong, S.}, \bibinfo{year}{2013}.
\newblock \bibinfo{title}{Constraints of the topography, gravity and volcanism
  on {{Venusian}} mantle dynamics and generation of plate tectonics}.
\newblock \bibinfo{journal}{Earth and Planetary Science Letters}
  \bibinfo{volume}{362}, \bibinfo{pages}{207--214}.
\newblock \DOIprefix\doi{10.1016/j.epsl.2012.11.051}.
%Type = Article
\bibitem[{Ivanov and Head(2015)}]{ivanovHistoryTectonismVenus2015}
\bibinfo{author}{Ivanov, M.A.}, \bibinfo{author}{Head, J.W.},
  \bibinfo{year}{2015}.
\newblock \bibinfo{title}{The history of tectonism on {{Venus}}: {{A}}
  stratigraphic analysis}.
\newblock \bibinfo{journal}{Planetary and Space Science}
  \bibinfo{volume}{113--114}, \bibinfo{pages}{10--32}.
\newblock \DOIprefix\doi{10.1016/j.pss.2015.03.016}.
%Type = Article
\bibitem[{James et~al.(2013)James, Zuber and
  Phillips}]{jamesCrustalThicknessSupport2013}
\bibinfo{author}{James, P.B.}, \bibinfo{author}{Zuber, M.T.},
  \bibinfo{author}{Phillips, R.J.}, \bibinfo{year}{2013}.
\newblock \bibinfo{title}{Crustal thickness and support of topography on
  {{Venus}}}.
\newblock \bibinfo{journal}{Journal of Geophysical Research-Planets}
  \bibinfo{volume}{118}, \bibinfo{pages}{859--875}.
\newblock \DOIprefix\doi{10.1029/2012JE004237}.
%Type = Article
\bibitem[{{Jimenez-Diaz} et~al.(2015){Jimenez-Diaz}, Ruiz, Kirby, Romeo, Tejero
  and Capote}]{jimenez-diazLithosphericStructureVenus2015}
\bibinfo{author}{{Jimenez-Diaz}, A.}, \bibinfo{author}{Ruiz, J.},
  \bibinfo{author}{Kirby, J.F.}, \bibinfo{author}{Romeo, I.},
  \bibinfo{author}{Tejero, R.}, \bibinfo{author}{Capote, R.},
  \bibinfo{year}{2015}.
\newblock \bibinfo{title}{Lithospheric structure of {{Venus}} from gravity and
  topography}.
\newblock \bibinfo{journal}{Icarus} \bibinfo{volume}{260},
  \bibinfo{pages}{215--231}.
\newblock \DOIprefix\doi{10.1016/j.icarus.2015.07.020}.
%Type = Article
\bibitem[{Karato and Wu(1993)}]{karatoRheologyUpperMantle1993}
\bibinfo{author}{Karato, S.i.}, \bibinfo{author}{Wu, P.}, \bibinfo{year}{1993}.
\newblock \bibinfo{title}{Rheology of the {{Upper Mantle}}: {{A Synthesis}}}.
\newblock \bibinfo{journal}{Science} \bibinfo{volume}{260},
  \bibinfo{pages}{771--778}.
\newblock \DOIprefix\doi{10.1126/science.260.5109.771}.
%Type = Article
\bibitem[{Kiefer et~al.(2015)Kiefer, Filiberto, Sandu and
  Li}]{kieferEffectsMantleComposition2015}
\bibinfo{author}{Kiefer, W.S.}, \bibinfo{author}{Filiberto, J.},
  \bibinfo{author}{Sandu, C.}, \bibinfo{author}{Li, Q.}, \bibinfo{year}{2015}.
\newblock \bibinfo{title}{The effects of mantle composition on the peridotite
  solidus: {{Implications}} for the magmatic history of {{Mars}}}.
\newblock \bibinfo{journal}{Geochimica et Cosmochimica Acta}
  \bibinfo{volume}{162}, \bibinfo{pages}{247--258}.
\newblock \DOIprefix\doi{10.1016/j.gca.2015.02.010}.
%Type = Article
\bibitem[{Kreslavsky et~al.(2015)Kreslavsky, Ivanov and
  Head}]{kreslavskyResurfacingHistoryVenus2015}
\bibinfo{author}{Kreslavsky, M.A.}, \bibinfo{author}{Ivanov, M.A.},
  \bibinfo{author}{Head, J.W.}, \bibinfo{year}{2015}.
\newblock \bibinfo{title}{The resurfacing history of {{Venus}}: {{Constraints}}
  from buffered crater densities}.
\newblock \bibinfo{journal}{Icarus} \bibinfo{volume}{250},
  \bibinfo{pages}{438--450}.
\newblock \DOIprefix\doi{10.1016/j.icarus.2014.12.024}.
%Type = Article
\bibitem[{Le~Feuvre and Wieczorek(2011)}]{lefeuvreNonuniformCrateringMoon2011}
\bibinfo{author}{Le~Feuvre, M.}, \bibinfo{author}{Wieczorek, M.A.},
  \bibinfo{year}{2011}.
\newblock \bibinfo{title}{Nonuniform cratering of the {{Moon}} and a revised
  crater chronology of the inner {{Solar System}}}.
\newblock \bibinfo{journal}{Icarus} \bibinfo{volume}{214},
  \bibinfo{pages}{1--20}.
\newblock \DOIprefix\doi{10.1016/j.icarus.2011.03.010}.
%Type = Article
\bibitem[{Louren{\c c}o et~al.(2016)Louren{\c c}o, Rozel and
  Tackley}]{lourencoMeltinginducedCrustalProduction2016a}
\bibinfo{author}{Louren{\c c}o, D.L.}, \bibinfo{author}{Rozel, A.},
  \bibinfo{author}{Tackley, P.J.}, \bibinfo{year}{2016}.
\newblock \bibinfo{title}{Melting-induced crustal production helps plate
  tectonics on {{Earth-like}} planets}.
\newblock \bibinfo{journal}{Earth and Planetary Science Letters}
  \bibinfo{volume}{439}, \bibinfo{pages}{18--28}.
\newblock \DOIprefix\doi{10.1016/j.epsl.2016.01.024}.
%Type = Article
\bibitem[{Louren{\c c}o et~al.(2020)Louren{\c c}o, Rozel, Ballmer and
  Tackley}]{lourencoPlutonicSquishyLidNew2020}
\bibinfo{author}{Louren{\c c}o, D.L.}, \bibinfo{author}{Rozel, A.B.},
  \bibinfo{author}{Ballmer, M.D.}, \bibinfo{author}{Tackley, P.J.},
  \bibinfo{year}{2020}.
\newblock \bibinfo{title}{Plutonic-{{Squishy Lid}}: {{A New Global Tectonic
  Regime Generated}} by {{Intrusive Magmatism}} on {{Earth-Like Planets}}}.
\newblock \bibinfo{journal}{Geochemistry, Geophysics, Geosystems}
  \bibinfo{volume}{21}, \bibinfo{pages}{e2019GC008756}.
\newblock \DOIprefix\doi{10.1029/2019GC008756}.
%Type = Article
\bibitem[{Louren{\c c}o et~al.(2018)Louren{\c c}o, Rozel, Gerya and
  Tackley}]{lourencoEfficientCoolingRocky2018}
\bibinfo{author}{Louren{\c c}o, D.L.}, \bibinfo{author}{Rozel, A.B.},
  \bibinfo{author}{Gerya, T.}, \bibinfo{author}{Tackley, P.J.},
  \bibinfo{year}{2018}.
\newblock \bibinfo{title}{Efficient cooling of rocky planets by intrusive
  magmatism}.
\newblock \bibinfo{journal}{Nature Geoscience} \bibinfo{volume}{11},
  \bibinfo{pages}{322--327}.
\newblock \DOIprefix\doi{10.1038/s41561-018-0094-8}.
%Type = Article
\bibitem[{Maal{\o}e(2004)}]{maaloeSolidusHarzburgiteGPa2004}
\bibinfo{author}{Maal{\o}e, S.}, \bibinfo{year}{2004}.
\newblock \bibinfo{title}{The solidus of harzburgite to 3\,{{GPa}} pressure:
  The compositions of primary abyssal tholeiite}.
\newblock \bibinfo{journal}{Mineralogy and Petrology} \bibinfo{volume}{81},
  \bibinfo{pages}{1--17}.
\newblock \DOIprefix\doi{10.1007/s00710-004-0028-6}.
%Type = Article
\bibitem[{Mackwell et~al.(1998)Mackwell, Zimmerman and
  Kohlstedt}]{mackwellHightemperatureDeformationDry1998}
\bibinfo{author}{Mackwell, S.J.}, \bibinfo{author}{Zimmerman, M.E.},
  \bibinfo{author}{Kohlstedt, D.L.}, \bibinfo{year}{1998}.
\newblock \bibinfo{title}{High-temperature deformation of dry diabase with
  application to tectonics on {{Venus}}}.
\newblock \bibinfo{journal}{Journal of Geophysical Research: Solid Earth}
  \bibinfo{volume}{103}, \bibinfo{pages}{975--984}.
\newblock \DOIprefix\doi{10.1029/97JB02671}.
%Type = Article
\bibitem[{Maia and Wieczorek(2022)}]{maiaLithosphericStructureVenusian2022}
\bibinfo{author}{Maia, J.S.}, \bibinfo{author}{Wieczorek, M.A.},
  \bibinfo{year}{2022}.
\newblock \bibinfo{title}{Lithospheric {{Structure}} of {{Venusian Crustal
  Plateaus}}}.
\newblock \bibinfo{journal}{Journal of Geophysical Research: Planets}
  \bibinfo{volume}{127}, \bibinfo{pages}{e2021JE007004}.
\newblock \DOIprefix\doi{10.1029/2021JE007004}.
%Type = Article
\bibitem[{Marcq et~al.(2013)Marcq, Bertaux, Montmessin and
  Belyaev}]{marcqVariationsSulphurDioxide2013}
\bibinfo{author}{Marcq, E.}, \bibinfo{author}{Bertaux, J.L.},
  \bibinfo{author}{Montmessin, F.}, \bibinfo{author}{Belyaev, D.},
  \bibinfo{year}{2013}.
\newblock \bibinfo{title}{Variations of sulphur dioxide at the cloud top of
  {{Venus}}'s dynamic atmosphere}.
\newblock \bibinfo{journal}{Nature Geoscience} \bibinfo{volume}{6},
  \bibinfo{pages}{25--28}.
\newblock \DOIprefix\doi{10.1038/ngeo1650}.
%Type = Incollection
\bibitem[{McKinnon et~al.(1997)McKinnon, Zahnle, Ivanov and
  Melosh}]{mckinnonCrateringVenusModels1997}
\bibinfo{author}{McKinnon, W.B.}, \bibinfo{author}{Zahnle, K.J.},
  \bibinfo{author}{Ivanov, B.A.}, \bibinfo{author}{Melosh, H.J.},
  \bibinfo{year}{1997}.
\newblock \bibinfo{title}{Cratering on {{Venus}}: {{Models}} and
  {{Observations}}}, in: \bibinfo{booktitle}{Venus {{II}} - {{Geology}},
  {{Geophysics}}, {{Atmosphere}}, and {{Solar Wind Experiment}}}.
  \bibinfo{publisher}{{The University of Arizona Press}}, pp.
  \bibinfo{pages}{1047--1086}.
%Type = Article
\bibitem[{Moresi and Solomatov(1998)}]{moresiMantleConvectionBrittle1998}
\bibinfo{author}{Moresi, L.}, \bibinfo{author}{Solomatov, V.},
  \bibinfo{year}{1998}.
\newblock \bibinfo{title}{Mantle convection with a brittle lithosphere:
  Thoughts on the global tectonic styles of the {{Earth}} and {{Venus}}}.
\newblock \bibinfo{journal}{Geophysical Journal International}
  \bibinfo{volume}{133}, \bibinfo{pages}{669--682}.
\newblock \DOIprefix\doi{10.1046/j.1365-246X.1998.00521.x}.
%Type = Article
\bibitem[{Nakagawa and Buffett(2005)}]{nakagawaMassTransportMechanism2005}
\bibinfo{author}{Nakagawa, T.}, \bibinfo{author}{Buffett, B.A.},
  \bibinfo{year}{2005}.
\newblock \bibinfo{title}{Mass transport mechanism between the upper and lower
  mantle in numerical simulations of thermochemical mantle convection with
  multicomponent phase changes}.
\newblock \bibinfo{journal}{Earth and Planetary Science Letters}
  \bibinfo{volume}{230}, \bibinfo{pages}{11--27}.
\newblock \DOIprefix\doi{10.1016/j.epsl.2004.11.005}.
%Type = Article
\bibitem[{Nakagawa and Tackley(2004)}]{nakagawaEffectsThermochemicalMantle2004}
\bibinfo{author}{Nakagawa, T.}, \bibinfo{author}{Tackley, P.J.},
  \bibinfo{year}{2004}.
\newblock \bibinfo{title}{Effects of thermo-chemical mantle convection on the
  thermal evolution of the {{Earth}}'s core}.
\newblock \bibinfo{journal}{Earth and Planetary Science Letters}
  \bibinfo{volume}{220}, \bibinfo{pages}{107--119}.
\newblock \DOIprefix\doi{10.1016/S0012-821X(04)00055-X}.
%Type = Article
\bibitem[{Nakagawa and Tackley(2005)}]{nakagawaCoreEvo2005}
\bibinfo{author}{Nakagawa, T.}, \bibinfo{author}{Tackley, P.J.},
  \bibinfo{year}{2005}.
\newblock \bibinfo{title}{Deep mantle heat flow and thermal evolution of the
  {{Earth}}'s core in thermochemical multiphase models of mantle convection}.
\newblock \bibinfo{journal}{Geochemistry, Geophysics, Geosystems}
  \bibinfo{volume}{6}.
\newblock \DOIprefix\doi{10.1029/2005GC000967}.
%Type = Article
\bibitem[{Nakagawa and Tackley(2010)}]{nakagawaInfluenceInitialCMB2010}
\bibinfo{author}{Nakagawa, T.}, \bibinfo{author}{Tackley, P.J.},
  \bibinfo{year}{2010}.
\newblock \bibinfo{title}{Influence of initial {{CMB}} temperature and other
  parameters on the thermal evolution of {{Earth}}'s core resulting from
  thermochemical spherical mantle convection}.
\newblock \bibinfo{journal}{Geochemistry, Geophysics, Geosystems}
  \bibinfo{volume}{11}.
\newblock \DOIprefix\doi{10.1029/2010GC003031}.
%Type = Article
\bibitem[{Nakagawa and Tackley(2014)}]{nakagawaCoreEvo2014}
\bibinfo{author}{Nakagawa, T.}, \bibinfo{author}{Tackley, P.J.},
  \bibinfo{year}{2014}.
\newblock \bibinfo{title}{Influence of combined primordial layering and
  recycled morb on the coupled thermal evolution of earth's mantle and core}.
\newblock \bibinfo{journal}{Geochem. Geophys. Geosyst.} \bibinfo{volume}{15},
  \bibinfo{pages}{619--633}.
\newblock \DOIprefix\doi{10.1002/2013GC005128}.
%Type = Article
\bibitem[{Nakagawa and Tackley(2015)}]{nakagawaInfluencePlateTectonic2015}
\bibinfo{author}{Nakagawa, T.}, \bibinfo{author}{Tackley, P.J.},
  \bibinfo{year}{2015}.
\newblock \bibinfo{title}{Influence of plate tectonic mode on the coupled
  thermochemical evolution of {{Earth}}'s mantle and core}.
\newblock \bibinfo{journal}{Geochemistry, Geophysics, Geosystems}
  \bibinfo{volume}{16}, \bibinfo{pages}{3400--3413}.
\newblock \DOIprefix\doi{10.1002/2015GC005996}.
%Type = Article
\bibitem[{Nakagawa et~al.(2009)Nakagawa, Tackley, Deschamps and
  Connolly}]{nakagawaMineralPhysics2009}
\bibinfo{author}{Nakagawa, T.}, \bibinfo{author}{Tackley, P.J.},
  \bibinfo{author}{Deschamps, F.}, \bibinfo{author}{Connolly, J.A.D.},
  \bibinfo{year}{2009}.
\newblock \bibinfo{title}{Incorporating self-consistently calculated mineral
  physics into thermo-chemical mantle convection simulations in a 3d spherical
  shell and its influence on seismic anomalies in earth’s mantle}.
\newblock \bibinfo{journal}{Geochem. Geophys. Geosyst.} \bibinfo{volume}{10},
  \bibinfo{pages}{doi:10.1029/2008GC002280}.
\newblock \DOIprefix\doi{10.1029/2008GC002280}.
%Type = Article
\bibitem[{Nakagawa et~al.(2010)Nakagawa, Tackley, Deschamps and
  Connolly}]{nakagawaSelfConsistentMineralPhysics2010}
\bibinfo{author}{Nakagawa, T.}, \bibinfo{author}{Tackley, P.J.},
  \bibinfo{author}{Deschamps, F.}, \bibinfo{author}{Connolly, J.A.D.},
  \bibinfo{year}{2010}.
\newblock \bibinfo{title}{The influence of {{MORB}} and harzburgite composition
  on thermo-chemical mantle convection in a 3-{{D}} spherical shell with
  self-consistently calculated mineral physics}.
\newblock \bibinfo{journal}{Earth and Planetary Science Letters}
  \bibinfo{volume}{296}, \bibinfo{pages}{403--412}.
\newblock \DOIprefix\doi{10.1016/j.epsl.2010.05.026}.
%Type = Article
\bibitem[{Nimmo and McKenzie(1998)}]{nimmoVolcanismTectonicsVenus1998}
\bibinfo{author}{Nimmo, F.}, \bibinfo{author}{McKenzie, D.},
  \bibinfo{year}{1998}.
\newblock \bibinfo{title}{Volcanism and {{Tectonics}} on {{Venus}}}.
\newblock \bibinfo{journal}{Annual Review of Earth and Planetary Sciences}
  \bibinfo{volume}{26}, \bibinfo{pages}{23--51}.
\newblock \DOIprefix\doi{10.1146/annurev.earth.26.1.23}.
%Type = Article
\bibitem[{Ogawa(2003)}]{ogawaChemicalStratificationTwodimensional2003}
\bibinfo{author}{Ogawa, M.}, \bibinfo{year}{2003}.
\newblock \bibinfo{title}{Chemical stratification in a two-dimensional
  convecting mantle with magmatism and moving plates}.
\newblock \bibinfo{journal}{Journal of Geophysical Research: Solid Earth}
  \bibinfo{volume}{108}.
\newblock \DOIprefix\doi{10.1029/2002JB002205}.
%Type = Article
\bibitem[{Ogawa and Yanagisawa(2014)}]{ogawaMantleEvolutionVenus2014}
\bibinfo{author}{Ogawa, M.}, \bibinfo{author}{Yanagisawa, T.},
  \bibinfo{year}{2014}.
\newblock \bibinfo{title}{Mantle evolution in {{Venus}} due to magmatism and
  phase transitions: {{From}} punctuated layered convection to whole-mantle
  convection}.
\newblock \bibinfo{journal}{Journal of Geophysical Research: Planets}
  \bibinfo{volume}{119}, \bibinfo{pages}{867--883}.
\newblock \DOIprefix\doi{10.1002/2013JE004593}.
%Type = Article
\bibitem[{O'Reilly and Davies(1981)}]{oreillyIo1981}
\bibinfo{author}{O'Reilly, T.C.}, \bibinfo{author}{Davies, G.F.},
  \bibinfo{year}{1981}.
\newblock \bibinfo{title}{Magma transport of heat on io: A mechanism allowing a
  thick lithosphere}.
\newblock \bibinfo{journal}{Geophys. Res. Lett.} \bibinfo{volume}{8},
  \bibinfo{pages}{313--316}.
%Type = Article
\bibitem[{O'Rourke et~al.(2019)O'Rourke, Buz, Fu and
  Lillis}]{orourkeDetectabilityRemanentMagnetism2019}
\bibinfo{author}{O'Rourke, J.G.}, \bibinfo{author}{Buz, J.},
  \bibinfo{author}{Fu, R.R.}, \bibinfo{author}{Lillis, R.J.},
  \bibinfo{year}{2019}.
\newblock \bibinfo{title}{Detectability of {{Remanent Magnetism}} in the
  {{Crust}} of {{Venus}}}.
\newblock \bibinfo{journal}{Geophysical Research Letters} \bibinfo{volume}{46},
  \bibinfo{pages}{5768--5777}.
\newblock \DOIprefix\doi{10.1029/2019GL082725}.
%Type = Article
\bibitem[{O'Rourke and Korenaga(2015)}]{orourkeThermalEvolutionVenus2015}
\bibinfo{author}{O'Rourke, J.G.}, \bibinfo{author}{Korenaga, J.},
  \bibinfo{year}{2015}.
\newblock \bibinfo{title}{Thermal evolution of {{Venus}} with argon degassing}.
\newblock \bibinfo{journal}{Icarus} \bibinfo{volume}{260},
  \bibinfo{pages}{128--140}.
\newblock \DOIprefix\doi{10.1016/j.icarus.2015.07.009}.
%Type = Article
\bibitem[{Orth and Solomatov(2012)}]{orthConstraintsVenusianCrustal2012}
\bibinfo{author}{Orth, C.P.}, \bibinfo{author}{Solomatov, V.S.},
  \bibinfo{year}{2012}.
\newblock \bibinfo{title}{Constraints on the {{Venusian}} crustal thickness
  variations in the isostatic stagnant lid approximation}.
\newblock \bibinfo{journal}{Geochemistry, Geophysics, Geosystems}
  \bibinfo{volume}{13}.
\newblock \DOIprefix\doi{10.1029/2012GC004377}.
%Type = Incollection
\bibitem[{Palme and O'Neill(2007)}]{palmeTreatise2003}
\bibinfo{author}{Palme, H.}, \bibinfo{author}{O'Neill, H.S.C.},
  \bibinfo{year}{2007}.
\newblock \bibinfo{title}{2.01 - {{Cosmochemical Estimates}} of {{Mantle
  Composition}}}, in: \bibinfo{editor}{Holland, H.D.},
  \bibinfo{editor}{Turekian, K.K.} (Eds.), \bibinfo{booktitle}{Treatise on
  {{Geochemistry}}}. \bibinfo{publisher}{{Pergamon}},
  \bibinfo{address}{{Oxford}}, pp. \bibinfo{pages}{1--38}.
\newblock \DOIprefix\doi{10.1016/B0-08-043751-6/02177-0}.
%Type = Article
\bibitem[{Papuc and Davies(2012)}]{papucBasaltBarrierVenus2012}
\bibinfo{author}{Papuc, A.M.}, \bibinfo{author}{Davies, G.F.},
  \bibinfo{year}{2012}.
\newblock \bibinfo{title}{Transient mantle layering and the episodic behaviour
  of {{Venus}} due to the `basalt barrier' mechanism}.
\newblock \bibinfo{journal}{Icarus} \bibinfo{volume}{217},
  \bibinfo{pages}{499--509}.
\newblock \DOIprefix\doi{10.1016/j.icarus.2011.09.024}.
%Type = Book
\bibitem[{Ranalli(1995)}]{ranalliRheologyEarth1995}
\bibinfo{author}{Ranalli, G.}, \bibinfo{year}{1995}.
\newblock \bibinfo{title}{Rheology of the {{Earth}}}.
\newblock \bibinfo{edition}{Second} ed., \bibinfo{publisher}{{Springer
  Netherlands}}.
%Type = Article
\bibitem[{Resor et~al.(2021)Resor, Gilmore, Straley, Senske and
  Herrick}]{resorFelsicTesseraeVenus2021}
\bibinfo{author}{Resor, P.G.}, \bibinfo{author}{Gilmore, M.S.},
  \bibinfo{author}{Straley, B.}, \bibinfo{author}{Senske, D.A.},
  \bibinfo{author}{Herrick, R.R.}, \bibinfo{year}{2021}.
\newblock \bibinfo{title}{Felsic {{Tesserae}} on {{Venus Permitted}} by
  {{Lithospheric Deformation Models}}}.
\newblock \bibinfo{journal}{Journal of Geophysical Research: Planets}
  \bibinfo{volume}{126}, \bibinfo{pages}{e2020JE006642}.
\newblock \DOIprefix\doi{10.1029/2020JE006642}.
%Type = Article
\bibitem[{Rolf et~al.(2018)Rolf, Steinberger, Sruthi and
  Werner}]{rolfInferencesMantleViscosity2018}
\bibinfo{author}{Rolf, T.}, \bibinfo{author}{Steinberger, B.},
  \bibinfo{author}{Sruthi, U.}, \bibinfo{author}{Werner, S.C.},
  \bibinfo{year}{2018}.
\newblock \bibinfo{title}{Inferences on the mantle viscosity structure and the
  post-overturn evolutionary state of {{Venus}}}.
\newblock \bibinfo{journal}{Icarus} \bibinfo{volume}{313},
  \bibinfo{pages}{107--123}.
\newblock \DOIprefix\doi{10.1016/j.icarus.2018.05.014}.
%Type = Article
\bibitem[{Rolf et~al.(2022)Rolf, Weller, G{\"u}lcher, Byrne, O'Rourke, Herrick,
  Bjonnes, Davaille, Ghail, Gillmann, Plesa and
  Smrekar}]{rolfDynamicsEvolutionVenus2022}
\bibinfo{author}{Rolf, T.}, \bibinfo{author}{Weller, M.},
  \bibinfo{author}{G{\"u}lcher, A.}, \bibinfo{author}{Byrne, P.},
  \bibinfo{author}{O'Rourke, J.G.}, \bibinfo{author}{Herrick, R.},
  \bibinfo{author}{Bjonnes, E.}, \bibinfo{author}{Davaille, A.},
  \bibinfo{author}{Ghail, R.}, \bibinfo{author}{Gillmann, C.},
  \bibinfo{author}{Plesa, A.C.}, \bibinfo{author}{Smrekar, S.},
  \bibinfo{year}{2022}.
\newblock \bibinfo{title}{Dynamics and {{Evolution}} of {{Venus}}' {{Mantle
  Through Time}}}.
\newblock \bibinfo{journal}{Space Science Reviews} \bibinfo{volume}{218},
  \bibinfo{pages}{70}.
\newblock \DOIprefix\doi{10.1007/s11214-022-00937-9}.
%Type = Article
\bibitem[{Romeo and Turcotte(2010)}]{romeoResurfacingVenus2010}
\bibinfo{author}{Romeo, I.}, \bibinfo{author}{Turcotte, D.L.},
  \bibinfo{year}{2010}.
\newblock \bibinfo{title}{Resurfacing on {{Venus}}}.
\newblock \bibinfo{journal}{Planetary and Space Science} \bibinfo{volume}{58},
  \bibinfo{pages}{1374--1380}.
\newblock \DOIprefix\doi{10.1016/j.pss.2010.05.022}.
%Type = Article
\bibitem[{Rozel(2012)}]{rozelGrainSize2012}
\bibinfo{author}{Rozel, A.}, \bibinfo{year}{2012}.
\newblock \bibinfo{title}{Impact of grain size on the convection of terrestrial
  planets}.
\newblock \bibinfo{journal}{Geochemistry, Geophysics, Geosystems}
  \bibinfo{volume}{13}.
\newblock \DOIprefix\doi{10.1029/2012GC004282}.
%Type = Article
\bibitem[{Rozel et~al.(2017)Rozel, Golabek, Jain, Tackley and
  Gerya}]{rozelContinentalCrustFormation2017}
\bibinfo{author}{Rozel, A.B.}, \bibinfo{author}{Golabek, G.J.},
  \bibinfo{author}{Jain, C.}, \bibinfo{author}{Tackley, P.J.},
  \bibinfo{author}{Gerya, T.}, \bibinfo{year}{2017}.
\newblock \bibinfo{title}{Continental crust formation on early {{Earth}}
  controlled by intrusive magmatism}.
\newblock \bibinfo{journal}{Nature} \bibinfo{volume}{545},
  \bibinfo{pages}{332--335}.
\newblock \DOIprefix\doi{10.1038/nature22042}.
%Type = Article
\bibitem[{Russell and Johnson(2021)}]{russellEvidenceLocallyThinned2021}
\bibinfo{author}{Russell, M.B.}, \bibinfo{author}{Johnson, C.L.},
  \bibinfo{year}{2021}.
\newblock \bibinfo{title}{Evidence for a {{Locally Thinned Lithosphere
  Associated With Recent Volcanism}} at {{Aramaiti Corona}}, {{Venus}}}.
\newblock \bibinfo{journal}{Journal of Geophysical Research: Planets}
  \bibinfo{volume}{126}, \bibinfo{pages}{e2020JE006783}.
\newblock \DOIprefix\doi{10.1029/2020JE006783}.
%Type = Article
\bibitem[{Schierjott et~al.(2020)Schierjott, Rozel and
  Tackley}]{schierjottGrainSize2020}
\bibinfo{author}{Schierjott, J.}, \bibinfo{author}{Rozel, A.},
  \bibinfo{author}{Tackley, P.}, \bibinfo{year}{2020}.
\newblock \bibinfo{title}{On the self-regulating effect of grain size evolution
  in mantle convection models: application to thermochemical piles}.
\newblock \bibinfo{journal}{Solid Earth} \bibinfo{volume}{11},
  \bibinfo{pages}{959--982}.
\newblock \DOIprefix\doi{10.5194/se-11-959-2020}.
%Type = Article
\bibitem[{Shah et~al.(2022)Shah, Helled, Alibert and Mezger}]{shahVenus2022}
\bibinfo{author}{Shah, O.}, \bibinfo{author}{Helled, R.},
  \bibinfo{author}{Alibert, Y.}, \bibinfo{author}{Mezger, K.},
  \bibinfo{year}{2022}.
\newblock \bibinfo{title}{Possible {{Chemical Composition And Interior
  Structure Models Of Venus Inferred From Numerical Modelling}}}.
\newblock \bibinfo{journal}{The Astrophysical Journal} \bibinfo{volume}{926},
  \bibinfo{pages}{217}.
\newblock \DOIprefix\doi{10.3847/1538-4357/ac410d}.
%Type = Article
\bibitem[{Smrekar et~al.(2022)Smrekar, Ostberg and O’Rourke}]{smrekar2022}
\bibinfo{author}{Smrekar, S.E.}, \bibinfo{author}{Ostberg, C.},
  \bibinfo{author}{O’Rourke, J.G.}, \bibinfo{year}{2022}.
\newblock \bibinfo{title}{{Earth-like lithospheric thickness and heat flow on
  Venus consistent with active rifting}}.
\newblock \bibinfo{journal}{Nature Geoscience} ,
  \bibinfo{pages}{1--6}\DOIprefix\doi{10.1038/s41561-022-01068-0}.
%Type = Article
\bibitem[{Smrekar et~al.(2010)Smrekar, Stofan, Mueller, Treiman,
  {Elkins-Tanton}, Helbert, Piccioni and
  Drossart}]{smrekarRecentHotspotVolcanism2010}
\bibinfo{author}{Smrekar, S.E.}, \bibinfo{author}{Stofan, E.R.},
  \bibinfo{author}{Mueller, N.}, \bibinfo{author}{Treiman, A.},
  \bibinfo{author}{{Elkins-Tanton}, L.}, \bibinfo{author}{Helbert, J.},
  \bibinfo{author}{Piccioni, G.}, \bibinfo{author}{Drossart, P.},
  \bibinfo{year}{2010}.
\newblock \bibinfo{title}{Recent {{Hotspot Volcanism}} on {{Venus}} from
  {{VIRTIS Emissivity Data}}}.
\newblock \bibinfo{journal}{Science} \bibinfo{volume}{328},
  \bibinfo{pages}{605--608}.
\newblock \DOIprefix\doi{10.1126/science.1186785}.
%Type = Article
\bibitem[{Solomatov and Moresi(1996)}]{solomatovVenusStagnantLid1996}
\bibinfo{author}{Solomatov, V.S.}, \bibinfo{author}{Moresi, L.N.},
  \bibinfo{year}{1996}.
\newblock \bibinfo{title}{Stagnant lid convection on {{Venus}}}.
\newblock \bibinfo{journal}{Journal of Geophysical Research: Planets}
  \bibinfo{volume}{101}, \bibinfo{pages}{4737--4753}.
\newblock \DOIprefix\doi{10.1029/95JE03361}.
%Type = Article
\bibitem[{Solomon and Head(1982)}]{solomonVenusTectonics1982}
\bibinfo{author}{Solomon, S.C.}, \bibinfo{author}{Head, J.W.},
  \bibinfo{year}{1982}.
\newblock \bibinfo{title}{Mechanisms for lithospheric heat transport on venus:
  Implications for tectonic style and volcanism}.
\newblock \bibinfo{journal}{J. Geophys. Res.} \bibinfo{volume}{87},
  \bibinfo{pages}{9236--9246}.
\newblock \DOIprefix\doi{10.1029/JB087iB11p09236}.
%Type = Article
\bibitem[{Stevenson et~al.(1983)Stevenson, Spohn and
  Schubert}]{stevensonMagnetism1983}
\bibinfo{author}{Stevenson, D.J.}, \bibinfo{author}{Spohn, T.},
  \bibinfo{author}{Schubert, G.}, \bibinfo{year}{1983}.
\newblock \bibinfo{title}{Magnetism and the thermal evolution of the
  terrestrial planets}.
\newblock \bibinfo{journal}{Icarus} \bibinfo{volume}{54},
  \bibinfo{pages}{466--489}.
%Type = Article
\bibitem[{Stixrude and Lithgow-Bertelloni(2022)}]{stixrude2022}
\bibinfo{author}{Stixrude, L.}, \bibinfo{author}{Lithgow-Bertelloni, C.},
  \bibinfo{year}{2022}.
\newblock \bibinfo{title}{Thermal expansivity, heat capacity and bulk modulus
  of the mantle}.
\newblock \bibinfo{journal}{Geophysical Journal International}
  \bibinfo{volume}{228}, \bibinfo{pages}{1119--1149}.
\newblock \DOIprefix\doi{10.1093/gji/ggab394}.
%Type = Article
\bibitem[{Stofan et~al.(2005)Stofan, Brian and
  Guest}]{stofanResurfacingStylesRates2005}
\bibinfo{author}{Stofan, E.R.}, \bibinfo{author}{Brian, A.W.},
  \bibinfo{author}{Guest, J.E.}, \bibinfo{year}{2005}.
\newblock \bibinfo{title}{Resurfacing styles and rates on {{Venus}}: Assessment
  of 18 venusian quadrangles}.
\newblock \bibinfo{journal}{Icarus} \bibinfo{volume}{173},
  \bibinfo{pages}{312--321}.
\newblock \DOIprefix\doi{10.1016/j.icarus.2004.08.004}.
%Type = Article
\bibitem[{Strom et~al.(1994)Strom, Schaber and
  Dawson}]{stromGlobalResurfacingVenus1994}
\bibinfo{author}{Strom, R.}, \bibinfo{author}{Schaber, G.},
  \bibinfo{author}{Dawson, D.}, \bibinfo{year}{1994}.
\newblock \bibinfo{title}{The {{Global Resurfacing}} of {{Venus}}}.
\newblock \bibinfo{journal}{Journal of Geophysical Research-Planets}
  \bibinfo{volume}{99}, \bibinfo{pages}{10899--10926}.
\newblock \DOIprefix\doi{10.1029/94JE00388}.
%Type = Article
\bibitem[{Surkov et~al.(1984)Surkov, Barsukov, Moskalyeva, Kharyukova and
  Kemurdzhian}]{surkovNewDataComposition1984}
\bibinfo{author}{Surkov, Y.A.}, \bibinfo{author}{Barsukov, V.L.},
  \bibinfo{author}{Moskalyeva, L.P.}, \bibinfo{author}{Kharyukova, V.P.},
  \bibinfo{author}{Kemurdzhian, A.L.}, \bibinfo{year}{1984}.
\newblock \bibinfo{title}{New data on the composition, structure, and
  properties of {{Venus}} rock obtained by {{Venera}} 13 and {{Venera}} 14}.
\newblock \bibinfo{journal}{Journal of Geophysical Research: Solid Earth}
  \bibinfo{volume}{89}, \bibinfo{pages}{B393--B402}.
\newblock \DOIprefix\doi{10.1029/JB089iS02p0B393}.
%Type = Article
\bibitem[{Tackley(2000)}]{tackleySelfconsistentGenerationTectonic2000}
\bibinfo{author}{Tackley, P.J.}, \bibinfo{year}{2000}.
\newblock \bibinfo{title}{Self-consistent generation of tectonic plates in
  time-dependent, three-dimensional mantle convection simulations}.
\newblock \bibinfo{journal}{Geochemistry, Geophysics, Geosystems}
  \bibinfo{volume}{1}.
\newblock \DOIprefix\doi{10.1029/2000GC000036}.
%Type = Article
\bibitem[{Tackley(2008)}]{tackleyModellingCompressibleMantle2008}
\bibinfo{author}{Tackley, P.J.}, \bibinfo{year}{2008}.
\newblock \bibinfo{title}{Modelling compressible mantle convection with large
  viscosity contrasts in a three-dimensional spherical shell using the yin-yang
  grid}.
\newblock \bibinfo{journal}{Physics of the Earth and Planetary Interiors}
  \bibinfo{volume}{171}, \bibinfo{pages}{7--18}.
\newblock \DOIprefix\doi{10.1016/j.pepi.2008.08.005}.
%Type = Article
\bibitem[{Tackley et~al.(2013)Tackley, Ammann, Brodholt, Dobson and
  Valencia}]{tackleyMantleDynamicsSuperEarths2013}
\bibinfo{author}{Tackley, P.J.}, \bibinfo{author}{Ammann, M.},
  \bibinfo{author}{Brodholt, J.P.}, \bibinfo{author}{Dobson, D.P.},
  \bibinfo{author}{Valencia, D.}, \bibinfo{year}{2013}.
\newblock \bibinfo{title}{Mantle dynamics in super-{{Earths}}:
  {{Post-perovskite}} rheology and self-regulation of viscosity}.
\newblock \bibinfo{journal}{Icarus} \bibinfo{volume}{225},
  \bibinfo{pages}{50--61}.
\newblock \DOIprefix\doi{10.1016/j.icarus.2013.03.013}.
%Type = Article
\bibitem[{Tackley et~al.(1993)Tackley, Stevenson, Glatzmaier and
  Schubert}]{tackleyavalanches1993}
\bibinfo{author}{Tackley, P.J.}, \bibinfo{author}{Stevenson, D.J.},
  \bibinfo{author}{Glatzmaier, G.A.}, \bibinfo{author}{Schubert, G.},
  \bibinfo{year}{1993}.
\newblock \bibinfo{title}{Effects of an endothermic phase transition at 670 km
  depth in a spherical model of convection in the earth's mantle}.
\newblock \bibinfo{journal}{Nature} \bibinfo{volume}{361},
  \bibinfo{pages}{699--704}.
%Type = Incollection
\bibitem[{Tackley et~al.(2005)Tackley, Xie, Nakagawa and
  Hernlund}]{tackleyNumericalLaboratoryStudies2005}
\bibinfo{author}{Tackley, P.J.}, \bibinfo{author}{Xie, S.},
  \bibinfo{author}{Nakagawa, T.}, \bibinfo{author}{Hernlund, J.W.},
  \bibinfo{year}{2005}.
\newblock \bibinfo{title}{Numerical and {{Laboratory Studies}} of {{Mantle
  Convection}}: {{Philosophy}}, {{Accomplishments}}, and {{Thermochemical
  Structure}} and {{Evolution}}}, in: \bibinfo{booktitle}{Earth's {{Deep
  Mantle}}: {{Structure}}, {{Composition}}, and {{Evolution}}}.
  \bibinfo{publisher}{{American Geophysical Union (AGU)}}, pp.
  \bibinfo{pages}{83--99}.
\newblock \DOIprefix\doi{10.1029/160GM07}.
%Type = Article
\bibitem[{Tsujino et~al.(2022)Tsujino, Yamazaki, Nishihara, Yoshino, Higo and
  Tange}]{tsujinoRheology2022}
\bibinfo{author}{Tsujino, N.}, \bibinfo{author}{Yamazaki, D.},
  \bibinfo{author}{Nishihara, Y.}, \bibinfo{author}{Yoshino, T.},
  \bibinfo{author}{Higo, Y.}, \bibinfo{author}{Tange, Y.},
  \bibinfo{year}{2022}.
\newblock \bibinfo{title}{Viscosity of bridgmanite determined by in situ stress
  and strain measurements in uniaxial deformation experiments}.
\newblock \bibinfo{journal}{Science Advances} \bibinfo{volume}{8},
  \bibinfo{pages}{eabm1821}.
\newblock \URLprefix \url{https://doi.org/10.1126/sciadv.abm1821},
  \DOIprefix\doi{10.1126/sciadv.abm1821}.
%Type = Article
\bibitem[{Uppalapati et~al.(2020)Uppalapati, Rolf, Crameri and
  Werner}]{uppalapatiDynamicsLithosphericOverturns2020}
\bibinfo{author}{Uppalapati, S.}, \bibinfo{author}{Rolf, T.},
  \bibinfo{author}{Crameri, F.}, \bibinfo{author}{Werner, S.C.},
  \bibinfo{year}{2020}.
\newblock \bibinfo{title}{Dynamics of {{Lithospheric Overturns}} and
  {{Implications}} for {{Venus}}'s {{Surface}}}.
\newblock \bibinfo{journal}{Journal of Geophysical Research: Planets}
  \bibinfo{volume}{125}, \bibinfo{pages}{e2019JE006258}.
\newblock \DOIprefix\doi{10.1029/2019JE006258}.
%Type = Article
\bibitem[{Vesterholt et~al.(2021)Vesterholt, Petersen and
  Nagel}]{vesterholtMantleOverturnThermochemical2021}
\bibinfo{author}{Vesterholt, A.L.}, \bibinfo{author}{Petersen, K.D.},
  \bibinfo{author}{Nagel, T.J.}, \bibinfo{year}{2021}.
\newblock \bibinfo{title}{Mantle overturn and thermochemical evolution of a
  non-plate tectonic mantle}.
\newblock \bibinfo{journal}{Earth and Planetary Science Letters}
  \bibinfo{volume}{569}, \bibinfo{pages}{117047}.
\newblock \DOIprefix\doi{10.1016/j.epsl.2021.117047}.
%Type = Article
\bibitem[{Weller and Kiefer(2020)}]{wellerPhysicsChangingTectonic2020}
\bibinfo{author}{Weller, M.B.}, \bibinfo{author}{Kiefer, W.S.},
  \bibinfo{year}{2020}.
\newblock \bibinfo{title}{The {{Physics}} of {{Changing Tectonic Regimes}}:
  {{Implications}} for the {{Temporal Evolution}} of {{Mantle Convection}} and
  the {{Thermal History}} of {{Venus}}}.
\newblock \bibinfo{journal}{Journal of Geophysical Research: Planets}
  \bibinfo{volume}{125}, \bibinfo{pages}{e2019JE005960}.
\newblock \DOIprefix\doi{10.1029/2019JE005960}.
%Type = Incollection
\bibitem[{Wieczorek(2015)}]{wieczorekGravityTopographyTerrestrial2015}
\bibinfo{author}{Wieczorek, M.A.}, \bibinfo{year}{2015}.
\newblock \bibinfo{title}{Gravity and {{Topography}} of the {{Terrestrial
  Planets}}}, in: \bibinfo{editor}{Schubert, G.} (Ed.),
  \bibinfo{booktitle}{Treatise on {{Geophysics}} ({{Second Edition}})}.
  \bibinfo{publisher}{{Elsevier}}, \bibinfo{address}{{Oxford}}, pp.
  \bibinfo{pages}{153--193}.
\newblock \DOIprefix\doi{10.1016/B978-0-444-53802-4.00169-X}.
%Type = Article
\bibitem[{Xie and Tackley(2004)}]{xieEvolutionUPbSmNd2004}
\bibinfo{author}{Xie, S.}, \bibinfo{author}{Tackley, P.J.},
  \bibinfo{year}{2004}.
\newblock \bibinfo{title}{Evolution of {{U-Pb}} and {{Sm-Nd}} systems in
  numerical models of mantle convection and plate tectonics}.
\newblock \bibinfo{journal}{Journal of Geophysical Research: Solid Earth}
  \bibinfo{volume}{109}.
\newblock \DOIprefix\doi{10.1029/2004JB003176}.
%Type = Article
\bibitem[{Xu et~al.(2008)Xu, Lithgow-Bertelloni, Stixrude and Ritsema}]{Xu2008}
\bibinfo{author}{Xu, W.}, \bibinfo{author}{Lithgow-Bertelloni, C.},
  \bibinfo{author}{Stixrude, L.}, \bibinfo{author}{Ritsema, J.},
  \bibinfo{year}{2008}.
\newblock \bibinfo{title}{The effect of bulk composition and temperature on
  mantle seismic structure}.
\newblock \bibinfo{journal}{Earth and Planetary Science Letters}
  \bibinfo{volume}{275}, \bibinfo{pages}{70--79}.
\newblock \DOIprefix\doi{10.1016/j.epsl.2008.08.012}.
%Type = Article
\bibitem[{Yamazaki and Karato(2001)}]{yamazakiMineralPhysicsConstraints2001}
\bibinfo{author}{Yamazaki, D.}, \bibinfo{author}{Karato, S.i.},
  \bibinfo{year}{2001}.
\newblock \bibinfo{title}{Some mineral physics constraints on the rheology and
  geothermal structure of {{Earth}}'s lower mantle}.
\newblock \bibinfo{journal}{American Mineralogist} \bibinfo{volume}{86},
  \bibinfo{pages}{385--391}.
\newblock \DOIprefix\doi{10.2138/am-2001-0401}.
%Type = Article
\bibitem[{Yan et~al.(2020)Yan, Ballmer and
  Tackley}]{yanEvolutionDistributionRecycled2020}
\bibinfo{author}{Yan, J.}, \bibinfo{author}{Ballmer, M.D.},
  \bibinfo{author}{Tackley, P.J.}, \bibinfo{year}{2020}.
\newblock \bibinfo{title}{The evolution and distribution of recycled oceanic
  crust in the {{Earth}}'s mantle: {{Insight}} from geodynamic models}.
\newblock \bibinfo{journal}{Earth and Planetary Science Letters}
  \bibinfo{volume}{537}, \bibinfo{pages}{116171}.
\newblock \DOIprefix\doi{10.1016/j.epsl.2020.116171}.
%Type = Article
\bibitem[{Zerr et~al.(1998)Zerr, Diegeler and
  Boehler}]{zerrSolidusEarthDeep1998}
\bibinfo{author}{Zerr, A.}, \bibinfo{author}{Diegeler, A.},
  \bibinfo{author}{Boehler, R.}, \bibinfo{year}{1998}.
\newblock \bibinfo{title}{Solidus of {{Earth}}'s {{Deep Mantle}}}.
\newblock \bibinfo{journal}{Science} \bibinfo{volume}{281},
  \bibinfo{pages}{243--246}.
\newblock \DOIprefix\doi{10.1126/science.281.5374.243}.
%Type = Article
\bibitem[{Čížková et~al.(2012)Čížková, van~den Berg, Spakman and
  Matyska}]{cizkova2012}
\bibinfo{author}{Čížková, H.}, \bibinfo{author}{van~den Berg, A.P.},
  \bibinfo{author}{Spakman, W.}, \bibinfo{author}{Matyska, C.},
  \bibinfo{year}{2012}.
\newblock \bibinfo{title}{The viscosity of earth’s lower mantle inferred from
  sinking speed of subducted lithosphere}.
\newblock \bibinfo{journal}{Physics of the Earth and Planetary Interiors}
  \bibinfo{volume}{200–201}, \bibinfo{pages}{56--62}.
\newblock \DOIprefix\doi{10.1016/j.pepi.2012.02.010}.

\end{thebibliography}





% \end{thebibliography}
\end{document}